\documentclass[doublespacing]{bmcart}

\usepackage[utf8]{inputenc} 
\usepackage{graphicx}
\usepackage{subfigure}
\usepackage{multirow} 
\usepackage[sort]{natbib}
\usepackage{multicol}
\usepackage{algorithm} 
\usepackage{algpseudocode}
\usepackage{amssymb,amsmath,mathtools}

\startlocaldefs
\endlocaldefs

\begin{document}

\begin{frontmatter}

\begin{fmbox}
\dochead{Research}


\title{Compressive Closeness in Networks}


\author[
   addressref={aff1},
   corref={aff1},
   noteref={n1},
   email={hmahyar@bu.edu}
]{\inits{H}\fnm{Hamidreza} \snm{Mahyar}}
\author[
   addressref={aff2},
   noteref={n1},
   email={shashemi@student.ethz.ch}
]{\inits{R}\fnm{Rouzbeh} \snm{Hasheminezhad}}
\author[
   addressref={aff1},
   email={hes@bu.edu}
]{\inits{HE}\fnm{H Eugene} \snm{Stanley}}


\address[id=aff1]{%
  \orgname{Boston University},
  \city{Boston},
  \cny{USA}
}
\address[id=aff2]{%
  \orgname{ETH},
  \city{Zurich},
  \cny{Switzerland}
}


\begin{artnotes}
\note[id=n1]{Equal contribution} 
\end{artnotes}

\end{fmbox}


\begin{abstractbox}

\begin{abstract} 
Distributed algorithms for network science applications are of great importance due to today's large real-world networks. In such algorithms, a node is allowed only to have local interactions with its immediate neighbors. This is because the whole network topological structure is often unknown to each node. Recently, distributed detection of central nodes, concerning different notions of importance, within a network has received much attention. Closeness centrality is a prominent measure to evaluate the importance (influence) of nodes, based on their accessibility, in a given network. In this paper, first, we introduce a local (ego-centric) metric that correlates well with the global closeness centrality; however, it has very low computational complexity. Second, we propose a compressive sensing (CS)-based framework to accurately recover high closeness centrality nodes in the network utilizing the proposed local metric. Both ego-centric metric computation and its aggregation via CS are efficient and distributed, using only local interactions between neighboring nodes. Finally, we evaluate the performance of the proposed method through extensive experiments on various synthetic and real-world networks. The results show that the proposed local metric correlates with the global closeness centrality, better than the current local metrics. Moreover, the results demonstrate that the proposed CS-based method outperforms the state-of-the-art methods with notable improvement.
\end{abstract}




\end{abstractbox}
%

\end{frontmatter}



\section{Introduction}\label{Introduction}
Many real-world systems can be modeled by a network $G=(V, E)$ of interacting actors. The actors are demonstrated by a set of nodes $V$ with cardinality $|V|$ that are connected via the set of edges (links) $E$ with cardinality $|E|$. The edges can be directed or undirected, depending on the type of interactions. Some well-known examples of such real-world systems include technological and transportation infrastructures, communication systems, biological networks, and social interactions. Centrality measures are means of quantifying the importance of a node within the given network. Some notions of centrality only consider local properties of the network; however some of them reflect global properties. Appropriate quantification of importance should be done given the application context. To address applications in which reachability of a node to the entire network is of importance, researchers have introduced the \textit{closeness centrality} measure. For an arbitrary node $u$, its closeness centrality $C(u)$ is defined as the inverse of its average distance to the other nodes in the network. More formally:
\begin{eqnarray}\label{Closeness}
C(u)=\frac{|V|-1}{\sum_{v\neq u \in V} d(u,v)}
\end{eqnarray}
where $d(u,v)$ is the shortest distance between $u$ and $v$. Locating public facilities over a transportation network such that they are easily accessible to everyone or identifying people with ideal social network location for information dissemination or network influence can be mentioned as scenarios in which identifying high closeness centralities is of great interest~\cite{SaxenaGI17,taheri2017hellrank,taheri2017extracting}. In these scenarios, we are mainly interested in efficiently and accurately detecting top-$k$ high closeness centrality nodes in the network, while their exact relative order compared to each other, as well as the actual closeness centrality values,  are not so important.

A trivial approach to identify top-$k$ closeness centrality nodes consists of the following steps: (1) Utilizing breadth-first search (BFS) for calculating closeness centrality for each node in $O\left(|V|+|E|\right)$  with a total computational cost of $O\left(|V||E|+|V|^2\right)$; (2) Sorting the computed values via a sorting algorithm in $O\left(|V|\log(|V|)\right)$, then report the top-$k$ nodes. The high computational cost of $O(|V||E|+|V|^2)$ and the requirement of full knowledge of the network topology may prevent such a method from being applied on large real-world networks~\cite{wehmuth2012distributed}. To address this issue, developing scalable distributed algorithms is of great importance, where each node is only interacting with its immediate neighbors~\cite{you2017distributed}. 

To the best of our knowledge, there is no distributed and decentralized algorithm for the task of detecting top-$k$ high closeness centrality nodes that operates while requiring each node only to have local interactions with its immediate neighbors.\newpage However, several algorithms are satisfying these properties and compute exact or approximated closeness centrality of each node in the network. Approximation approaches compute an alternative centrality score that highly correlates with the global closeness centrality. An efficient sorting algorithm can then be utilized on top of these methods to identify top-$k$ high closeness centrality nodes.
There are two major shortcomings with such approaches: (1) Not exploiting the fact that the vector consisting of closeness centrality values has a few large coefficients ($k$) and many small coefficients so that it can be well approximated by a $k$-sparse vector (signal). In general, a centrality measure (\textit{e.g.} closeness centrality) must have a right-skewed probability distribution to be useful in selecting important nodes. (2) Requiring \textit{direct measurement} (query) from each node, which is not always possible due to log-in requirements, API query limits, and treating user data as proprietary.

To address these issues, we transform the problem of detecting top-$k$ closeness central nodes to the problem of sparse recovery in networks. The breakthrough of the sparse recovery problem is compressive sensing (aka compressive sampling) which performs a few indirect end-to-end measurements on a signal $x$ and recovers a good sparse approximation of that signal. However, two additional requirements must be taken into account when these measurements are performed over a graph, rather than an arbitrary signal. Creating feasible measurements that satisfy these constraints (will be discussed in section~\ref{CompressiveSensingOverNetworks}) has initiated the field of compressive sampling over graphs.

Our contributions in this paper are two-fold: (1) We propose a local (ego-centric) metric which can be computed in a distributed manner at each node. The computation can be carried out requiring each node to have only local knowledge of its immediate neighborhood. In section~\ref{experiments}, we experimentally show that the suggested local metric is highly correlated with the global closeness centrality on many real-world and synthetic networks. (2) We propose a general compressive sensing framework for distributed identification of central nodes in networks based on the introduced local metric using indirect end-to-end (aggregated) measurements. We experimentally show the superiority of our approach in terms of accuracy for the prediction of high closeness central nodes compared to the best existing competing methods.

The rest of this paper is organized as follows.  In section~\ref{Prelim}, we briefly explain the preliminary notations and definitions. We review the related works on distributed detection of central nodes requiring only local interactions with the neighbors from each node, in section~\ref{related}. In section~\ref{proposed}, we introduce our novel approach in detail and analyze its time and space complexity. Later in section ~\ref{experiments}, the settings and results of our experimental evaluations are presented. We conclude the paper in section~\ref{conclusions}.

A preliminary version of this paper has appeared in~\cite{mahyar2018closeness}. Here, we explain the backgrounds and the intuitions behind the idea in more details. Also, we comprehensively review the related work and describe their limitations with our corresponding solutions. Moreover, we add three different types of real datasets and a several test scenarios to our extensive experimental evaluations in order to show the generalization of the proposed method.

\section{Preliminaries}
\label{Prelim}
\subsection{Compressive Sampling}
As an alternative to direct measurements, one can utilize sampling-based approaches. Based on the Nyquist-Shannon theorem, a general signal $x$ can be completely recovered by sampling it with the Nyquist rate. However, sampling with the Nyquist rate can be costly or impossible due to a massive scale in many real-world networks we are facing today. If the underlying signal is sparse in a suitable basis, sampling with the Nyquist rate only to recover a relatively small fraction of non-zero elements results in loss of system resources and induces two sources of error, sampling (collection) error and identification (compression) error.

The state-of-the-art approach for recovery of sparse signals is Compressive Sensing/Sampling (CS) which addresses these drawbacks. In compressive sampling, one can simultaneously sample and compress a signal $x_{n\times 1}$ through a measurement matrix $\mathcal{A}_{m\times n}$ where $m \ll n$ to acquire the following linear system:
\begin{eqnarray}
y_{m\times 1}=\mathcal{A}_{m\times n}~x_{n\times 1}
\end{eqnarray}
The resulting system is under-determined and does not have a unique solution in general.
$\mathcal{A}$ is said to satisfy the $2k$-restricted isometry property (RIP) if there exists $0<\delta_{2k}<1$, such that for all $2k$-sparse signals $x'$, it holds:
\begin{eqnarray}\label{rip}
(1-\delta_{2k})||x'||_2 \leq ||Ax'||_2 \leq (1+\delta_{2k})||x'||_2
\end{eqnarray}

In case the measurement matrix satisfies the $2k$-RIP one can prove uniqueness of a $k$-sparse solution to the above linear system ($y=\mathcal{A}x$). To see this, assume $x_1$ and $x_2$ are both $k$-sparse signals and $\mathcal{A}x_1=\mathcal{A}x_2$, so vector $x'=x_1-x_2$ is a $2k$-sparse signal (has at most $2k$ non-zero entries). Since $\mathcal{A}$ satisfies the $2k$-RIP, Equation (\ref{rip}) can be rewritten for some $0<\delta'_{2k}<1$ which ensures $x_1=x_2$, as:
\begin{eqnarray}
(1-\delta'_{2k})||x_1-x_2||_2 \leq 0 \leq (1+\delta'_{2k})||x_1-x_2||_2
\end{eqnarray} 
Let $x^*$ be any arbitrary $k$-sparse vector, and $\mathcal{A}$ be an arbitrary measurement matrix that satisfies the $2k$-RIP property. Then given what we have discussed so far, it is easy to see that $x^*$ can be recovered by solving:
\begin{eqnarray}\label{NP-hard}
\min_{x} \Vert x \Vert_0 ~~~\text{s.t.}~~~ y = \mathcal{A} x
\end{eqnarray}
where $\Vert x \Vert_0$ indicates the number of non-zero entries in $x$. Unfortunately, solving this optimization problem is NP-hard. 
Thus the following relaxation is considered which utilizes the sparsity inducing $\ell_1$-norm and is referred to as Basis Pursuit (BP):
\begin{eqnarray}
\label{BP}
\min_{x} \Vert x \Vert_1 ~~~\text{s.t.}~~~ y = \mathcal{A} x
\end{eqnarray}

It has been shown when the $2k$-restricted isometry is satisfied for $\mathcal{A}$, the solution of BP is $x^*$. In this case, by utilizing the convexity of BP, the recovery is very efficient and computationally fast. Note that the strict condition $y=\mathcal{A}x$ within the Basis Pursuit formulation is very sensitive to imperfect sparsity or noise. The following formulation, known as LASSO, addresses this by removing the exact constraint and penalizing its violation:
\begin{eqnarray}
\label{LASSO}
\min_{x} \Vert x \Vert_1 + \Vert \mathcal{A} x - y \Vert_{2}^{2}
\end{eqnarray}
This objective has extremely fast distributed numerical solvers and will be utilized for the optimization step in this paper.

\subsection{Compressive Sensing over Networks}
\label{CompressiveSensingOverNetworks}
In case the signal to be recovered is defined over a graph (network), three additional constraints must be taken into account~\cite{xu2011related,mahyar2013ucsnt} in CS problems: (1) Each element $\mathcal{A}_{i,j}$ would be $1$ if the node $j$ is visited by measurement $i$ and $0$ otherwise; (2) The nodes visited by a measurement must correspond to a connected induced sub-graph~\cite{Mahyar2017MLG,Mahyar2015CScomdet,mahyar2018compressive,Mahyar2017ICML}; (3) The signal $x$ which contains a graph property, defined for each node, is almost always non-negative ($x\geq0$). 

Based on the compressive sensing framework, we would like to efficiently recover $k$ highest closeness centrality nodes from $m$ indirect end-to-end measurements, in a way that $m \ll n$.
In the linear system $y_{m \times 1} = \mathcal{A}_{m \times n} ~x_{n \times 1}$, let $\mathcal{A}$ be an $m \times n$ measurement matrix, where its $i$-th row corresponds to the $i$-th feasible measurement. For $i = 1, ..., m$ and $j = 1, ..., n$, $\mathcal{A}_{ij} = 1$ if and only if node $j$ is visited by the $i$-th measurement, otherwise $\mathcal{A}_{ij} = 0$. Let $x$ be an $n \times 1$ non-negative vector whose $j$-th entry is the value of a certain type of network characteristic (\textit{e.g.} a global/local centrality metric) over node $j \in V$, and $y \in \mathcal{R}^m$ denotes the measurements vector whose $i$-th entry represents the additive aggregation values of network nodes in the $i$-th row of the measurement matrix $\mathcal{A}$ that induces a \textit{connected sub-graph} over $G$. Note that this way of measurements construction already satisfies the network topological constraints of the feasibility conditions mentioned in the beginning of this section.

\begin{figure}[t!]
  \centering
   \includegraphics[trim = 0mm 80mm 40mm 0mm, scale=0.45]{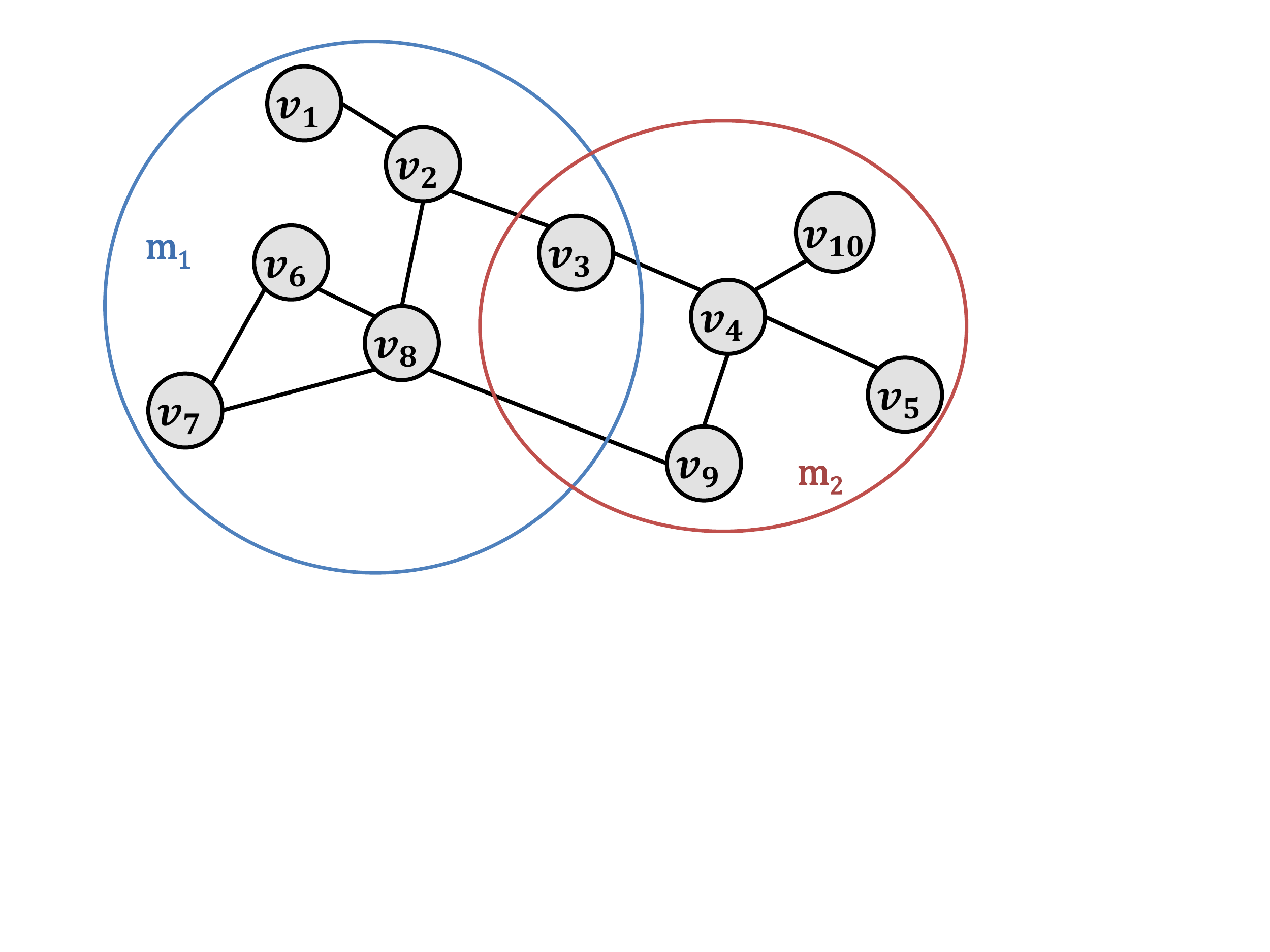}
  \caption{A network with 10 nodes and 11 links. The measurements $m_1$ and $m_2$ are feasible considering the network topological constraints (each of them induces a connected sub-graph over the network).}
  \label{nettopol}
\vspace{-5pt}
\end{figure}

For the example network shown in Figure~\ref{nettopol} with $n=10$ nodes and $\vert E \vert = 11$ links, each of two measurements $m_1$ and $m_2$ includes a different subset of connected nodes. The corresponding feasible measurement matrix $\mathcal{A}$ with these measurements is:
\vspace{-5pt}
\begin{equation}
\label{measurementmatrix}
\mathcal{A} = \bordermatrix{~ & v_1 & v_2 & v_3 & v_4 & v_5 & v_6 & v_7 & v_8 & v_9  & v_{10}\cr
m_1 & 1 & 1 & 1 & 0 & 0 & 1 & 1 & 1 & 0 & 0 \cr
m_2 & 0 & 0 & 1 & 1 & 1 & 0 & 0 & 0 & 1 & 1\cr
}
\end{equation}

To understand how the additive aggregation over connected induced sub-graphs is motivated for each measurement in practice, we mention an example from~\cite{wang2012related}. Consider a network where the nodes represent sensors, and the links represent communications between sensors. For the set $T$ of active nodes within an arbitrary feasible measurement that induce a connected sub-graph, a node $u \in T$ monitors the total values corresponding to nodes in $T$. Every node in $T$ obtains values from its children, if any, and aggregates them with its value on the spanning tree rooted at $u$, then sends the sum to its parent. After that, the fusion center can obtain the sum of values corresponding to all the nodes in $T$ by only communicating with $u$. The explained paradigm in data acquisition and aggregation is highly utilized within the wireless sensor network literature for applications such as air quality monitoring, volcanic activity detection, and object localization~\cite{middya2017compressive}. Some recent work has applied a similar acquisition and aggregation paradigm in network tomography~\cite{mahyar2013ucsnt}, community detection~\cite{Mahyar2015CScomdet} and finding key actors in social networks~\cite{Mahyar2015TopK, Mahyar2015LSRweighted,grosu2018compressed}.

Based on the above idea, a straight forward approach utilized in practice to construct measurement matrices satisfying these properties, is to create a correspondence between every single measurement and a \textit{random walk} on the graph. Each random walk additively aggregates values computed by the nodes during the walk. The random walk strategy and the values computed by the nodes are what separate a method from the others. Performance of these methods and RIP satisfaction can then be verified theoretically or experimentally~\cite{mahyar2018closeness, Mahyar2015TopK,mahyar2018compressive,xu2011related}. An alternative approach~\cite{mahyar2018dicenod} employs a well-known randomized method in compressive sensing literature which satisfies the restricted isometry property with very high probability and makes deriving theoretical recovery guarantees straightforward. Also, it is possible to show that each constructed measurement will almost surely correspond to an induced connected sub-graph.

\section{Related Work}\label{related}
In this section, we first review local metrics that highly correlate with the global closeness centrality and can be computed in a distributed manner relying only on interactions of neighboring nodes. After that, we review compressive sensing (CS)-based methods that can be utilized to recover top-$k$ central nodes, using the mentioned local metrics by constructing a feasible measurement matrix.

\subsection{Local Closeness Metrics}\label{RelatedLocalMetric}
\textbf{Dist-Exact~\cite{you2017distributed}:}
They proposed a distributed method to compute and collect the set of nodes with an exact distance of $h$ from an arbitrary node $u$. The parameter $h$ varies from $1$ to $\mathcal{D}$, where $\mathcal{D}$ denotes the diameter of the network. The collected sets can then be utilized to compute the closeness centrality at each node.

\noindent\textbf{Dist-Est~\cite{wang2015distributed}:}
They derived a set of affine constraints which are distributed in nature and characterize closeness centrality according to its original definition. The derived constraints are used to develop an algorithm, which enables nodes in a network to cooperatively estimate their closeness centrality.

\noindent\textbf{DACCER~\cite{wehmuth2012distributed}:}
Let $\text{vol}_h(u)$ denote the sum of degrees for all nodes in the $h$-hop neighborhood of $u$. In this work, the authors showed a high correlation between $\text{vol}_h(u), \forall u \in V$ and the closeness centrality distribution for $h>0$. The correlation is shown to become stronger as $h$ grows.

\noindent\textbf{Weight-Vol~\cite{Kim2012WeightedVol}:}
This work was an extension to the metric in DACCER, based on two simple observations. First, closer nodes to a node have more contributions than farther nodes in the dissemination of the node's information. Second, the nodes with low clustering coefficients are hubs linking neighboring network parts.

\subsection{CS-based Methods for Data Aggregation}
\textbf{RW~\cite{xu2011related}:} This work is one of the state-of-the-art method in compressive sensing over graphs that constructs random-walk based measurements. Each measurement in the measurement matrix can be used to aggregate a metric of choice additively.

\noindent\textbf{TopCent~\cite{Mahyar2015TopK}:} This method constructs a measurement matrix to recover top-$k$ degree central nodes in networks. Since degree centrality is highly correlated with the closeness centrality in some real-world networks, this method is expected to perform well for the task of detecting closeness centralities, as well.

\noindent\textbf{DICeNod~\cite{mahyar2018dicenod}:} This approach does not perform walks to create a measurement matrix, instead it utilizes a well-known randomized matrix construction technique in compressive sensing. They showed that the constructed measurements correspond to induced connected sub-graphs in networks with high probability.

\section{Proposed Method}\label{proposed}
In this section, we introduce the proposed framework in the following steps: (1) defining a new ego-centric centrality measure; (2) introducing a subroutine, called \textsc{CS-HiClose-ScoreCompute}, which calculates the proposed ego-centric centrality metric in a distributed and decentralized manner; (3) introducing a subroutine, called \textsc{CS-HiClose-Aggregate}, which aggregates the local scores via decentralized measurements construction in compressive sensing. This will be executed only after the execution of the previous subroutine; and (4) analyzing the overall time and space complexity of the proposed approach. The pseudo-code of the proposed approach, \textsc{CS-HiClose}, is in Algorithm~\ref{alg:CS-HiClose}, which mainly calls the two subroutines mentioned in steps (2) and (3).
\begin{algorithm}[t!]
\renewcommand{\algorithmicrequire}{\textbf{Input:}}
\renewcommand{\algorithmicensure}{\textbf{Output:}}
\renewcommand{\algorithmicforall}{\textbf{Foreach}}
\newcommand*\rfrac[2]{{}^{#1}\!/_{#2}}
\caption{The Proposed Method: \textsc{CS-HiClose}}
\label{alg:CS-HiClose}
\begin{algorithmic}
    \Require $V, m, l, h$
    \State $V$: set of network nodes
    \State $m$: number of required measurements
    \State $l$: measurements length
    \State $h$: neighbourhood radius size at each node
    \State $\textsc{CS-HiClose-ScoreCompute}(V,h)$
    \State $\hat{x}=\textsc{CS-HiClose-Aggregate}(V,m,l)$
 \Ensure sparse approximation $\hat{x}$
\end{algorithmic}
\end{algorithm}

\subsection{Proposed Local Metric}\label{ProposedLocalMetric}
We introduce the $h$-hop ego-centric (local) closeness centrality of node $v$ as:
\begin{equation}\label{egoCloseness}
egoC_h(v)=\sum_{\tau=1}^h |B_\tau(v)|/\tau    
\end{equation}
where $B_\tau(v)$ indicates the set of nodes that have an exact shortest distance of length $\tau$ from node $v$. The intuition behind this metric is that, the farther nodes from $v$ have lower effect in dissemination of goods (\textit{e.g.} information) emerged from it.

\subsection{Score Computation Subroutine}
The computation of the sets $B_\tau(v)$ for $\tau\leq h, \forall v\in V$ can be done by executing a breadth-first search (BFS) process at each node in parallel, with exploration radius of $h$. This will require computational cost of at most $O(\Delta^h)$ where $\Delta$ is the maximum degree of the network. The required memory storage at each node is also $O(\Delta^h)$. The computed sets can be utilized to evaluate ego closeness centrality at each node in a distributed and decentralized manner, with $O(1)$ computational and storage cost per node.  Thus we will have the following steps for ego-closeness computation:
\begin{enumerate}
\item[$\MakeLowercase{(\romannumeral 1)}$] For each node $v \in V$ in the network, run $BFS_h(v)$ to calculate the number of nodes in its $i$-hop neighborhood denoted as $B_i(v)$ where $i$ ranges from $1$ to $h$. This step can be executed in a decentralized manner for each node independently from the others.
\item[$\MakeLowercase{(\romannumeral 2)}$] Once $B_i(v)$ is available for each node $v\in V$, $i$ ranging from $1$ to $h$, one can easily compute the ego-closeness centrality metric based on Equation (\ref{egoCloseness}). This step can be also executed in a decentralized fashion for each node independently. The pseudo-code for this subroutine is in Algorithm~\ref{Compute}.
\end{enumerate}
\begin{algorithm}[t!]
\renewcommand{\algorithmicrequire}{\textbf{Input:}}
\renewcommand{\algorithmicensure}{\textbf{Output:}}
\renewcommand{\algorithmicforall}{\textbf{Foreach}}
\newcommand*\rfrac[2]{{}^{#1}\!/_{#2}}
\caption{$\textsc{CS-HiClose-ScoreCompute}(V,h)$}
\label{Compute}
\begin{algorithmic}
    \Require $V, h$
    \State $V$: set of network nodes
    \State $h$: neighborhood radius size at each node
    \ForAll{$v \in V$}
    \Comment In a distributed manner
    \State Calculate $BFS_h(v)$ to initialize $B_{i}(v)$ for $i=1\dots h$
    \State $egoC_h(v) = \sum_{\tau=1}^h |B_{\tau}(v)|/\tau$   
    \EndFor
 \Ensure For each node $v\in V$, its $h$-hop ego-centric measure $egoC_h(v)$ is computed
\end{algorithmic}
\end{algorithm}

\subsection{Score Aggregation Subroutine}\label{ProposedAlg}
 The proposed compressive sensing-based method for aggregating the computed ego-centric metric is depicted in Algorithm~\ref{Aggregate}, which contains fours steps:
\begin{algorithm}[t!]
\renewcommand{\algorithmicrequire}{\textbf{Input:}}
\renewcommand{\algorithmicensure}{\textbf{Output:}}
\renewcommand{\algorithmicforall}{\textbf{Foreach}}
\newcommand*\rfrac[2]{{}^{#1}\!/_{#2}}
\caption{$\textsc{CS-HiClose-Aggregate}(V,m,l)$}
\label{Aggregate}
\begin{algorithmic}
    \Require $V, m, l$
    \State $V$: set of network nodes
    \State $m$: number of required measurements
    \State $l$: measurements length
    \State $\mathcal{A} = \mathbf{0}_{m\times n}$ 
    \State $y = \mathbf{0}_{m \times 1}$
\For{$i = 1 \to m$}
    \Comment In a distributed manner
    \State Choose $v_{first}$ uniformly at random from $V$
    \State $S = \lbrace v_{first} \rbrace$ 
    \State $\mathcal{N}(S) = \mathcal{N}(v_{first})$ 
    \State $\mathcal{A} [i,v_{first}] = 1$
    \State $y [i] = egoC_h(v_{first})$ 
    \For{$j = 1 \to l$}
    \State Choose $v_{next}$ relative to $egoC_h(v_{next})$ from $\mathcal{N}(S)$
    \State $S=S\bigcup \{v_{next}\}$ 
    \State $\mathcal{N}(S)=\mathcal{N}(S)\setminus \{v_{next}\}$
    \State $\mathcal{N}(S)=\mathcal{N}(S) \bigcup \mathcal{N}(v_{next})$
    \State $\mathcal{A} [i,v_{next}] = 1$
    \State $y [i] = y [i]+egoC_h(v_{next})$ 
    \EndFor
    \EndFor
     \State $\hat{x} = \min\limits_{x} \Vert x \Vert_1 + \Vert \mathcal{A} x - y \Vert_{2}^{2}$
     \Comment{See Equation~(\ref{LASSO})}
 \Ensure sparse approximation $\hat{x}$
\end{algorithmic}
\end{algorithm}

\begin{enumerate}
\item[$\MakeLowercase{(\romannumeral 1)}$] The first node $v_{first}$ is added to the visited set $S$ and all of its neighbors are added to the neighbor set $\mathcal{N}(S)$. 
\item[$\MakeLowercase{(\romannumeral 2)}$] The next node is selected relative to $egoC_h(v_{next})$ from the nodes in $\mathcal{N}(S)$, which are already computed in the previous subroutine. 
\item[$\MakeLowercase{(\romannumeral 3)}$] The selected next node is added to the visited set $S$ and it is removed from the neighbor set $\mathcal{N}(S)$, then its neighbors are added to the neighbor set $\mathcal{N}(S)$. 
\item[$\MakeLowercase{(\romannumeral 4)}$] The steps $\MakeLowercase{(\romannumeral 1)}- \MakeLowercase{(\romannumeral 3)}$ are fulfilled \lq $l$\rq ~times which is the length of a measurement, to generate a new row for the matrix $\mathcal{A}$ and the vector $y$.
\item[$\MakeLowercase{(\romannumeral 5)}$] Step $\MakeLowercase{(\romannumeral 4)}$ is repeated \lq $m$\rq ~times (in parallel) to construct a feasible measurement matrix $\mathcal{A}$ with \lq $m$\rq ~measurements 
 and the corresponding measurement vector $y$. 
\item[$\MakeLowercase{(\romannumeral 6)}$] To find the sparse approximation $\hat{x}$ of $x$, we optimize the LASSO objective function subject to the linear sketch of $y = \mathcal{A} x$, based on Equation~(\ref{LASSO}).
\end{enumerate}

In this algorithm, we have $m$ parallel aggregation processes, where each is to be started from a node selected uniformly at random from $V$. The random seeds to choose the starting point of each aggregating process can be fixed in time $O(m)$. A measurement corresponding to a process with a starting node will keep track of two sets $S$ and $\mathcal{N}(S)$. The set $S$ is initialized with $v_{first}$ and the set $\mathcal{N}(S)$ is initialized by its immediate neighbors, denoted by $\mathcal{N}(v_{first})$. Within $l$ sequential iterations, a candidate $v_{next}$ from $\mathcal{N}(S)$ will be selected relative to $egoC_h(v_{next})$, removed from $\mathcal{N}(S)$ and added to $S$. Moreover the neighbors of $v_{next}$ that are not already present in $\mathcal{N}(S)$ will be added to $\mathcal{N}(S)$. In other words $S$ is the set of visited nodes and $\mathcal{N}(S)$ is the set of candidate nodes that are not in $S$ but are connected to some node(s) in $S$. This ensures that the set of visited nodes $S$ at each single iteration corresponds to an induced connected sub-graph from the network. At iteration $i$ of total $l$ iterations, the maximum size of $\mathcal{N}(S)$ is $\min(i\Delta-i,|V|)$, thus selection of a member from $\mathcal{N}(S)$ relative to ego-closeness centralities using a binary search will be possible with computational cost of $\log\left(\min(i\Delta-i,|V|)\right)$. The total cost of applying this binary search method is $O(|V| \log (|V|))$ in total. To show this, we consider two different cases. If $l \le \left\lfloor {\frac{|V|}{{\Delta - 1}}} \right\rfloor$, then:
\begin{align*}
\sum_{i=1}^l \log\left(\min(i\Delta,|V|)\right) =&\sum\limits_{i = 1}^l {\log \big( i \Delta - i \big)}= \log \big(l ! \big) + l \log \big(\Delta - 1 \big) \\
\le& ~l\log(l) + l\log(\Delta )\le 2 |V| \log(|V|)
\end{align*}
Otherwise, if $l > \left\lfloor {\frac{|V|}{{\Delta - 1}}} \right\rfloor$, then:
\begin{align*}
&\sum_{i=1}^l \log\left(\min(i\Delta,|V|)\right)= \sum\limits_{i = 1}^{\left\lfloor {\frac{|V|}{{\Delta - 1}}}\right\rfloor}{\log\big(i\Delta- i\big)} + \big(|V|-\left\lfloor{\frac{|V|}{{\Delta-1}}}\right\rfloor\big)\log(|V|) \\
= & ~\log\big(\left\lfloor {\frac{|V|}{{\Delta - 1}}} \right\rfloor!\big)+ \left\lfloor {\frac{|V|}{{\Delta - 1}}} \right\rfloor \log \big(\Delta - 1 \big) +
\big( |V| - \left\lfloor {\frac{|V|}{{\Delta - 1}}} \right\rfloor \big) \log (|V|) \\
\le & ~\left\lfloor {\frac{|V|}{{\Delta - 1}}} \right\rfloor \big( \log \big( \frac{|V|}{{\Delta - 1}} \big) + \log \big( \Delta - 1\big)\big) +\big( |V| - \left\lfloor {\frac{|V|}{{\Delta - 1}}} \right\rfloor \big) \log (|V|) \\
= & ~|V| \log (|V|)
\end{align*}

Moreover, the number of deletions from and additions to $\mathcal{N}(S)$ are at most $|V|$. Each addition/deletion operation can be done efficiently in $O(1)$, using an array structure. Thus, the total time complexity for the aggregating stage is $O(m+|V|\log(|V|)+|V|)=O(|V|\log(|V|))$, where we have assumed $m \ll |V|$ aggregating processes (measurements). The required space for each aggregating process is $O(l)$ to save the visited nodes, and $O(1)$ for saving the aggregated values of the visited nodes. Also, a space of at most $O(|V|)$ is required for keeping track of the lists $S$ and $\mathcal{N}(S)$. Finally, global space storage of size $O(m)$ is needed to save the initial measurements seeds.

\subsection{The Complexity Analysis of \textsc{CS-HiClose}}
Overall, our approach requires a running time of $O(|V|\log(|V|)+\Delta^h)$, local storage of $O(\Delta^h)$ at each node and global storage of size $O(m)$ for the seeds. Besides, a local storage space of $O(l)$ is required for each aggregating process (measurement). In the next section, we will show a high correlation between the proposed ego-centric centrality with $h=2$ and the global closeness centrality of the nodes in various networks. The experiments indicate that one does not gain much more correlation by increasing $h$ to some number greater than two, although one will endure  $\Delta$ times higher computational and storage cost to do so, in the worst case. Thus, we suggest $h=2$ for satisfactory yet efficient utilization of our algorithm.  It is worth noting that in most real-world networks, in particular social networks, nodes are connected to a tiny portion of the whole network’s nodes, which means $\Delta$ (and in turn $\Delta^2$) is very small. For example, the maximum number of connections allowed on Twitter and Facebook is about 5000, that is much smaller than their network size~\cite{mahyar2018compressive}. This shows that our approach is practically efficient and scalable on real-world networks.

\section{Experimental Evaluation}\label{experiments}
In this section, we experimentally evaluate the performance of the proposed method in various scenarios over both synthetic and real-world networks. We first introduce the networks used for the evaluation. Then, we explain the settings of the experiments. Finally, the achieved results for each test scenario and their analyses are presented.

\subsection{Datasets}\label{Datasets}
For the evaluations of the proposed method, we considered both synthetic and real networks. We summarize the properties of the real-world networks used in experiments in Table~\ref{RealNetwork}. The four notations $\langle deg\rangle$, $\langle\mathcal{C}\rangle$, $\mathcal{D}$, and $\delta_{0.9}$ represent the \lq\lq average degree\rq\rq, \lq\lq average clustering coefficient\rq\rq, \lq\lq network diameter\rq\rq, and \lq\lq 90-percentile effective diameter\rq\rq, respectively. In the case of a disconnected network, we extracted the largest (strongly) connected component. 

We also considered three well-known models (\textit{i.e.} Barab\'{a}si-Albert (BA), Erd\H{o}s-R\'{e}nyi (ER), and Watts-Strogatz (SW)) for generating synthetic networks. We have summarized these networks in Table~\ref{SyntheticNetworks}. In ER network, the link existence probability $p=0.01$  ensures that the generated network is connected as $p > \frac{\ln |V|}{|V|}$ is a sharp threshold for connectedness of ER networks with $|V|$ vertices.

\begin{table}
\begin{center}
\caption{Real-World Networks}
\label{RealNetwork}
\begin{tabular}{|c||c|c|c|c|c|c|}
\hline
Network & $|V|$ & $|E|$ & $\langle deg\rangle$ & $\langle\mathcal{C}\rangle$ & $\mathcal{D}$ & $\delta_{0.9}$ \\
\hline
Facebook~\cite{ref:tore1} & 1893 & 6917 & 7.31 & 0.06 & 8 & 3.65 \\
\hline
Twitter~\cite{GephiTwitterDataset} & 3656 & 94356 & 51.62 & 0.3 & 6 & 2.89 \\
\hline
ca-AstroPh~\cite{Lescovec2007dataset} & 17903 & 197001 & 22.01 & 0.32 & 14 & 5.01 \\
\hline
ca-CondMat~\cite{Lescovec2007dataset} & 21363 & 91314 & 8.55 & 0.26 & 15 & 6.52 \\
\hline
ca-HepPh~\cite{Lescovec2007dataset} & 11204 & 117634 & 21 & 0.66 & 13 & 5.79 \\
\hline
ca-HepTh~\cite{Lescovec2007dataset} & 8638 & 24816 & 5.75 & 0.28 & 18 & 7.42 \\
\hline
email-Enron~\cite{leskovec2009community} & 33696 & 180811 & 10.73 & 0.09 & 13 & 4.79 \\
\hline
DBLP~\cite{yang2015defining} & 317080 & 524933 & 3.31 & 0.31 & 23 & 8.16 \\
\hline
wiki-Vote~\cite{Leskovec2010WikiVote} & 7066 & 51831 & 14.67 & 0.13 & 7 & 3.78 \\
\hline
\end{tabular}
\end{center}
\end{table}

\begin{table}
\begin{center}
\caption{Synthetic Network Models}
\label{SyntheticNetworks}
\begin{tabular}{|c||c|c|c|c|}
\hline
Network Model & $|V|$ & $|E|$ & Parameter & $\langle deg\rangle$ \\
\hline
Barab\'{a}si-Albert (BA)~\cite{ref:ba} & 500 & 2979 & 5 & 11.92 \\
\hline
Erd\H{o}s-R\'{e}nyi (ER)~\cite{ref:er} & 500 & 4000 & 0.01 & 16 \\
\hline
Watts-Strogatz (SW)~\cite{ref:watts} & 500 & 4466 & [0.2 ; 9] & 17.86 \\
\hline
\end{tabular}
\end{center}
\end{table}

\subsection{Settings}\label{Settings}
To evaluate the accuracy of the proposed method (\textsc{CS-HiClose}) compared to the competing methods in identifying top-$k$ closeness centrality nodes, we measured the \textit{precision} and \textit{recall} of the algorithms. Precision quantifies the number of correctly detected nodes in the list of $k$ highest closeness centrality nodes divided by the total number of detected nodes. Recall quantifies the number of correctly identified nodes divided by the total number of nodes in the network. The relevancy of the detected nodes (precision) and the portion of relevant nodes that are detected (recall) are both of importance. To take both into account, we utilized the popular F-measure metric, a harmonic mean of precision and recall, which is defined as:
\begin{eqnarray}
\text{F-measure} = 2 \times \frac{Precision \times Recall}{Precision + Recall}
\end{eqnarray} 

Since \textsc{CS-HiClose}, RW, TopCent, and DICeNod have a source of randomness, the experiments were repeated ten times to reduce the variance. The denoted points in the figures represent the mean value of these repetitions along with their asymmetric standard deviations, which quantifies the amount of variations of F-measure at each point in each figure. Implementation codes in Python can be found at \texttt{https://github.com/hamidreza-mahyar/CS-HiClose}.
We used POGS~\cite{POGS}, a fast and parallel optimization solver, for the optimization phase of \textsc{CS-HiClose}. POGS tries to minimize LASSO (Equation (\ref{LASSO})) as an objective function, and is extremely quick by leveraging the power of GPUs. For example~\cite{parikh2014block}, it can solve the LASSO objective on a graph of 100,000 nodes with 10,000 measurements in only 21s on a single Nvidia K40 GPU. For computations of the global closeness centrality in Equation (\ref{Closeness}), we used available tools in \textit{Python-iGraph} package.

\subsection{Evaluation Results}
\subsubsection{Correlation between Our ego-Closeness and the Global Closeness}
\label{CorrSubsec}
We experimentally analyzed the correlation between the proposed ego-centric (local) centrality metric and the global closeness centrality over several synthetic and real-world networks. To compare these two centrality metrics, we used Pearson product moment correlation coefficient ($\rho$), which in fact measures the strength of a linear association between two variables and is defined as~\cite{benesty2009pearson}:
\begin{eqnarray}
\rho = \frac{\sum_{i=1}^{|V|} (x_i-\overline{x})(y_i-\overline{y}) }{\sqrt{\sum_{i=1}^{|V|} (x_i-\overline{x})^2 \sum_i (y_i-\overline{y})^2}}
\end{eqnarray}
where $|V|$ is the number of network nodes and $x_i$, $y_i$ correspond to the local and global centrality measures of node $i$, respectively. $\overline{x}$ and $\overline{y}$ are mean of these variables. The Pearson coefficient $\rho$ can take a value in range $[-1 , +1]$. A value of $0$ shows that there is not any association, a value greater than $0$ indicates a positive association, and a value less than $0$ indicates a negative association.

Table~\ref{PearsonCorr} illustrates the correlation coefficients between the proposed ego-closeness and the global closeness centrality. As mentioned in Section~\ref{ProposedLocalMetric}, the computational and storage cost of the ego-closeness centrality is directly impacted by the choice of $h$. Thus, our aim is to yield good results in distributively assessing top-$k$ network centralities with a small value of $h$. We calculated the correlation for various sparsity levels $k$ and small values of $h$ (\textit{i.e.} 2 and 3) for different networks. It is worth noting that for $h = 1$, any local metric would be the same as the degree centrality. Overall, the results show that our proposed local metric and the global closeness centrality highly correlate on various types of networks. According to the problem addressed in this paper, we want to identify top-$k$ central nodes for $k \ll |V|$, so the results show that in this case choosing $h=2$ is sufficient yet efficient, in terms of having a good trade-off between computational complexity and accuracy.

\begin{table*}[t!]
\begin{center}
\caption{Pearson correlation coefficients between the proposed ego-centric closeness centrality with small exploration radius (\textit{i.e.} $h = \{2, 3\}$) and the global closeness centrality on synthetic and real-world networks.}
\label{PearsonCorr}
\resizebox{\textwidth}{!}{\begin{tabular}{|c|c||c|c|c|c|c|c|c|c|c|c|c|c|}
\cline{3-14}
\multicolumn{2}{c|}{} & \multirow{2}{*}{Facebook} & \multirow{2}{*}{Twitter} & \multirow{2}{*}{ca-AstroPh} & \multirow{2}{*}{ca-CondMat} & \multirow{2}{*}{ca-HepPh} & \multirow{2}{*}{ca-HepTh} & \multirow{2}{*}{email-Enron} & \multirow{2}{*}{DBLP} & \multirow{2}{*}{wikiVote} & \multirow{2}{*}{BA} & \multirow{2}{*}{ER} & \multirow{2}{*}{SW} \\
\cline{1-2}
$h$ & $k/|V|$ &  &  &  &  &  &  &  &  &  &  &  &  \\
\hline
\multirow{10}{*}{2} & 0.1 & 1.00 & 1.00 & 0.99 & 0.91 & 0.98 & 0.93 & 0.97 & 0.90 & 0.99 & 1.00 & 1.00 & 0.96 \\
 & 0.2 & 1.00 & 1.00 & 0.98 & 0.91 & 0.98 & 0.93 & 0.98 & 0.87 & 0.99 & 1.00 & 1.00 & 0.96 \\
 & 0.3 & 1.00 & 1.00 & 0.98 & 0.91 & 0.97 & 0.93 & 0.98 & 0.84 & 0.99 & 1.00 & 1.00 & 0.96 \\
 & 0.4 & 1.00 & 1.00 & 0.98 & 0.90 & 0.96 & 0.91 & 0.98 & 0.82 & 0.99 & 1.00 & 1.00 & 0.96 \\
 & 0.5 & 1.00 & 1.00 & 0.96 & 0.87 & 0.95 & 0.89 & 0.92 & 0.76 & 0.99 & 0.99 & 1.00 & 0.96 \\
 & 0.6 & 1.00 & 1.00 & 0.94 & 0.86 & 0.93 & 0.88 & 0.91 & 0.74 & 0.99 & 0.99 & 1.00 & 0.97 \\
 & 0.7 & 1.00 & 1.00 & 0.93 & 0.84 & 0.91 & 0.86 & 0.89 & 0.72 & 0.99 & 0.99 & 1.00 & 0.97 \\
 & 0.8 & 0.99 & 1.00 & 0.90 & 0.82 & 0.89 & 0.84 & 0.83 & 0.69 & 0.99 & 0.99 & 1.00 & 0.97 \\
 & 0.9 & 0.99 & 1.00 & 0.87 & 0.79 & 0.86 & 0.81 & 0.79 & 0.66 & 0.98 & 0.99 & 1.00 & 0.97 \\
 & 1.0 & 0.96 & 0.99 & 0.80 & 0.73 & 0.80 & 0.75 & 0.72 & 0.60 & 0.97 & 1.00 & 1.00 & 0.97 \\
\hline
\multirow{10}{*}{3} & 0.1 & 0.99 & 0.97 & 1.00 & 0.98 & 1.00 & 0.98 & 0.99 & 0.95 & 1.00 & 1.00 & 0.99 & 0.97 \\
 & 0.2 & 0.99 & 0.96 & 1.00 & 0.98 & 1.00 & 0.98 & 0.99 & 0.92 & 1.00 & 1.00 & 0.99 & 0.97 \\
 & 0.3 & 1.00 & 0.96 & 1.00 & 0.98 & 1.00 & 0.98 & 0.99 & 0.90 & 0.99 & 1.00 & 1.00 & 0.98 \\
 & 0.4 & 1.00 & 0.96 & 1.00 & 0.98 & 1.00 & 0.97 & 0.99 & 0.88 & 0.99 & 1.00 & 1.00 & 0.98 \\
 & 0.5 & 0.99 & 0.96 & 1.00 & 0.97 & 1.00 & 0.96 & 0.99 & 0.85 & 0.99 & 1.00 & 1.00 & 0.98 \\
 & 0.6 & 0.99 & 0.96 & 1.00 & 0.96 & 1.00 & 0.95 & 0.99 & 0.83 & 0.99 & 1.00 & 1.00 & 0.99 \\
 & 0.7 & 0.99 & 0.97 & 1.00 & 0.95 & 0.99 & 0.94 & 0.99 & 0.81 & 0.99 & 1.00 & 1.00 & 0.99 \\
 & 0.8 & 0.99 & 0.98 & 1.00 & 0.93 & 0.99 & 0.92 & 0.99 & 0.78 & 0.99 & 1.00 & 1.00 & 0.99 \\
 & 0.9 & 0.99 & 0.99 & 0.99 & 0.90 & 0.98 & 0.89 & 0.98 & 0.74 & 0.99 & 1.00 & 1.00 & 0.99 \\
 & 1.0 & 0.98 & 0.96 & 0.95 & 0.84 & 0.94 & 0.83 & 0.95 & 0.67 & 0.99 & 0.99 & 1.00 & 0.99 \\
\hline
\end{tabular}}
\end{center}
\end{table*}

\begin{table*}
\begin{center}
\caption{Pearson correlation coefficients between the existing local metrics (Dist-Exact, DACCER, Weighted-Vol, and our proposed ego-centric centrality) with $h = 2$ and the global closeness centrality on synthetic and real-world networks for varying percentage of sparsity.}
\label{CorrComparison}
\resizebox{\textwidth}{!}{\begin{tabular}{|c|l||c|c|c|c|c|c|c|c|c|c|c|c|}
\cline{3-14}
\multicolumn{2}{c|}{} & \multirow{2}{*}{Facebook} & \multirow{2}{*}{Twitter} & \multirow{2}{*}{ca-AstroPh} & \multirow{2}{*}{ca-CondMat} & \multirow{2}{*}{ca-HepPh} & \multirow{2}{*}{ca-HepTh} & \multirow{2}{*}{email-Enron} & \multirow{2}{*}{DBLP} & \multirow{2}{*}{wikiVote} & \multirow{2}{*}{BA} & \multirow{2}{*}{ER} & \multirow{2}{*}{SW} \\
\cline{1-2}
$k/|V|$ & Local Metric &  &  &  &  &  &  &  &  &  &  &  &  \\
\hline
\multirow{4}{*}{0.1} & Dist-Exact & -0.93 & -0.41 & -0.94 & -0.83 & -0.89 & -0.87 & -0.75 & -0.73 & -0.94 & -0.94 & -0.99 & -0.93 \\
 & DACCER & 0.91 & 0.47 & 0.96 & 0.87 & 0.77 & 0.91 & 0.96 & 0.87 & 0.83 & 0.97 & 0.96 & 0.95 \\
 & Weight-Vol & 0.97 & 0.80 & 0.97 & 0.89 & 0.93 & 0.90 & \bf{0.97} & \bf{0.90} & 0.97 & 0.99 & 0.97 & 0.93 \\
 & \bf{Our Metric} & \bf{1.00} & \bf{1.00} & \bf{0.99} & \bf{0.91} & \bf{0.98} & \bf{0.93} & \bf{0.97} & \bf{0.90} & \bf{0.99} & \bf{1.00} & \bf{1.00} & \bf{0.96} \\
\hline
\multirow{4}{*}{0.2} & Dist-Exact & -0.93 & -0.61 & -0.93 & -0.80 & -0.84 & -0.84 & -0.57 & -0.70 & -0.94 & -0.93 & -0.99 & -0.91 \\
 & DACCER & 0.92 & 0.58 & 0.97 & 0.89 & 0.79 & 0.92 & 0.97 & 0.84 & 0.89 & 0.97 & 0.98 & 0.94 \\
 & Weight-Vol & 0.97 & 0.77 & 0.97 & \bf{0.91} & 0.92 & 0.92 & \bf{0.98} & \bf{0.87} & 0.98 & 0.99 & 0.99 & 0.94 \\
 & \bf{Our Metric} & \bf{1.00} & \bf{1.00} & \bf{0.98} & \bf{0.91} & \bf{0.98} & \bf{0.93} & \bf{0.98} & \bf{0.87} & \bf{0.99} & \bf{1.00} & \bf{1.00} & \bf{0.96} \\
\hline
\multirow{4}{*}{0.3} & Dist-Exact & -0.93 & -0.73 & -0.91 & -0.78 & -0.82 & -0.81 & -0.61 & -0.67 & -0.94 & -0.92 & -0.99 & -0.93 \\
 & DACCER & 0.93 & 0.63 & 0.97 & 0.90 & 0,81 & 0.92 & \bf{0.98} & 0.81 & 0.92 & 0.97 & 0.99 & 0.95 \\
 & Weight-Vol & 0.97 & 0.75 & \bf{0.98} & \bf{0.91} & 0.92 & 0.92 & \bf{0.98} & \bf{0.84} & 0.98 & 0.99 & 0.99 & \bf{0.96} \\
 & \bf{Our Metric} & \bf{1.00} & \bf{1.00} & \bf{0.98} & \bf{0.91} & \bf{0.97} & \bf{0.93} & \bf{0.98} & \bf{0.84} & \bf{0.99} & \bf{1.00} & \bf{1.00} & \bf{0.96} \\
\hline
\multirow{4}{*}{0.4} & Dist-Exact & -0.92 & -0.79 & -0.90 & -0.74 & -0.76 & -0.75 & -0.59 & -0.61 & -0.94 & -0.91 & -0.99 & -0.94 \\
 & DACCER & 0.95 & 0.67 & 0.97 & 0.89 & 0.86 & \bf{0.91} & \bf{0.98} & 0.79 & 0.95 & 0.97 & 0.99 & 0.95 \\
 & Weight-Vol & 0.98 & 0.75 & \bf{0.98} & \bf{0.90} & 0.93 & \bf{0.91} & \bf{0.98} & \bf{0.82} & 0.98 & 0.99 & 0.99 & \bf{0.96} \\
 & \bf{Our Metric} & \bf{1.00} & \bf{1.00} & \bf{0.98} & \bf{0.90} & \bf{0.96} & \bf{0.91} & \bf{0.98} & \bf{0.82} & \bf{0.99} & \bf{1.00} & \bf{1.00} & \bf{0.96} \\
\hline
\end{tabular}}
\end{center}
\end{table*}

Table~\ref{CorrComparison} shows the Pearson correlation coefficients between the existing local metrics reviewed in Section~\ref{RelatedLocalMetric} (\textit{i.e.} Dist-Exact, DACCER, and Weight-Vol) and our proposed ego-centric centrality measure, all with $h=2$, and the global closeness centrality on synthetic and real-world networks. In this experiment, we mainly focus on high sparsity levels $k = \{ 0.1|V|, 0.2|V|, 0.3|V|, 0.4|V| \}$. After implementing DistEst~\cite{wang2015distributed}, we found that the computed values for this metric critically depend on parameters' initialization (\textit{e.g.} each node should have an estimation about its closeness value which is an unrealistic assumption). Moreover, this metric needs a very large number of iterations for message passing to converge. To have a fair comparison, we set the same number of iterations as our metric, but its correlation coefficients were around 0, so the results for this metric were excluded. 

The results show that Dist-Exact for $h=2$ has linear correlation, but negative association with the closeness centrality in networks with various levels of sparsity. One can observe that our proposed metric has almost always the best correlation coefficient compared to the other metrics. Another interesting observation in Tables~\ref{PearsonCorr} and~\ref{CorrComparison} is that our ego-centric metric has lower correlation coefficient with the global closeness centrality on the networks (\textit{i.e.} ca-CondMat, ca-HepTh, and DBLP) with relatively small average degree, small average clustering coefficient, and large network diameter (both full and 90-percentile).

\begin{figure}[t!]
\includegraphics[width=.95\textwidth]{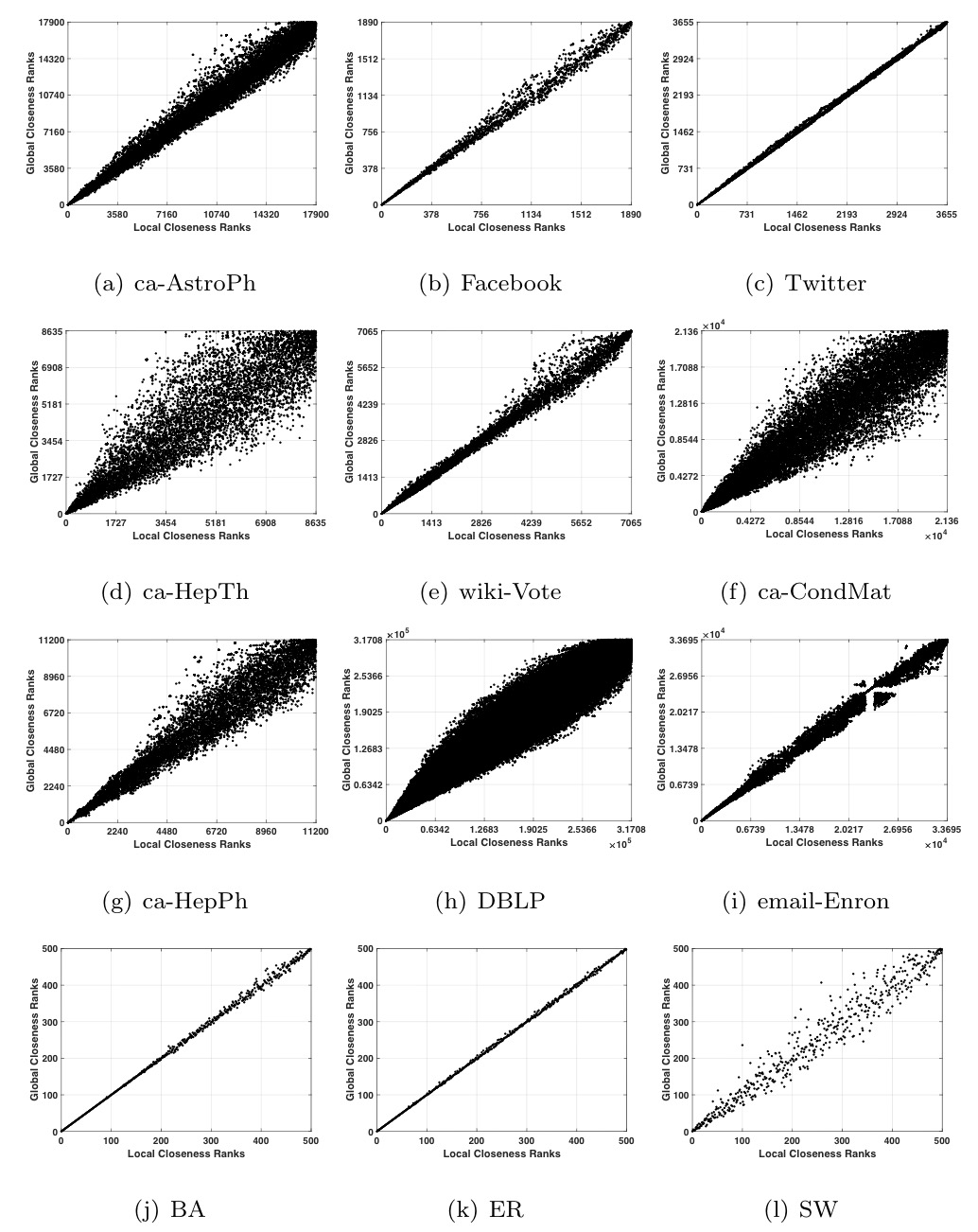}
\caption{Correlations between the nodes' ranks provided by the proposed local metric and the global closeness centrality on synthetic and real-world networks. These two metrics correlate very well.}
\label{CorrelationPlot}
\end{figure}

To have more analysis of the correlation between the proposed ego-centric (local) metric and the global closeness centrality, Figure~\ref{CorrelationPlot} shows the scatter plots of all nodes' ranks provided by one versus the other, on various networks. Each point in the figure corresponds to a node's rank using these two metrics. Based on the results of the previous test cases, we calculated our local measure for $h=2$ to have low computational complexity, yet high accuracy. One can easily observe the linear correlation and positive association (as the rank with respect to the local metric increases, so does the rank with respect to the global metric), especially for the top-$k$ nodes' ranks which is the target of this paper. One can easily see the similar observation, as in Tables~\ref{PearsonCorr} and~\ref{CorrComparison}, that our metric has relatively lower correlation with the global closeness centrality on ca-CondMat, ca-HepTh, and DBLP networks, that share properties like small average degree, small clustering coefficient, and large network diameter.

Although the Pearson product-moment correlation coefficient is the most common and almost exclusively used measure for correlation studies of centrality indices, non-linear dependencies are not adequately captured by it. Moreover, assuming only a linear correlation between two scores is very strong and maybe not realistic. A common workaround to depict some of the existing non-linear dependencies is to employ the Pearson correlation on the logarithm of the original scores, and it is mainly used for illustrative purposes~\cite{Schoch2015Posit-34821}. Table~\ref{LogarithmPearsonCorr} is similar to Table~\ref{PearsonCorr}, instead it shows the Pearson correlation on the logarithms of the proposed ego-closeness (with $h=2$) and the global closeness scores. The result suggests that our proposed ego-centric metric not only has a high positive linear association (as inferred by Table~\ref{PearsonCorr}) but also demonstrates a very high positive non-linear association with the global closeness centrality. 

\begin{table*}[t!]
\begin{center}
\caption{Pearson correlation coefficients between the \textbf{logarithms} of the proposed ego-centric score with small exploration radius (\textit{i.e.} $h = \{2, 3\}$) and the global closeness score on synthetic and real-world networks.}
\label{LogarithmPearsonCorr}
\resizebox{\textwidth}{!}{\begin{tabular}{|c|c||c|c|c|c|c|c|c|c|c|c|c|c|}
\cline{3-14}
\multicolumn{2}{c|}{} & \multirow{2}{*}{Facebook} & \multirow{2}{*}{Twitter} & \multirow{2}{*}{ca-AstroPh} & \multirow{2}{*}{ca-CondMat} & \multirow{2}{*}{ca-HepPh} & \multirow{2}{*}{ca-HepTh} & \multirow{2}{*}{email-Enron} & \multirow{2}{*}{DBLP} & \multirow{2}{*}{wikiVote} & \multirow{2}{*}{BA} & \multirow{2}{*}{ER} & \multirow{2}{*}{SW} \\
\cline{1-2}
$h$ & $k/|V|$ &  &  &  &  &  &  &  &  &  &  &  &  \\
\hline
\multirow{10}{*}{2}
 & 0.1 & 0.99 & 1.00 & 0.98 & 0.91 & 0.96 & 0.93 & 0.94 & 0.88 & 0.98 & 0.99 & 1.00 & 0.96 \\
 & 0.2 & 0.99 & 0.99 & 0.98 & 0.92 & 0.96 & 0.93 & 0.92 & 0.87 & 0.98 & 0.99 & 1.00 & 0.95 \\
 & 0.3 & 0.99 & 0.99 & 0.98 & 0.92 & 0.96 & 0.92 & 0.91 & 0.86 & 0.98 & 0.98 & 1.00 & 0.95 \\
 & 0.4 & 0.98 & 0.99 & 0.98 & 0.93 & 0.96 & 0.91 & 0.90 & 0.86 & 0.99 & 0.98 & 1.00 & 0.96 \\
 & 0.5 & 0.98 & 0.99 & 0.98 & 0.92 & 0.96 & 0.91 & 0.90 & 0.85 & 0.98 & 0.98 & 1.00 & 0.96 \\
 & 0.6 & 0.97 & 0.99 & 0.98 & 0.92 & 0.96 & 0.90 & 0.91 & 0.85 & 0.98 & 0.97 & 1.00 & 0.96 \\
 & 0.7 & 0.97 & 0.99 & 0.98 & 0.93 & 0.96 & 0.90 & 0.92 & 0.85 & 0.97 & 0.97 & 1.00 & 0.97 \\
 & 0.8 & 0.97 & 0.98 & 0.98 & 0.93 & 0.96 & 0.90 & 0.83 & 0.86 & 0.96 & 0.96 & 1.00 & 0.97 \\
 & 0.9 & 0.97 & 0.97 & 0.98 & 0.93 & 0.96 & 0.89 & 0.88 & 0.86 & 0.96 & 0.96 & 1.00 & 0.97 \\
 & 1.0 & 0.97 & 0.92 & 0.96 & 0.92 & 0.94 & 0.86 & 0.90 & 0.84 & 0.96 & 0.96 & 0.99 & 0.97 \\
\hline
\multirow{10}{*}{3}
 & 0.1 & 0.99 & 0.97 & 0.99 & 0.97 & 1.00 & 0.97 & 0.99 & 0.98 & 1.00 & 1.00 & 0.99 & 0.97 \\
 & 0.2 & 0.99 & 0.96 & 0.99 & 0.97 & 0.99 & 0.98 & 0.98 & 0.98 & 0.99 & 1.00 & 0.99 & 0.97 \\
 & 0.3 & 0.99 & 0.96 & 0.99 & 0.97 & 0.99 & 0.98 & 0.98 & 0.97 & 0.99 & 1.00 & 1.00 & 0.98 \\
 & 0.4 & 1.00 & 0.95 & 0.99 & 0.98 & 0.98 & 0.98 & 0.98 & 0.97 & 0.99 & 1.00 & 1.00 & 0.98 \\
 & 0.5 & 0.99 & 0.96 & 0.98 & 0.98 & 0.97 & 0.98 & 0.97 & 0.97 & 0.98 & 1.00 & 1.00 & 0.98 \\
 & 0.6 & 0.99 & 0.96 & 0.98 & 0.98 & 0.97 & 0.98 & 0.96 & 0.97 & 0.98 & 1.00 & 1.00 & 0.99 \\
 & 0.7 & 0.99 & 0.97 & 0.97 & 0.98 & 0.97 & 0.98 & 0.94 & 0.97 & 0.98 & 1.00 & 1.00 & 0.99 \\
 & 0.8 & 0.98 & 0.98 & 0.97 & 0.98 & 0.97 & 0.98 & 0.93 & 0.97 & 0.97 & 1.00 & 1.00 & 0.99 \\
 & 0.9 & 0.97 & 0.99 & 0.97 & 0.98 & 0.97 & 0.98 & 0.93 & 0.97 & 0.97 & 1.00 & 1.00 & 0.99 \\
 & 1.0 & 0.89 & 0.85 & 0.97 & 0.98 & 0.97 & 0.96 & 0.95 & 0.95 & 0.92 & 0.99 & 1.00 & 0.99 \\
\hline
\end{tabular}}
\end{center}
\end{table*}

\begin{table}
\begin{center}
\caption{Running time (in \textit{milliseconds}) comparison for different local metrics on synthetic networks in a simulated distributed environment.}
\label{RuntimeComparison}
\begin{scriptsize}
\begin{tabular}{|c||c|c|c|c|}
\hline
Network & Dist-Exact & DACCER & Weight-Vol & Our Metric \\
\hline
Barab\'{a}si-Albert (BA) & 3.77 & 4.26 & 13.78 & 3.74 \\
\hline
Erd\H{o}s-R\'{e}nyi (ER) & 1.20 & 1.45 & 8.16 & 1.18 \\
\hline
Watts-Strogatz (SW) & 1.10 & 1.02 & 6.65 & 0.80 \\
\hline
\end{tabular}
\end{scriptsize}
\end{center}
\end{table}

\subsubsection{Running Time Comparison}
\label{runtimeComp}
In Table~\ref{RuntimeComparison}, we empirically compare the running time for computation of the local metrics reviewed in Section~\ref{RelatedLocalMetric} (\textit{i.e.} Dist-Exact, DACCER, Weight-Vol, and our proposed ego-centric measure) over the synthetic networks. The running time of these metrics measured in a simulated distributed environment on a 2.5 GHz Intel Core i7 Apple
MacBook Pro laptop. We set the radius of the local neighborhood for each node to $h=2$, similar to the other experiments and for the same reasons.

Note that in the distributed and decentralized setting that we considered here, each node in the network begins executing a process to compute its corresponding local metric based on its visible neighborhood radius. Each node's process runs independent of the other nodes' processes. The distributed running time that we report for a metric on a network is equal to the longest execution time among all network nodes' processes for computation of the desired local metric. Table~\ref{RuntimeComparison} shows that our proposed metric is the fastest local measure to be calculated locally in a decentralized manner over all synthetic networks.

\subsubsection{Effect of Sparsity Level $k$ on Accuracy:}
\begin{figure}[t!]
\begin{center}
{
\subfigure[
ca-AstroPh
]{\includegraphics[trim = 3mm 0mm 10mm 0mm, clip, scale =0.18]{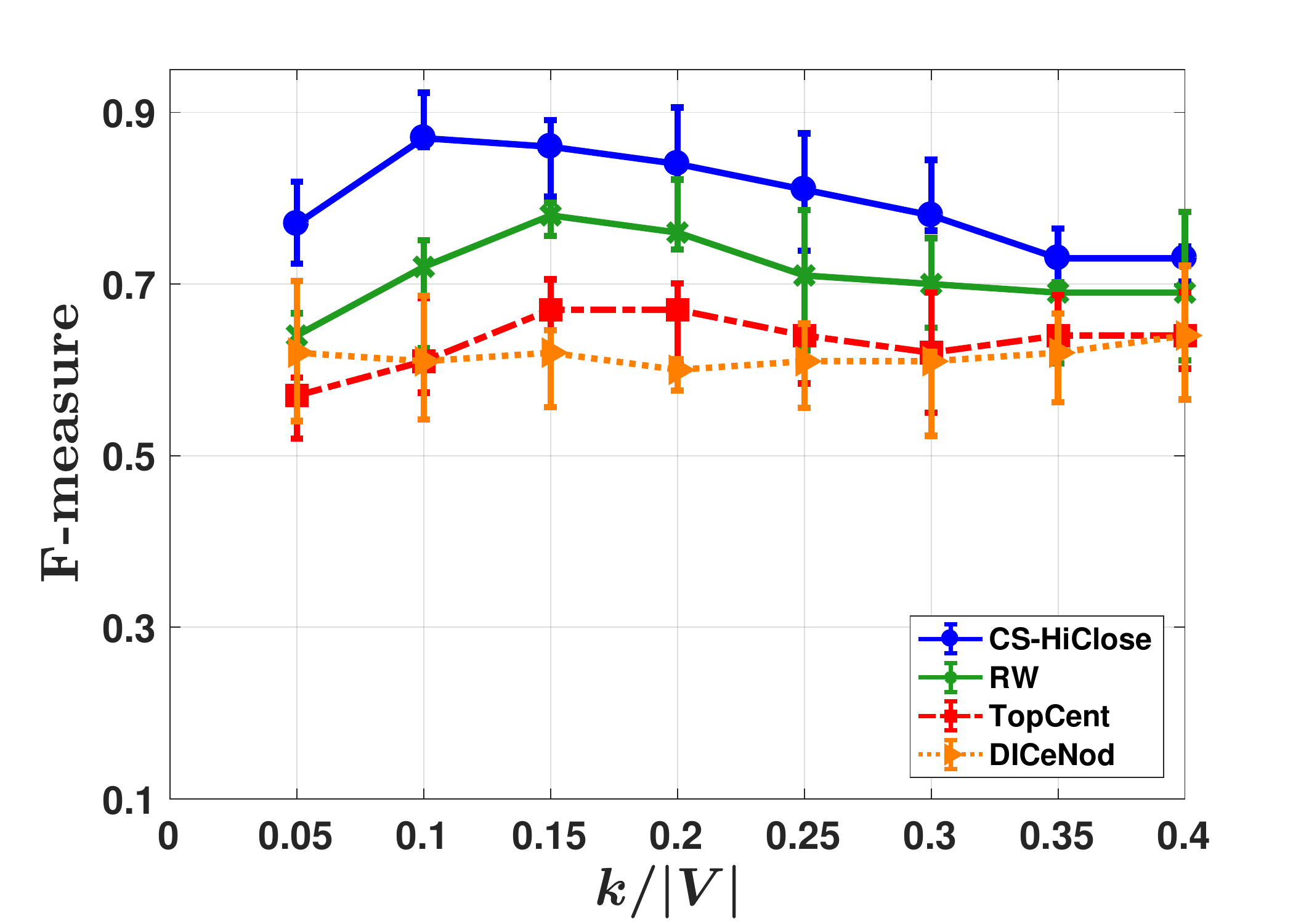}}
\subfigure[
Facebook
]{\includegraphics[trim = 3mm 0mm 10mm 0mm, clip, scale =0.18]{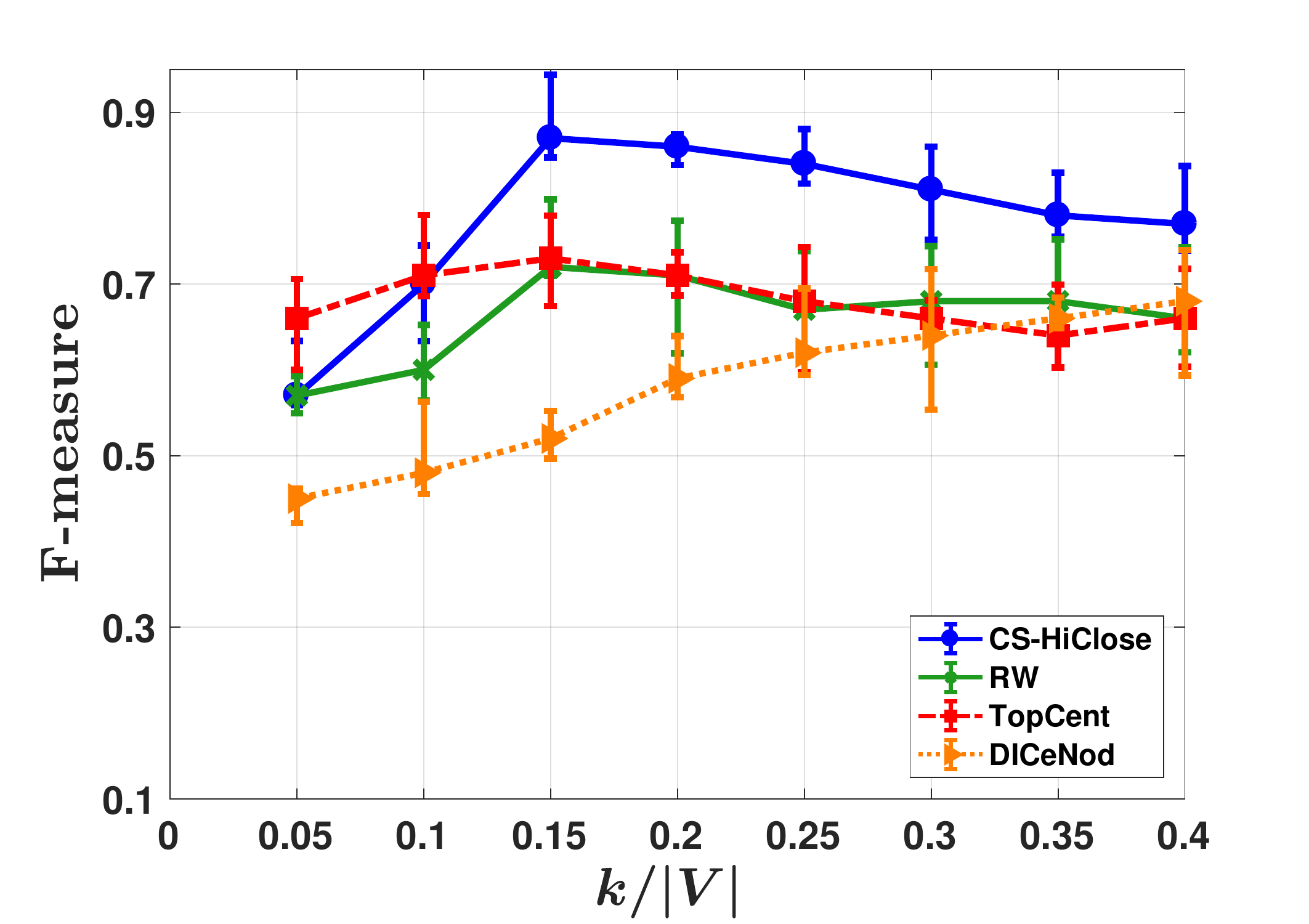}}
\subfigure[
Twitter
]{\includegraphics[trim = 3mm 0mm 10mm 0mm, clip, scale =0.18]{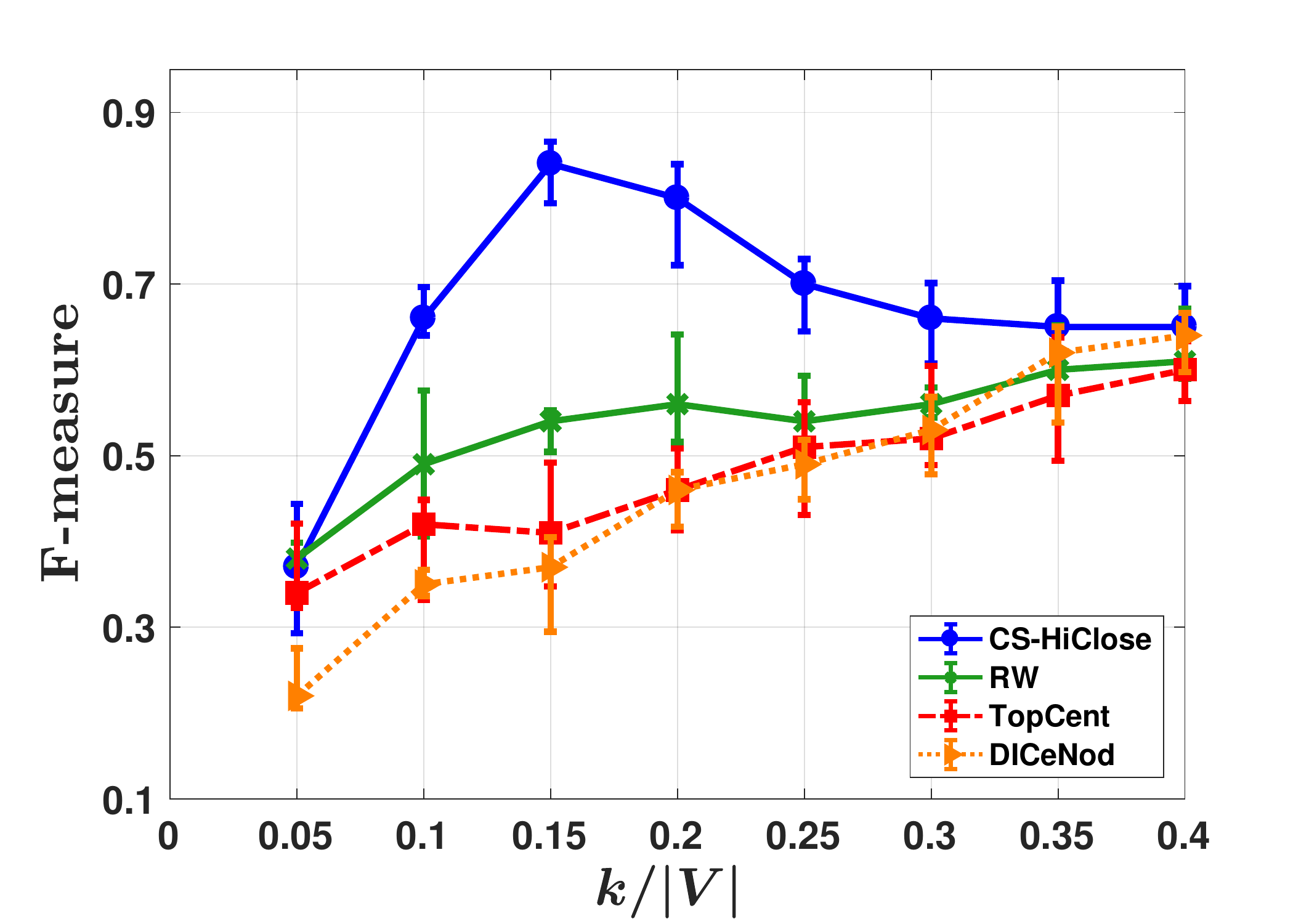}}
}
{
\subfigure[
ca-HepTh
]{\includegraphics[trim = 3mm 0mm 10mm 0mm, clip, scale =0.18]{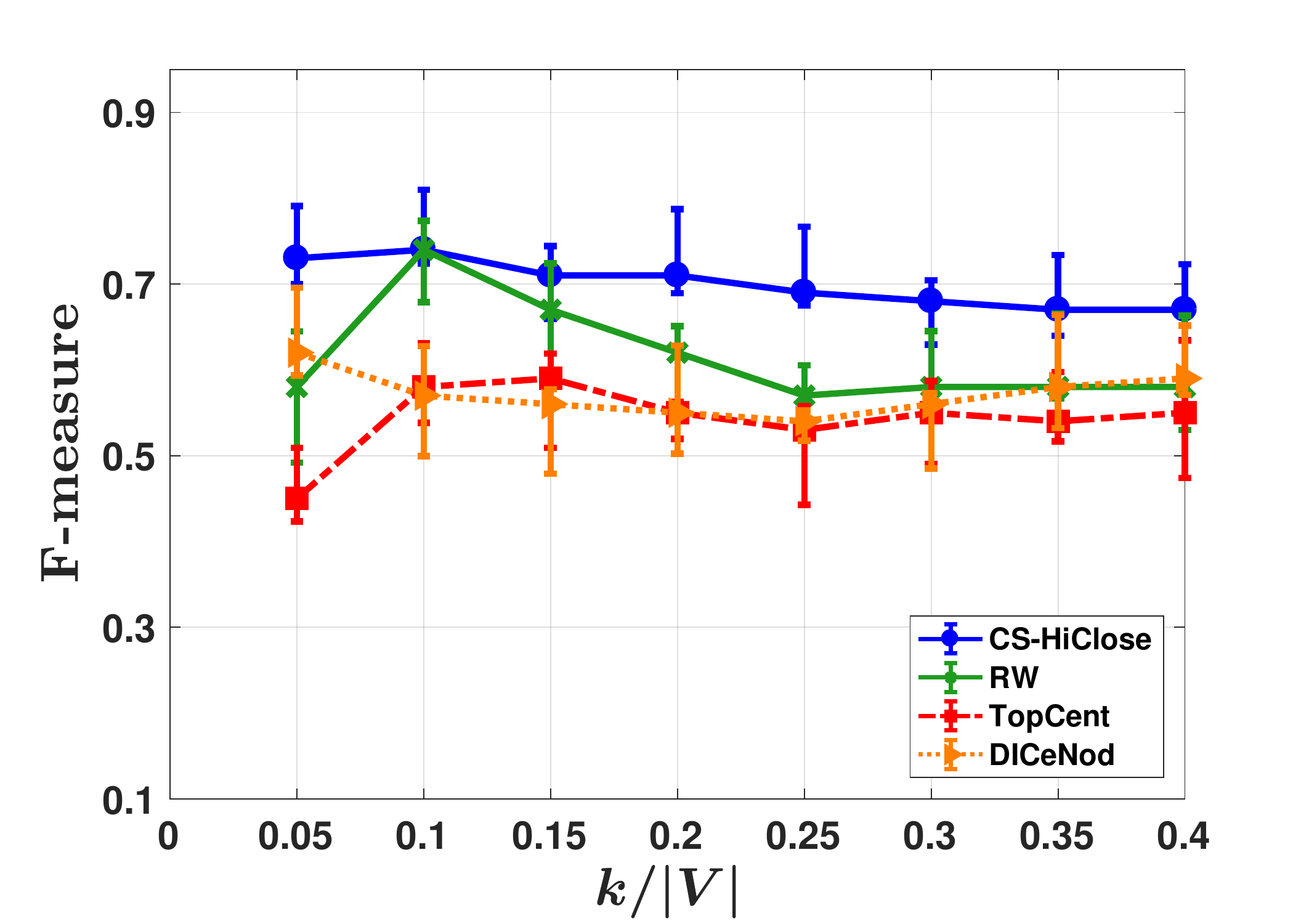}}
\subfigure[
wiki-Vote
]{\includegraphics[trim = 3mm 0mm 10mm 0mm, clip, scale =0.18]{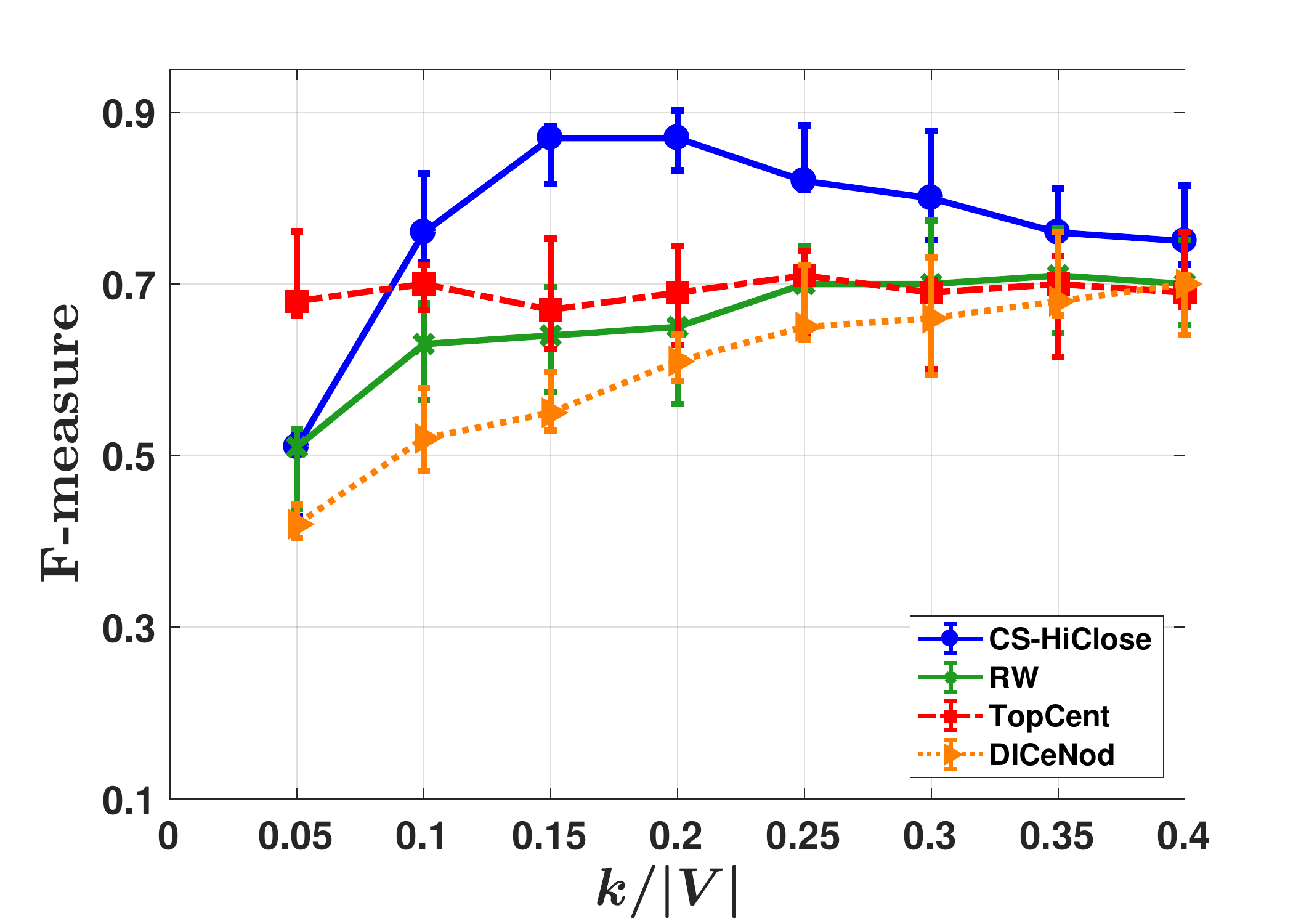}}
\subfigure[
ca-CondMat
]{\includegraphics[trim = 3mm 0mm 10mm 0mm, clip, scale =0.18]{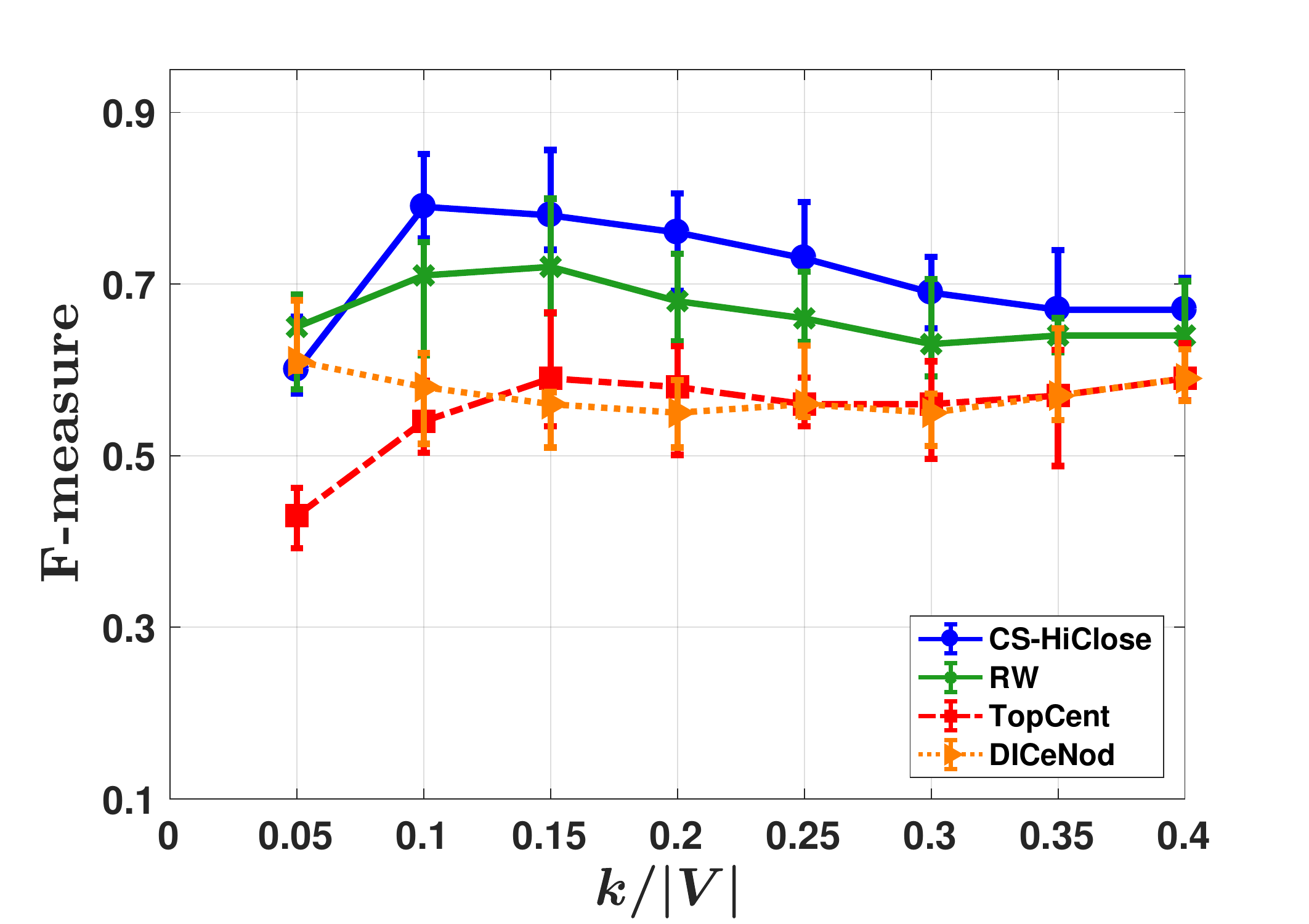}}
}
{
\subfigure[
ca-HepPh
]{\includegraphics[trim = 3mm 0mm 10mm 0mm, clip, scale =0.18]{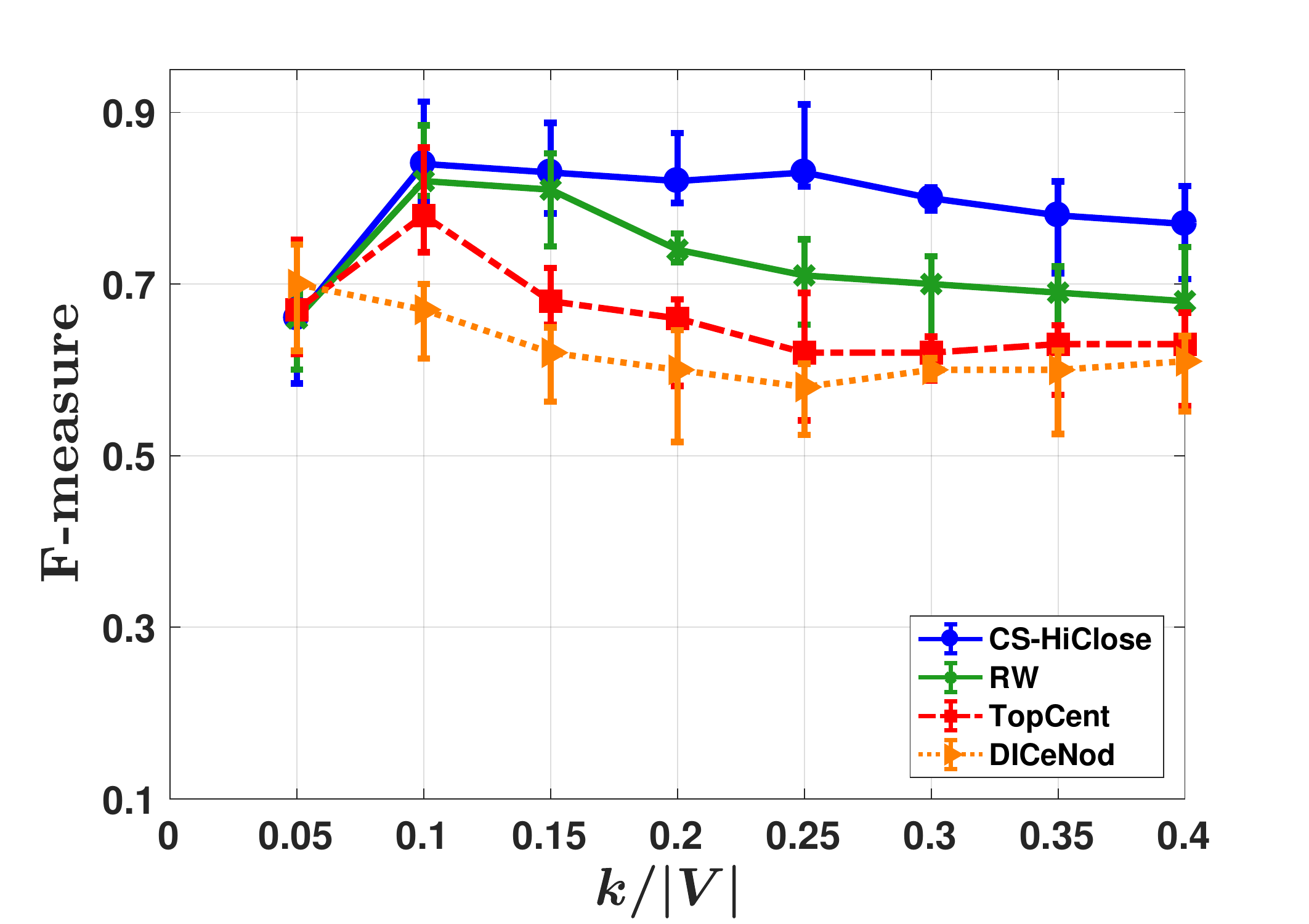}}
\subfigure[
DBLP
]{\includegraphics[trim = 3mm 0mm 10mm 0mm, clip, scale =0.18]{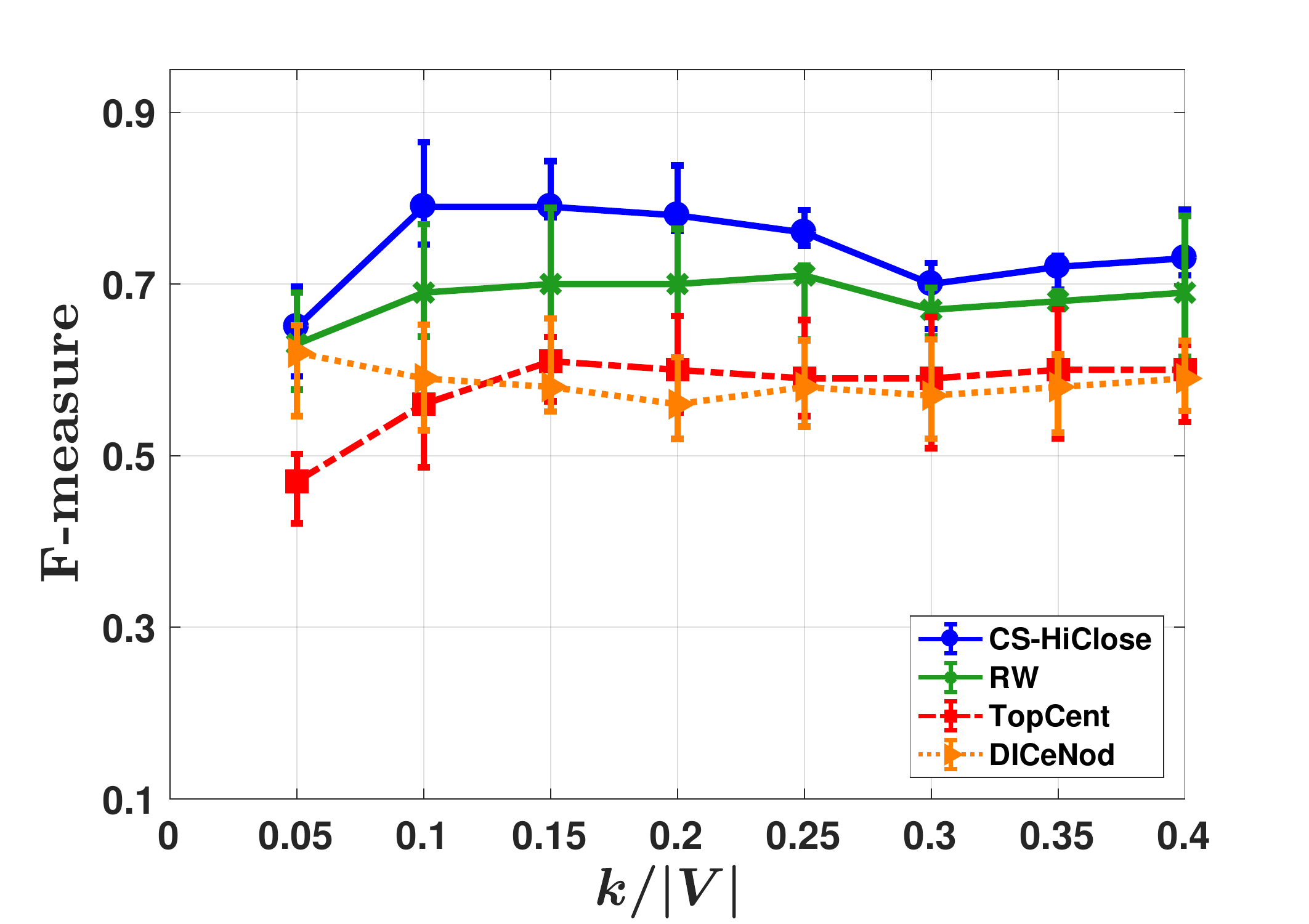}}
\subfigure[
email-Enron
]{\includegraphics[trim = 3mm 0mm 10mm 0mm, clip, scale =0.18]{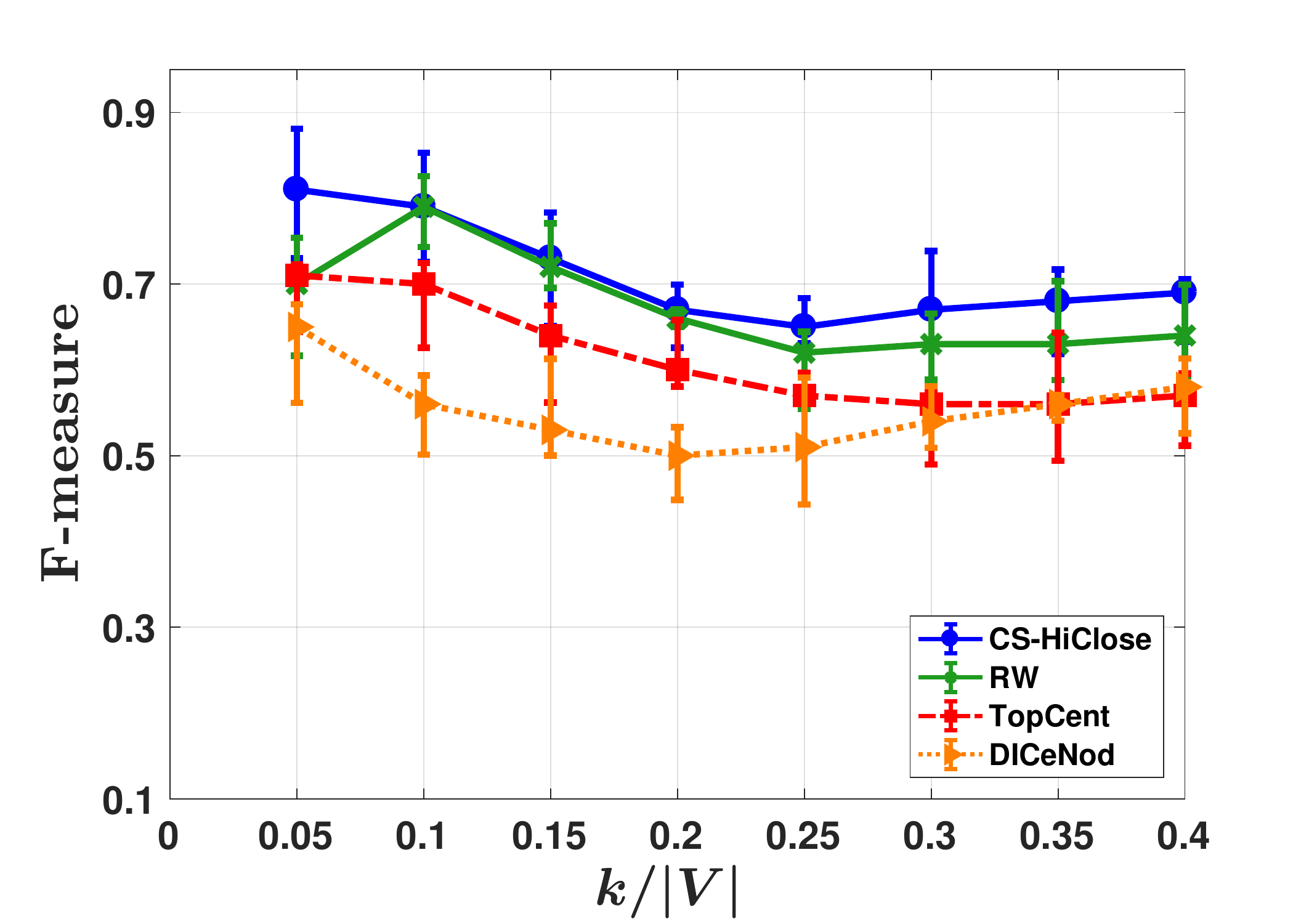}}
}
{
\subfigure[
BA
]{\includegraphics[trim = 3mm 0mm 10mm 0mm, clip, scale =0.18]{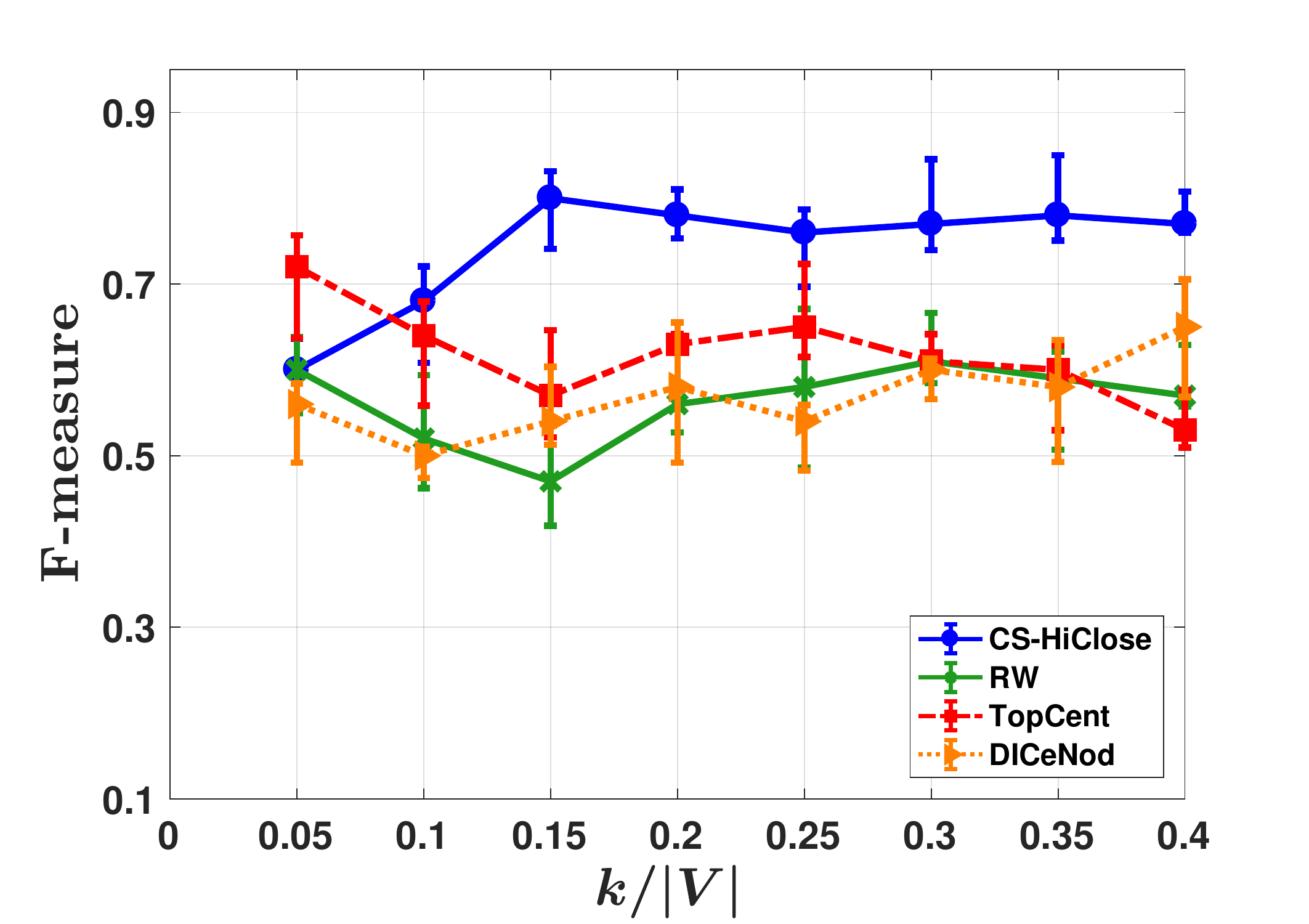}}
\subfigure[
ER
]{\includegraphics[trim = 3mm 0mm 10mm 0mm, clip, scale =0.18]{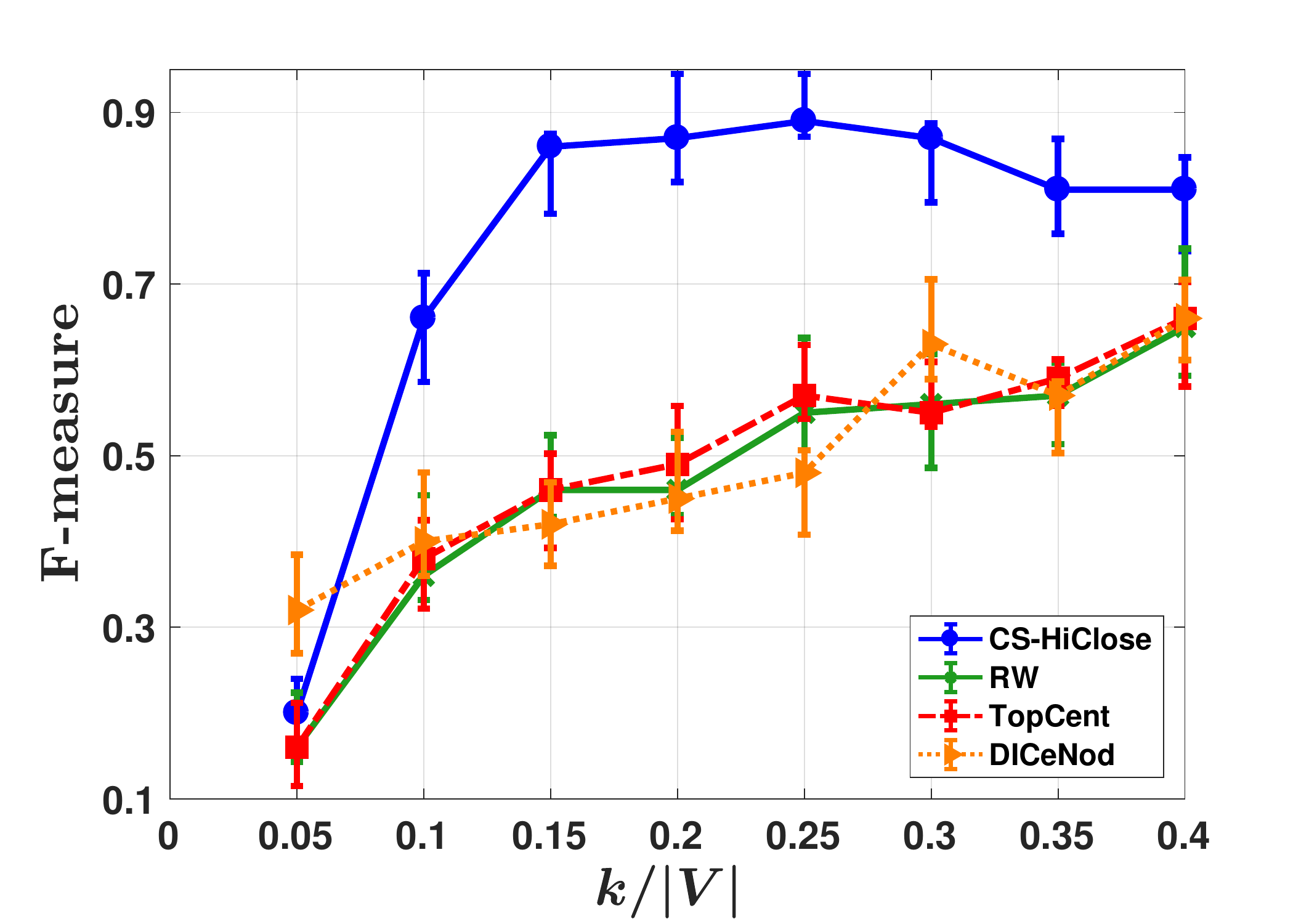}}
\subfigure[
SW
]{\includegraphics[trim = 3mm 0mm 10mm 0mm, clip, scale =0.18]{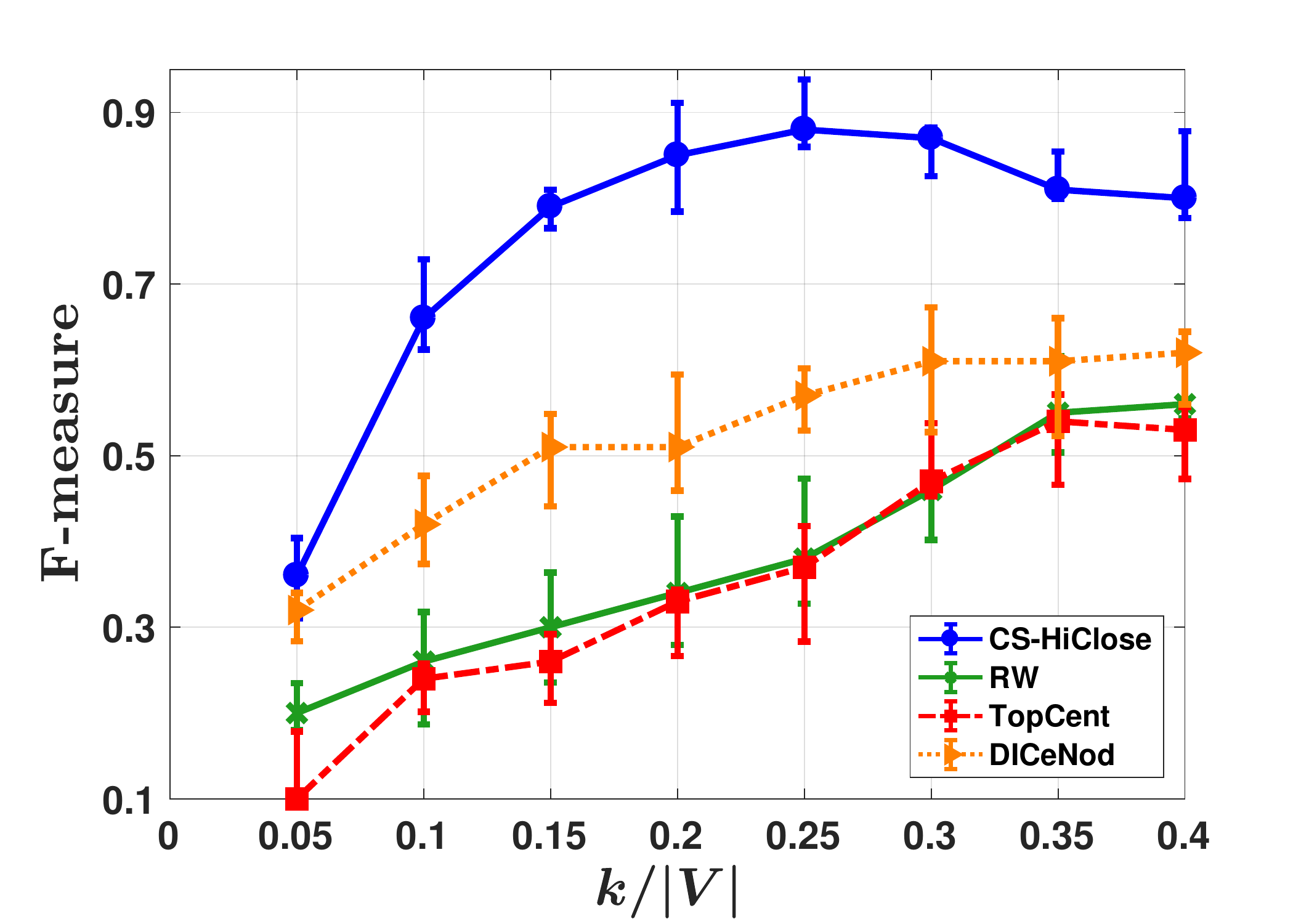}}
}
\caption{Effect of sparsity level $k$ on the accuracy of \textsc{CS-HiClose} and the competing methods for the number of correctly detected top-$k$ closeness centrality nodes. For all methods, we set the number of measurements to $0.4 |V|$ and the measurements length to $0.25 |V|$. The higher the value of F-measure is, the more correlation between the top-$k$ nodes list identified by a method and the global closeness centrality will be.}
\label{effectK}
\end{center}
\end{figure}

Figure~\ref{effectK} shows the effect of sparsity level $k$ on the accuracy of \textsc{CS-HiClose} in comparison with the CS-based competing methods in the case where the number of measurements set to $0.4 |V|$ and the measurements length set to $0.25 |V|$. The measurements length in DICeNod is defined according to another parameter $d = \frac{\varepsilon}{C k}m$, where $\varepsilon \in (0, \frac{1}{6})$ and $C > 1$. To have a fair comparison, we chose $\varepsilon$ and $C$ in a way that the average measurement length in this method and the other methods are the same. The higher the value of F-measure is, the more correlation between the top-$k$ nodes identified by a method and the global closeness centrality will be.

\subsubsection{Effect of Number of Measurements $m$ on Accuracy:}
The accuracy of \textsc{CS-HiClose} is compared to the existing CS-based methods in terms of F-measure for varying number of measurements, while the measurements length ($l$) set to $0.25 |V|$ and the sparsity ($k$) set to $0.15 |V|$ in a network with $|V|$ nodes. For DICeNod, $l$ is determined based on $m$ and $k$.
\begin{figure}[t!]
\begin{center}
{
\subfigure[
ca-AstroPh
]{\includegraphics[trim = 3mm 0mm 10mm 0mm, clip, scale =0.18]{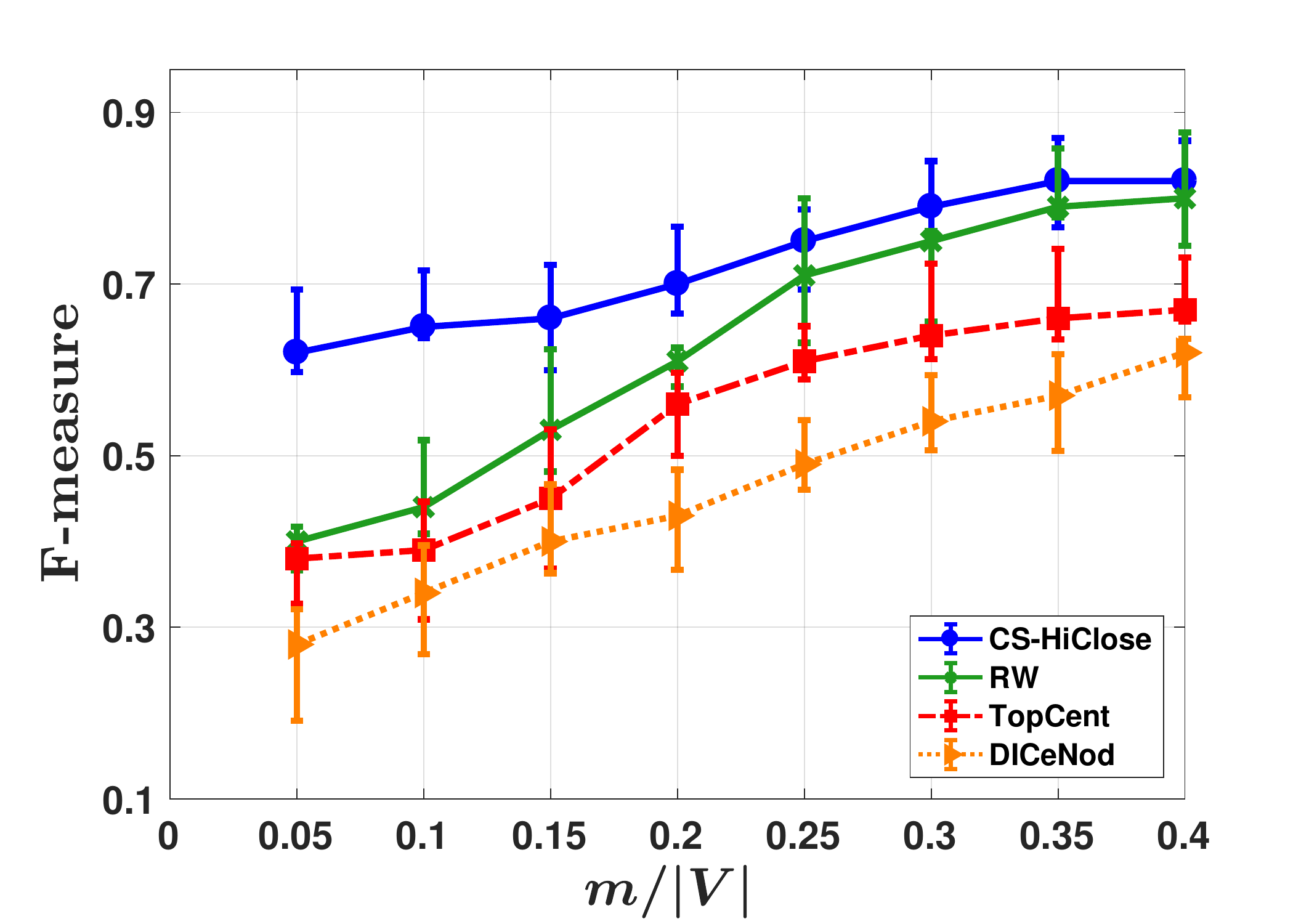}}
\subfigure[
Facebook
]{\includegraphics[trim = 3mm 0mm 10mm 0mm, clip, scale =0.18]{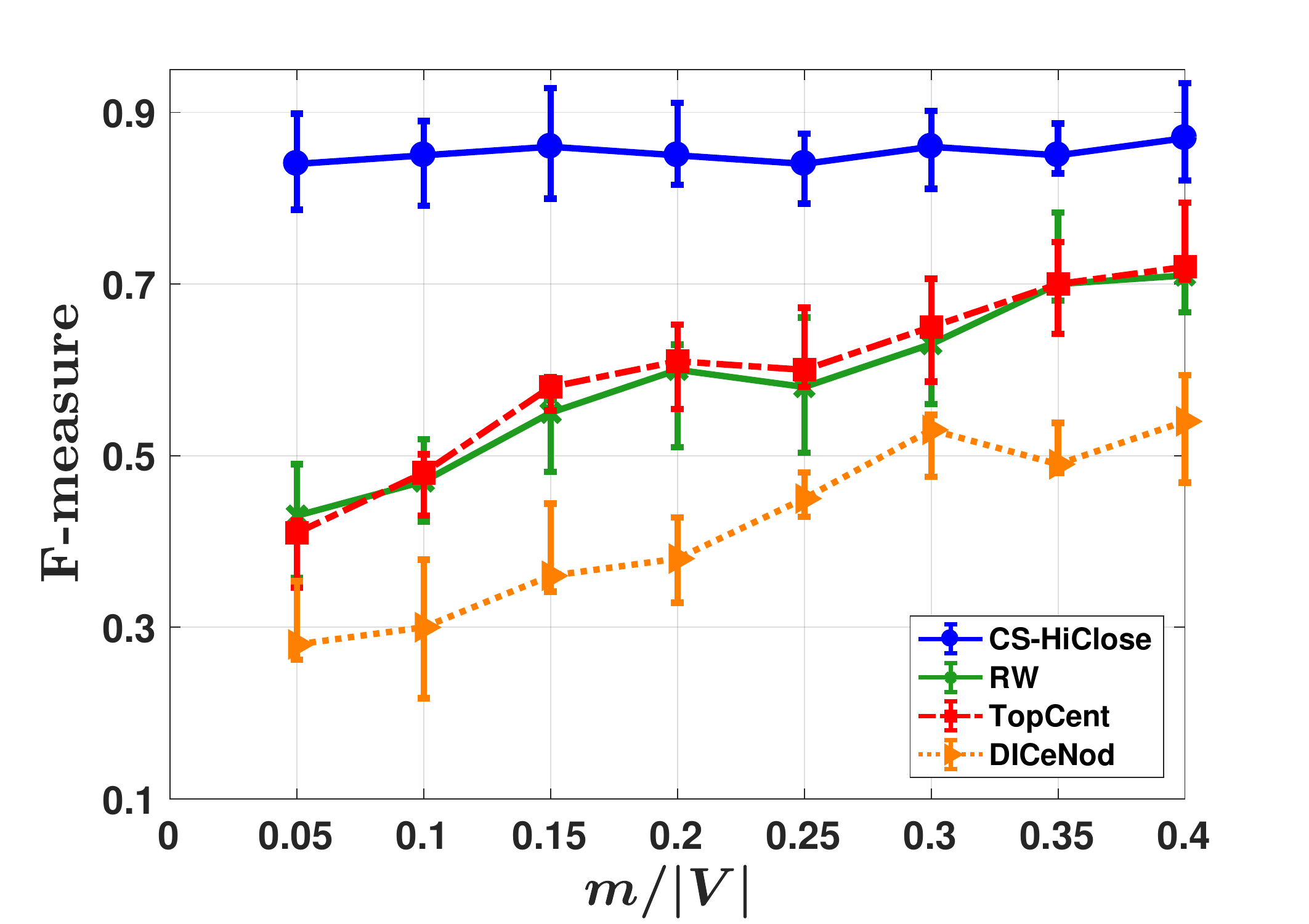}}
\subfigure[
Twitter
]{\includegraphics[trim = 3mm 0mm 10mm 0mm, clip, scale =0.18]{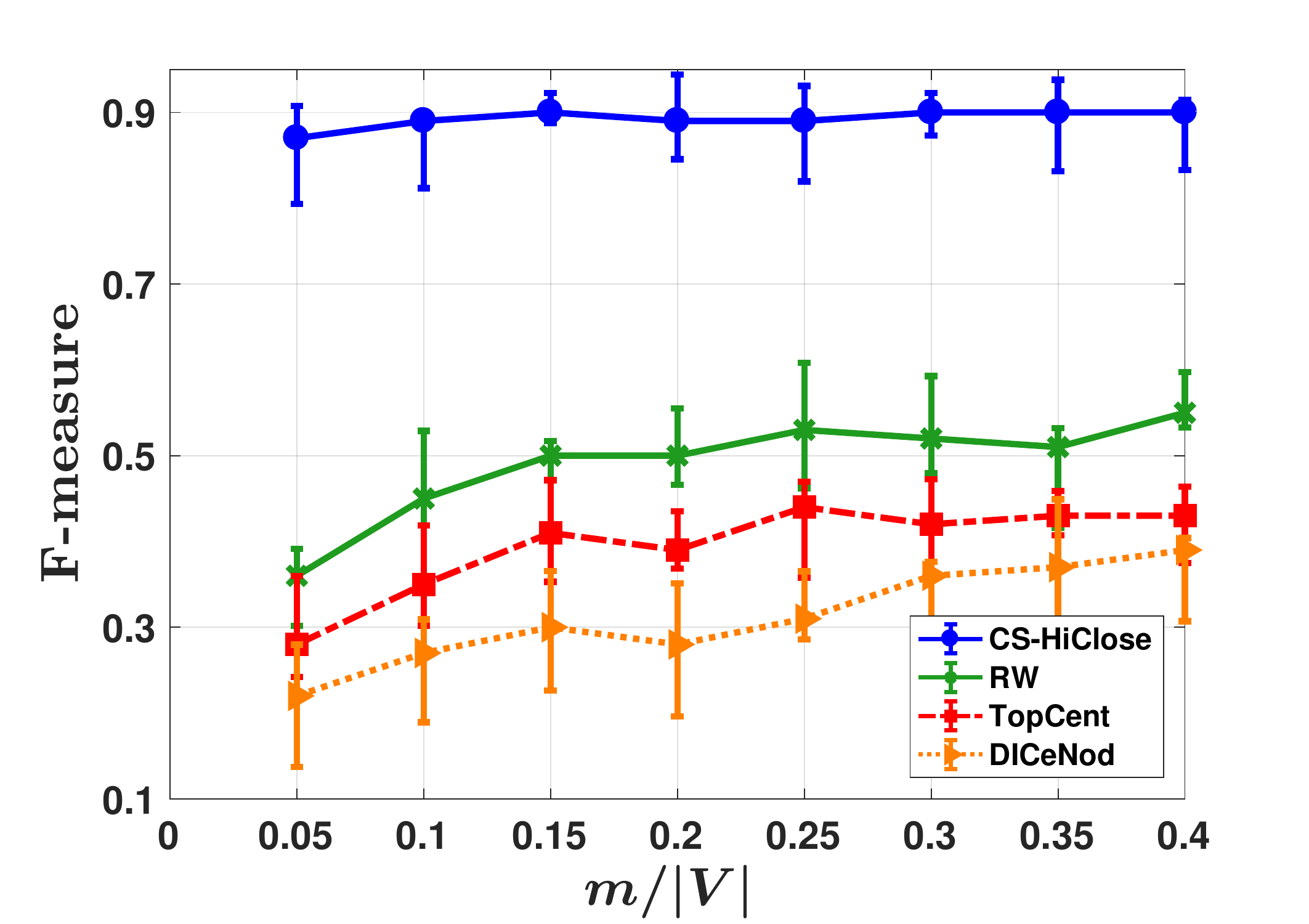}}
}
{
\subfigure[
ca-HepTh
]{\includegraphics[trim =3mm 0mm 10mm 0mm, clip, scale =0.18]{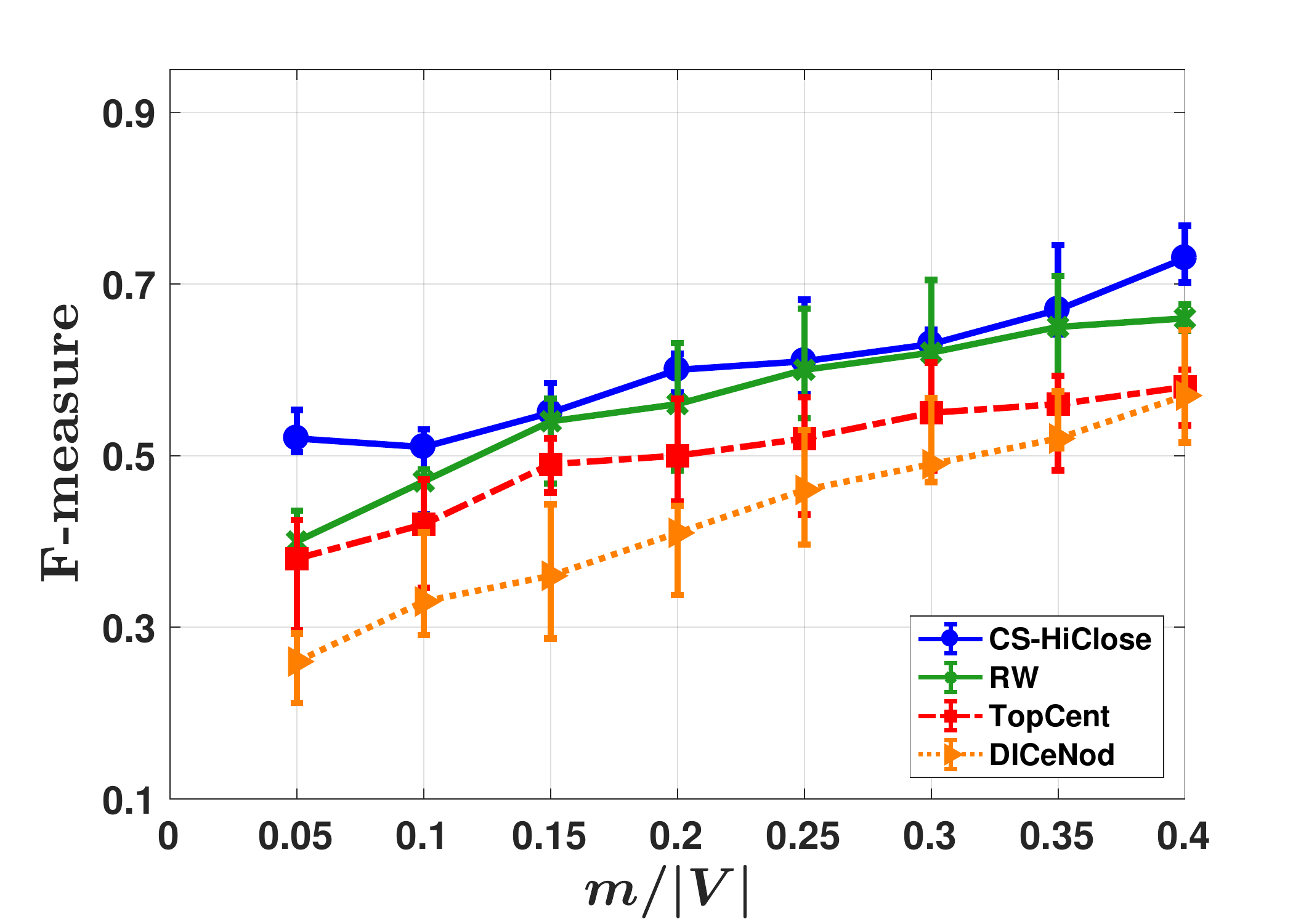}}
\subfigure[
wiki-Vote
]{\includegraphics[trim = 3mm 0mm 10mm 0mm, clip, scale =0.18]{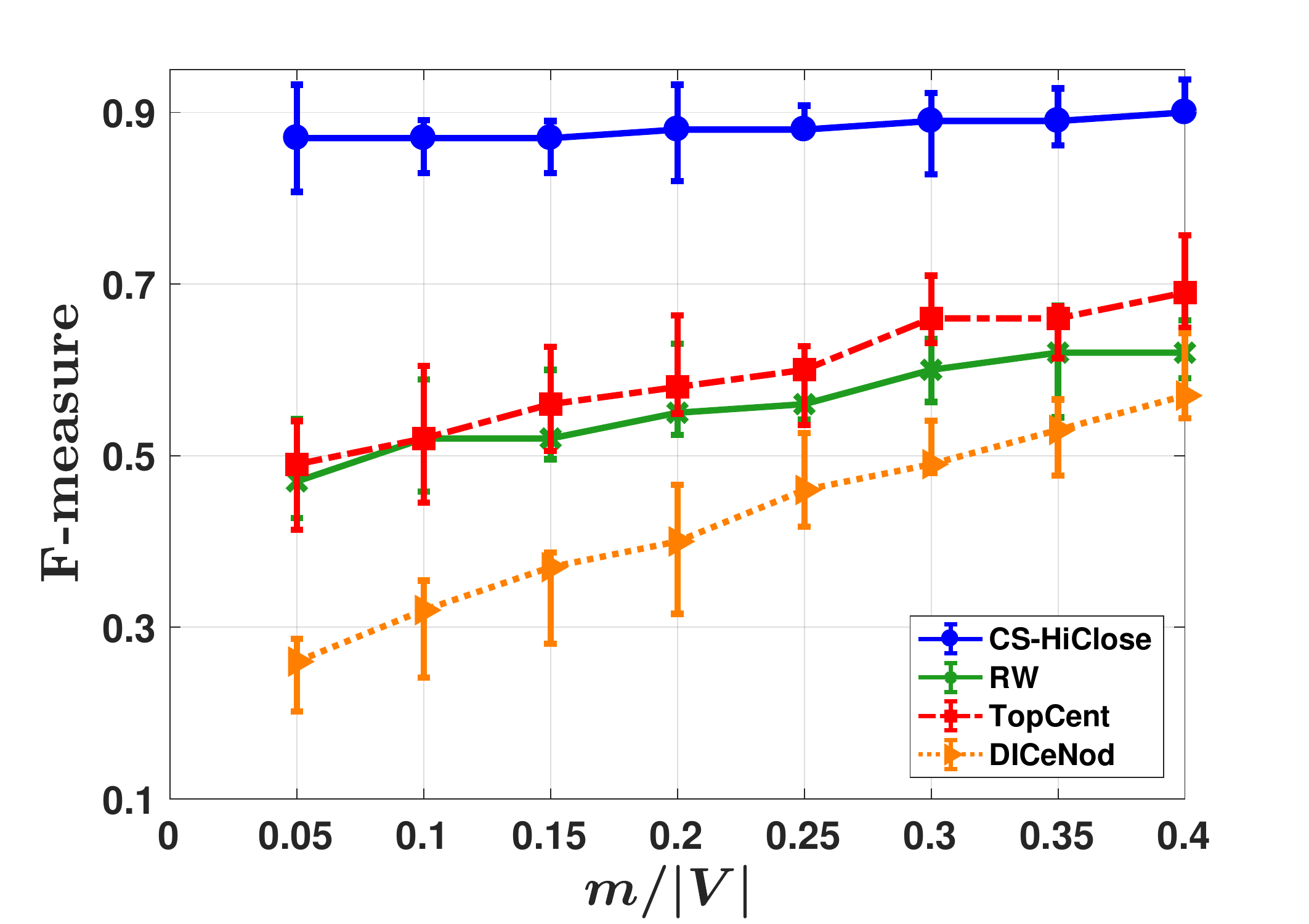}}
\subfigure[
ca-CondMat
]{\includegraphics[trim =3mm 0mm 10mm 0mm, clip, scale =0.18]{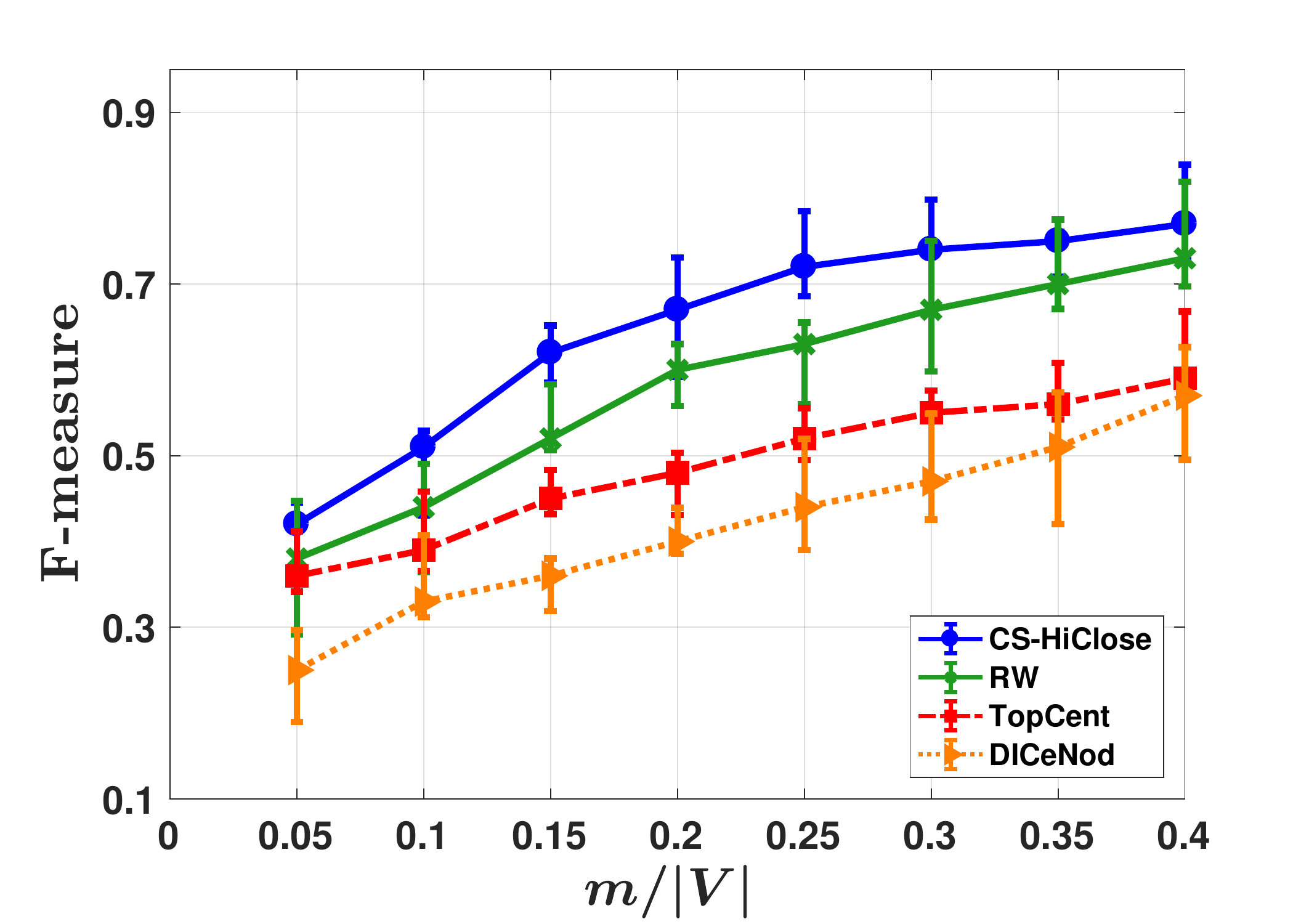}}
}
{
\subfigure[
ca-HepPh
]{\includegraphics[trim = 3mm 0mm 10mm 0mm, clip, scale =0.18]{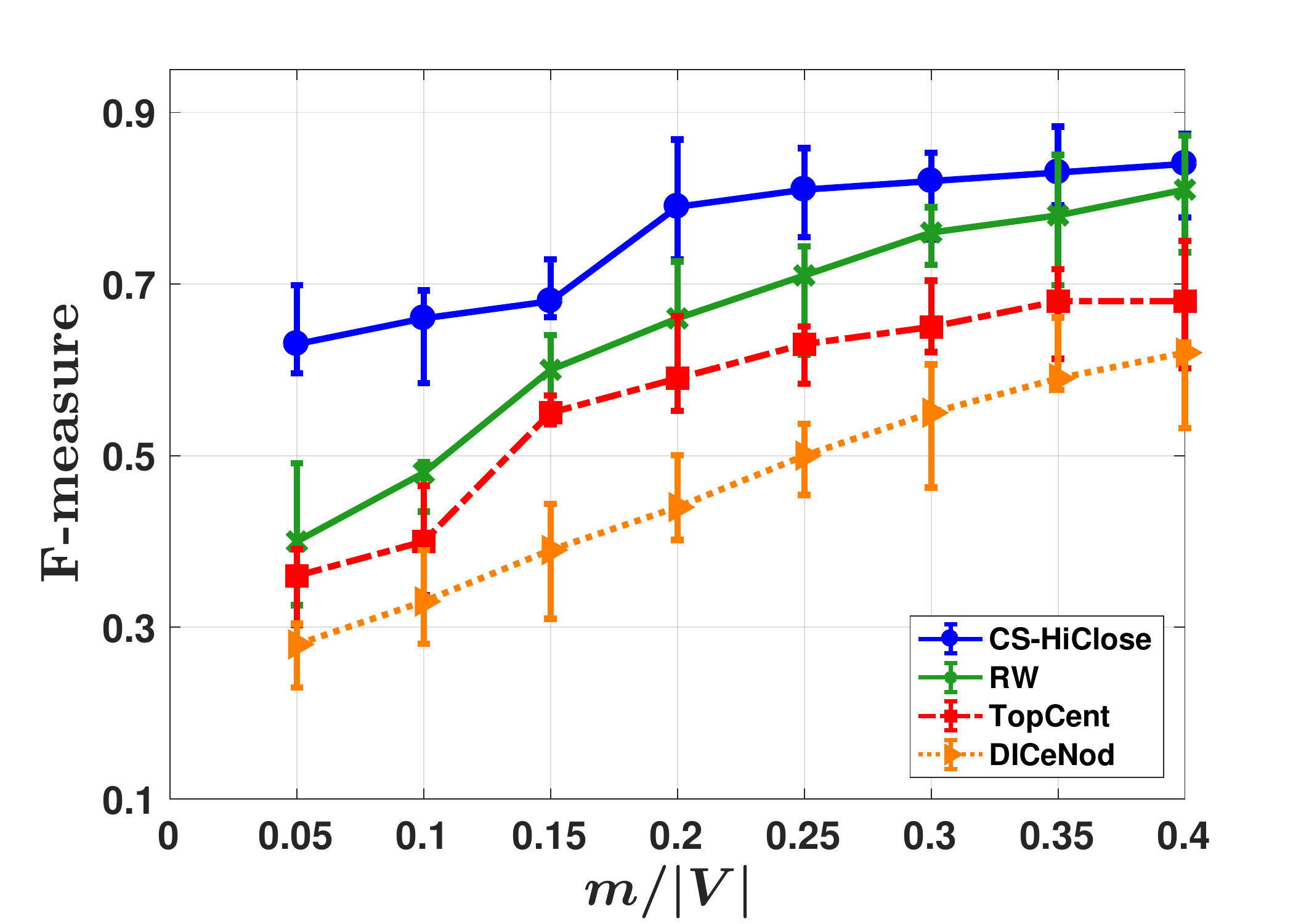}}
\subfigure[
DBLP
]{\includegraphics[trim = 3mm 0mm 10mm 0mm, clip, scale =0.18]{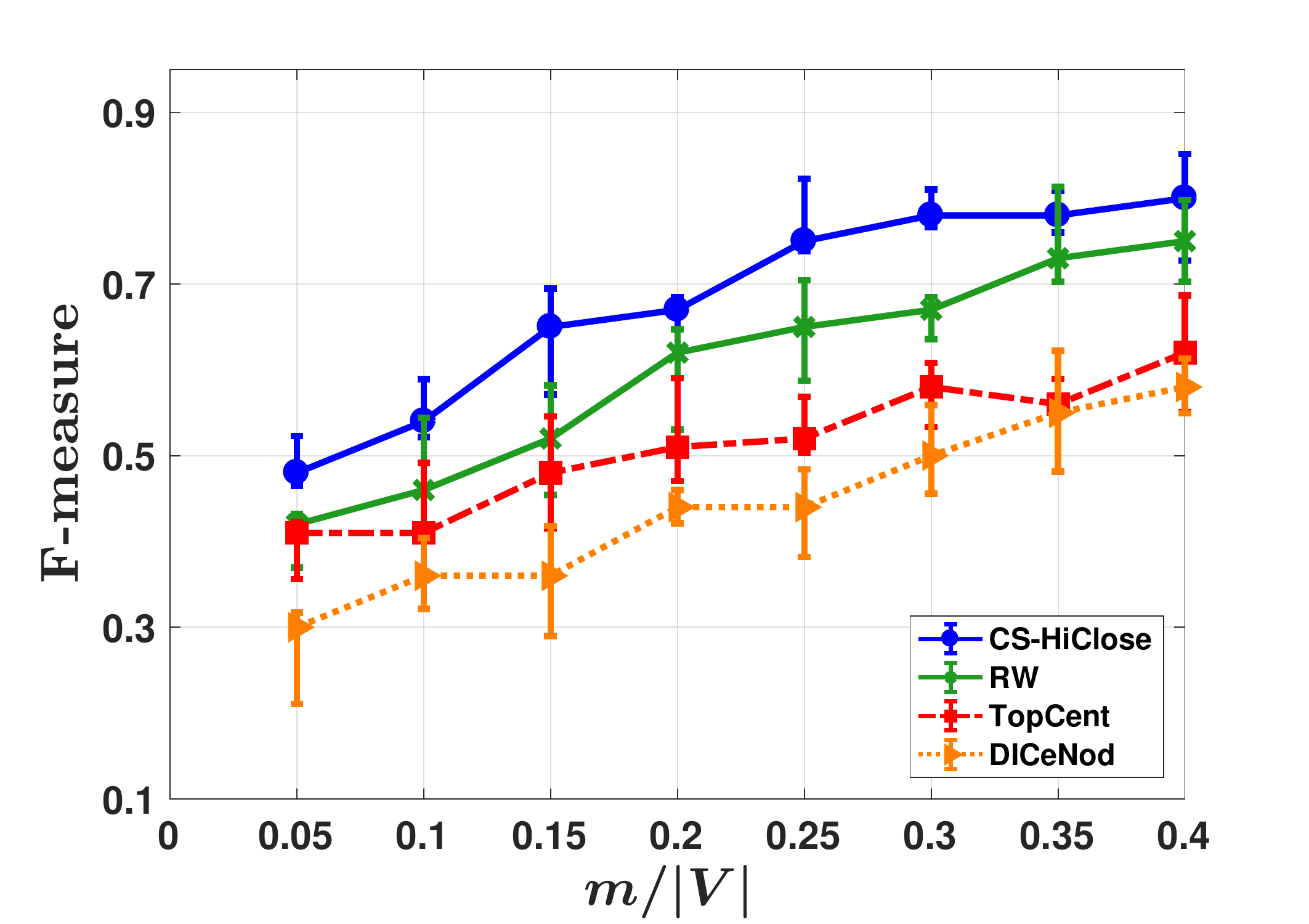}}
\subfigure[
email-Enron
]{\includegraphics[trim = 3mm 0mm 10mm 0mm, clip, scale =0.18]{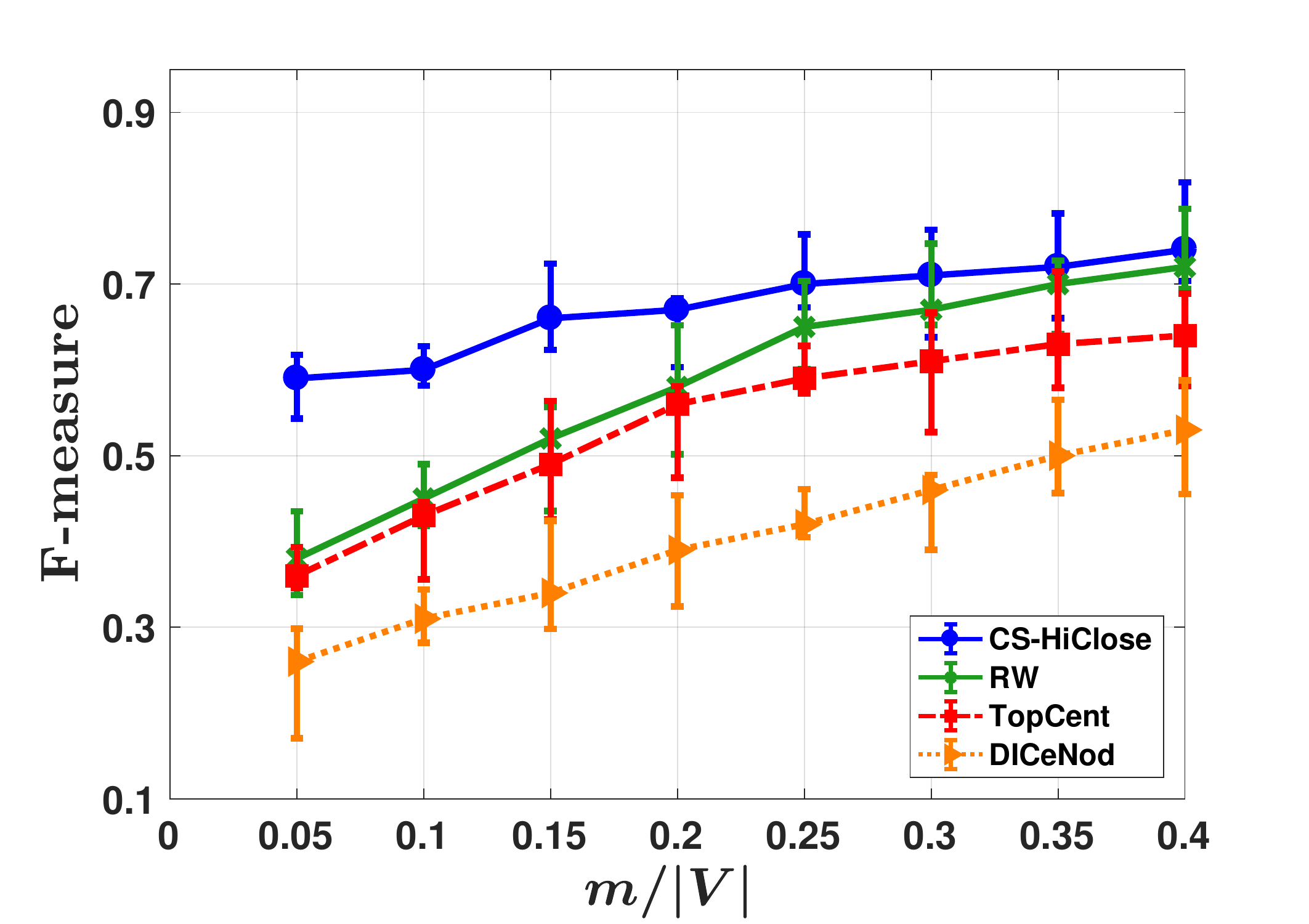}}
}
{
\subfigure[
BA
]{\includegraphics[trim = 3mm 0mm 10mm 0mm, clip, scale =0.18]{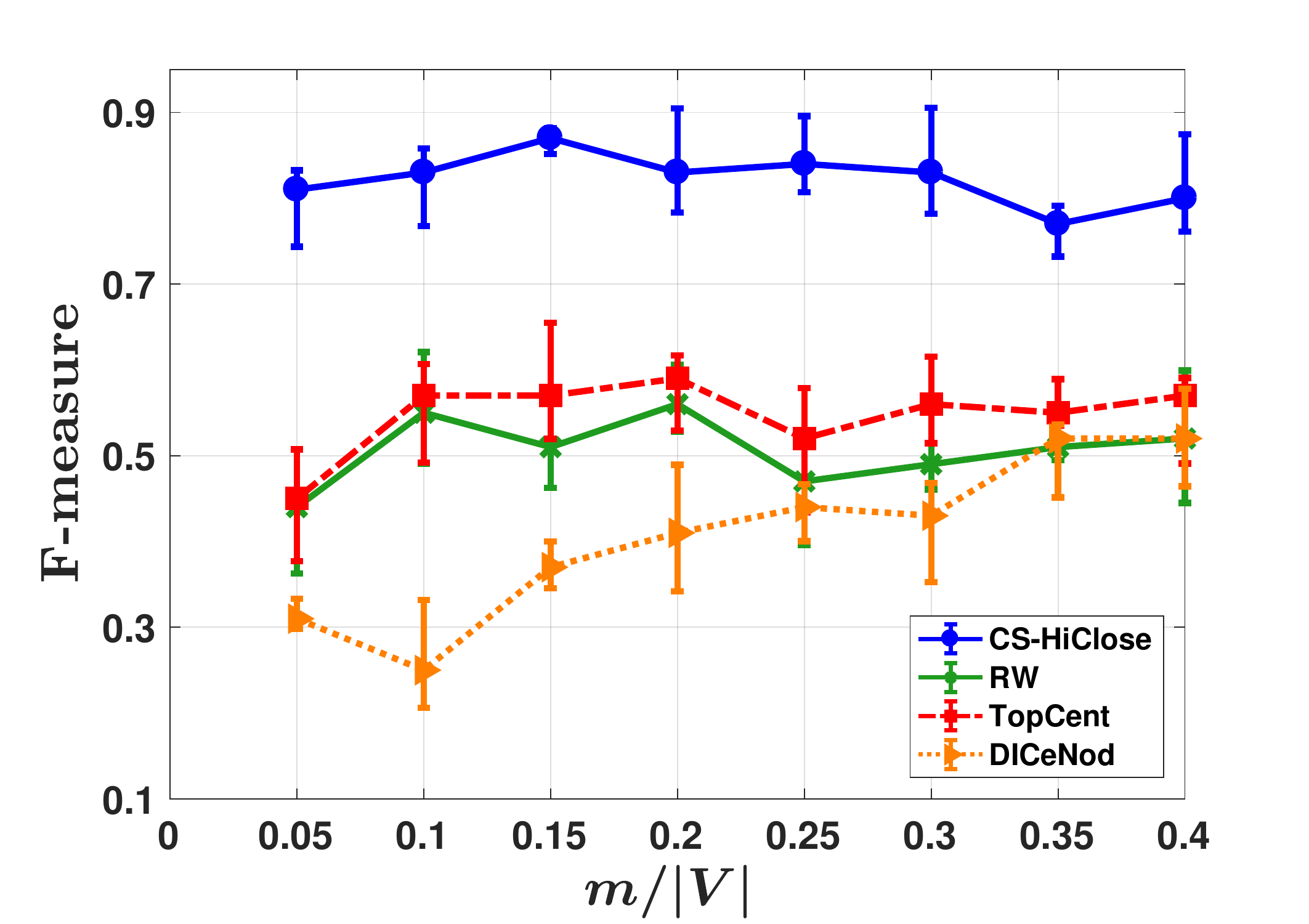}}
\subfigure[
ER
]{\includegraphics[trim = 3mm 0mm 10mm 0mm, clip, scale =0.18]{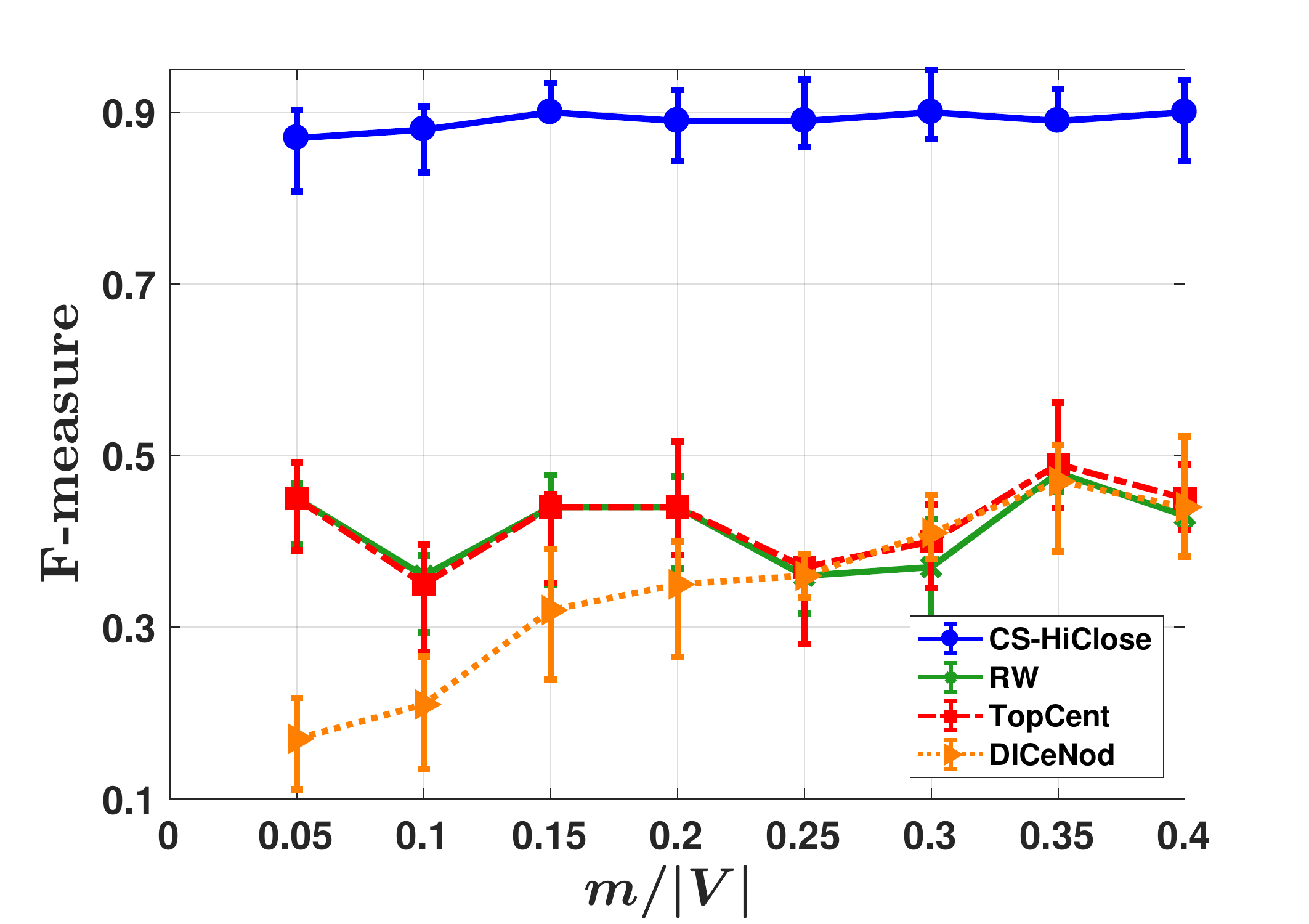}}
\subfigure[
SW
]{\includegraphics[trim = 3mm 0mm 10mm 0mm, clip, scale =0.18]{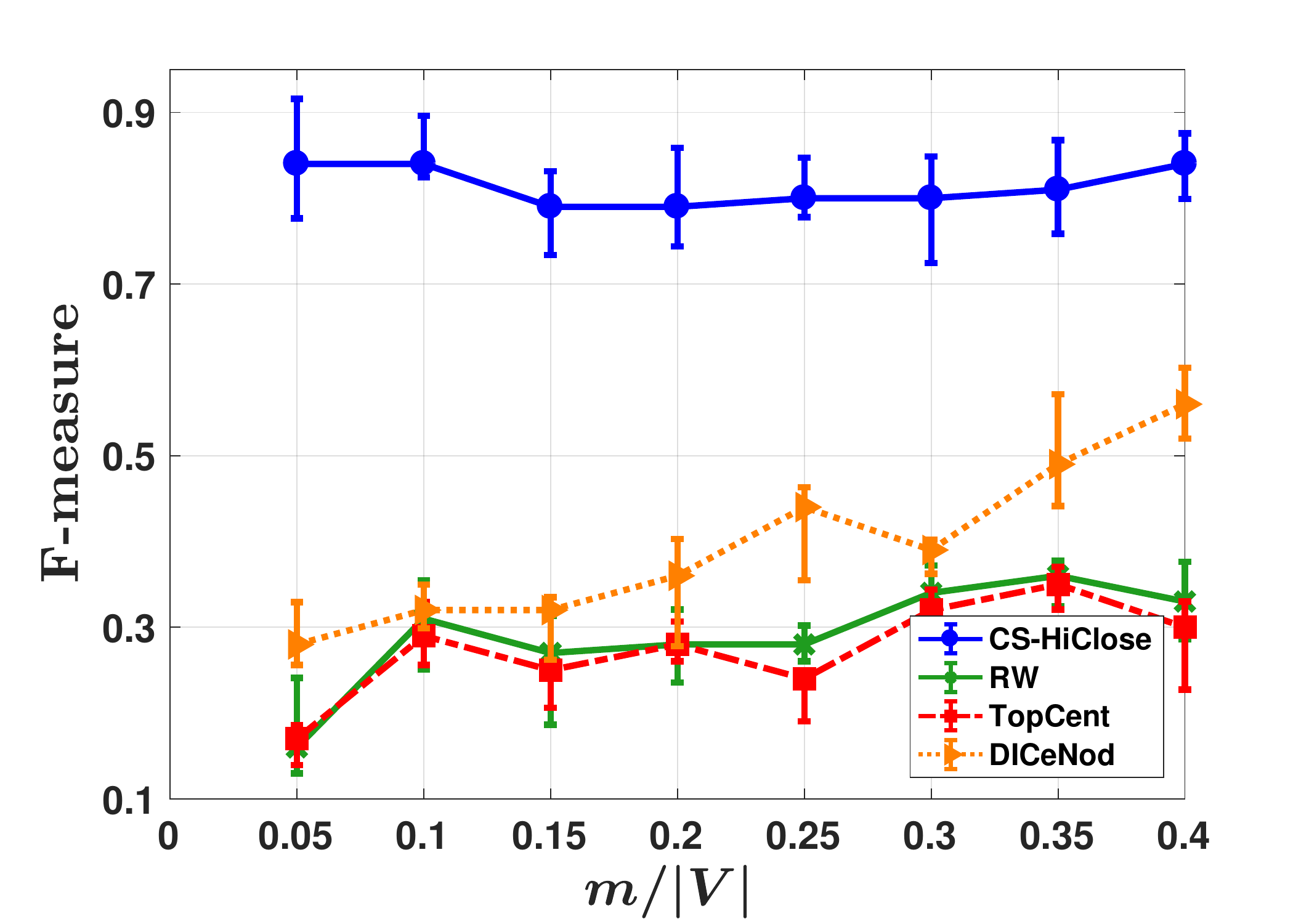}}
}
\caption{Effect of the required number of measurements $m$ on the accuracy of \textsc{CS-HiClose} in terms of F-measure, compared to RW, TopCent, and DICeNod. For each method, we set the measurements length to $0.25 |V|$ and the sparsity to $0.15 |V|$ in a network with $|V|$ nodes.}
\label{EffectM}
\end{center}
\end{figure}
In Figure~\ref{EffectM}, it is clearly depicted that \textsc{CS-HiClose} outperforms the competing methods in terms of having higher F-measure for almost all number of measurements. Moreover, our method has better accuracy even in small number of measurements. This improvement can be very important in the situations where performing measurements has a high computational cost~\cite{Mahyar2015LSRweighted,mahyar2013ucswn}. 

\begin{figure}[t!]
\begin{center}
{
\subfigure[
ca-AstroPh
]{\includegraphics[trim = 3mm 0mm 10mm 0mm, clip, scale =0.18]{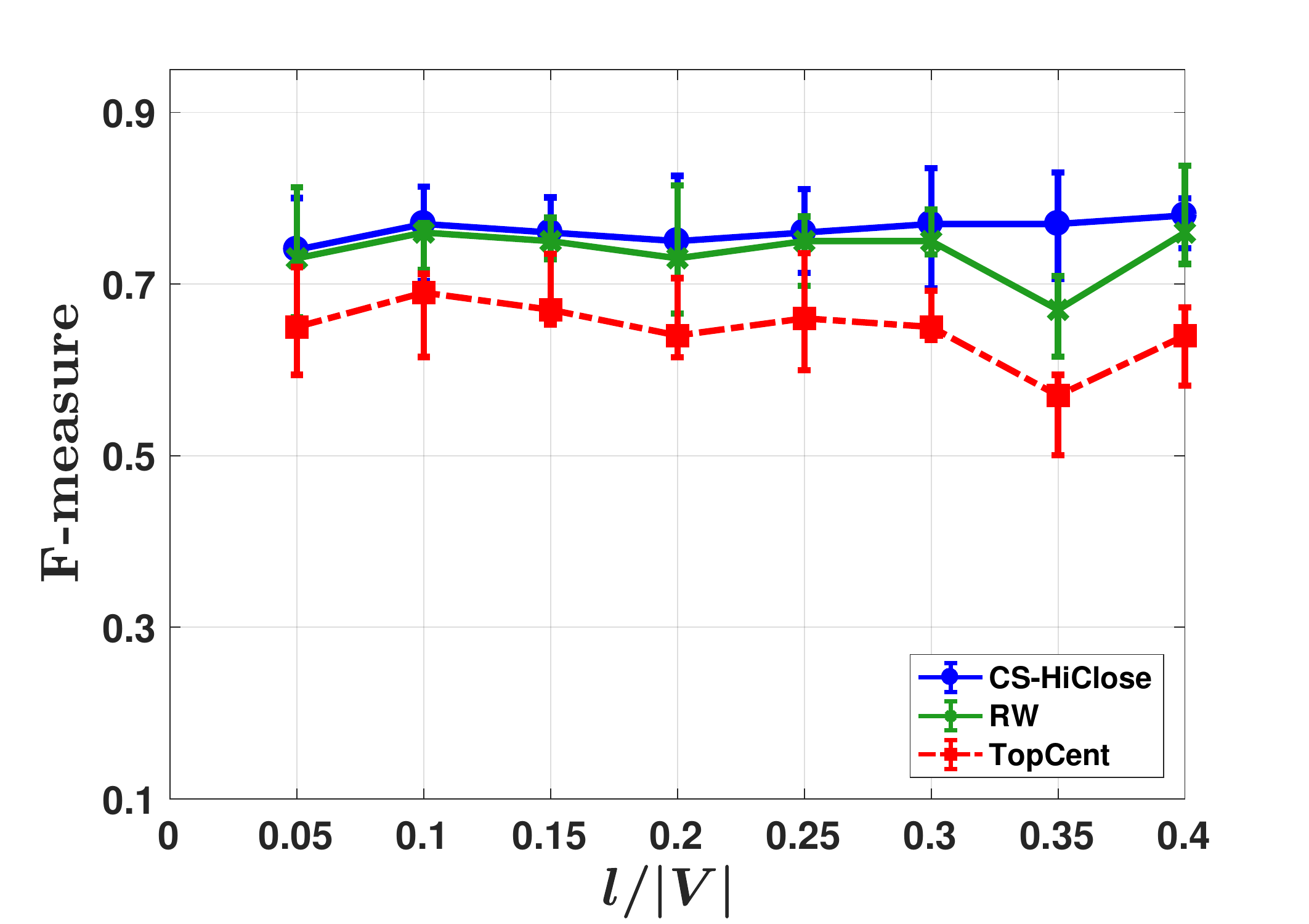}}
\subfigure[
Facebook
]{\includegraphics[trim = 3mm 0mm 10mm 0mm, clip, scale =0.18]{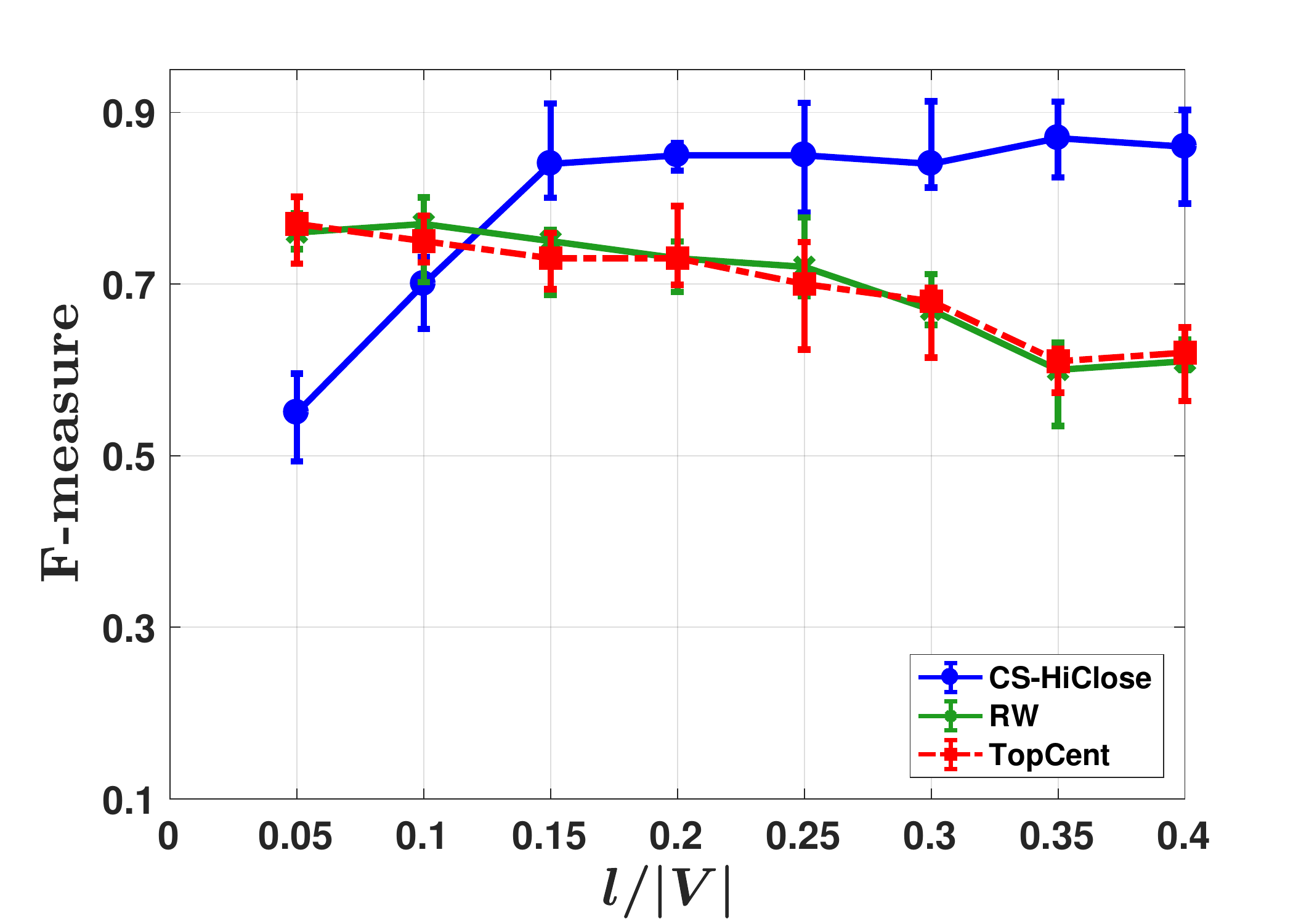}}
\subfigure[
Twitter
]{\includegraphics[trim = 3mm 0mm 10mm 0mm, clip, scale =0.18]{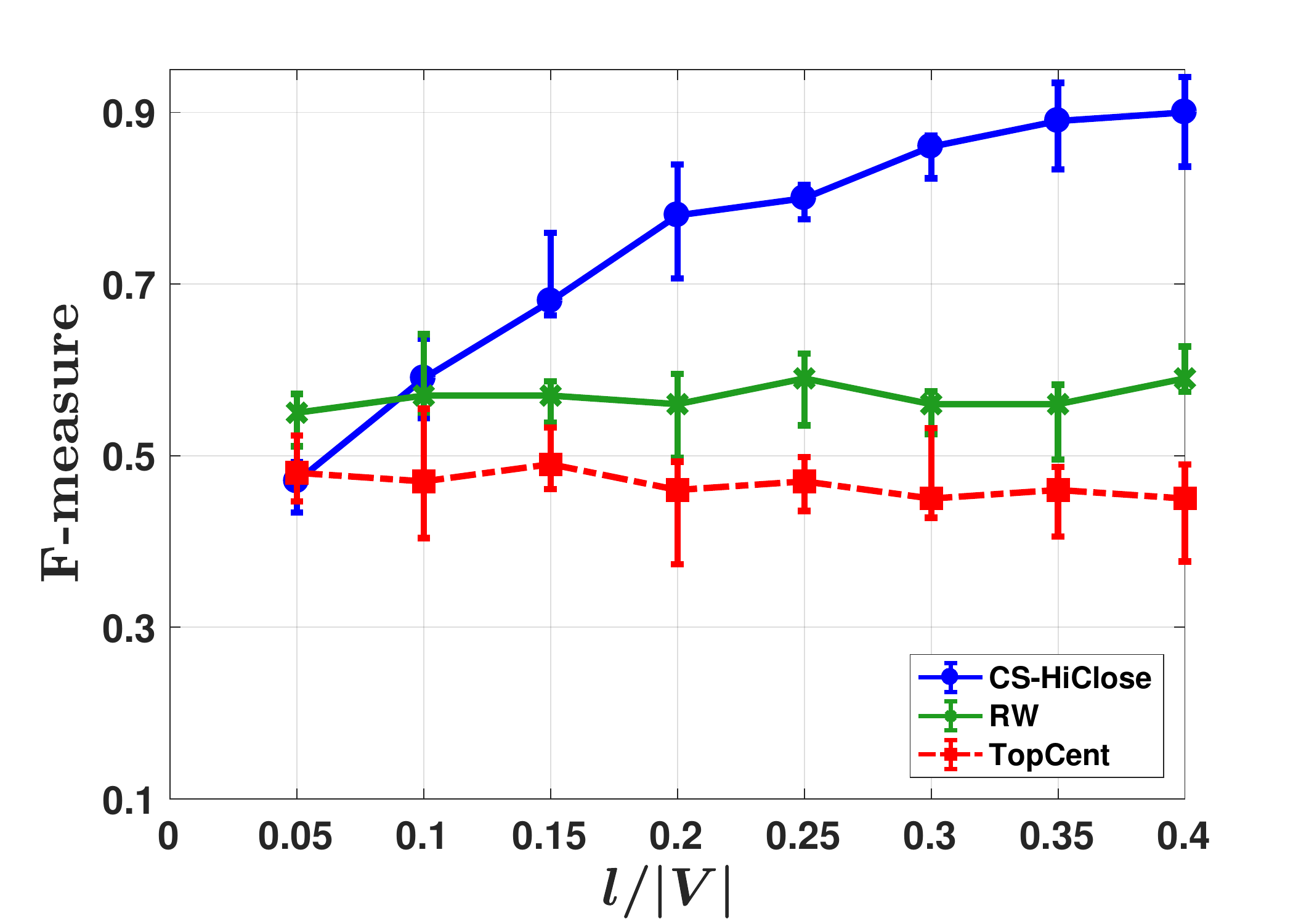}}
}
{
\subfigure[
ca-HepTh
]{\includegraphics[trim =3mm 0mm 10mm 0mm, clip, scale =0.18]{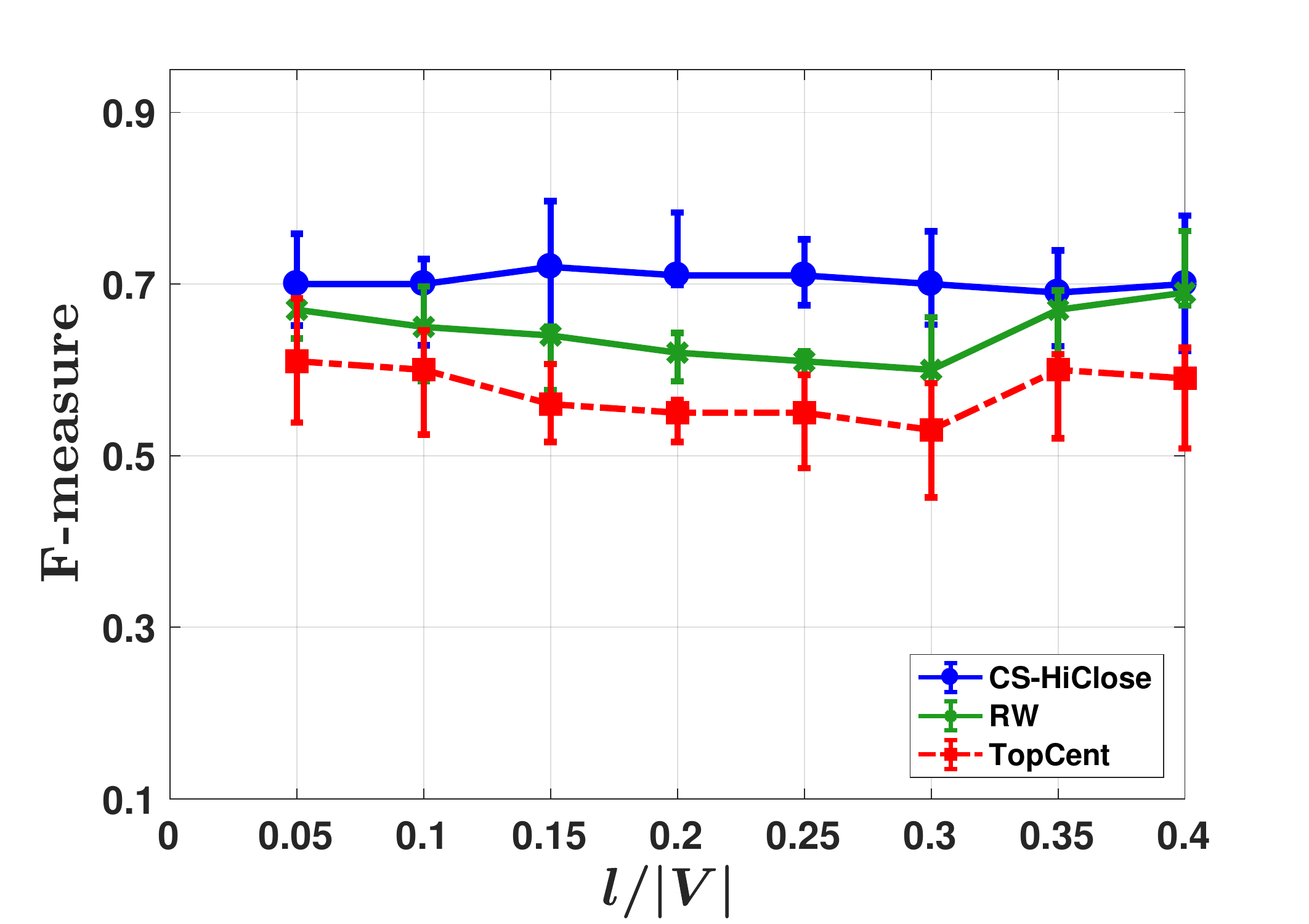}}
\subfigure[
wiki-Vote
]{\includegraphics[trim = 3mm 0mm 10mm 0mm, clip, scale =0.18]{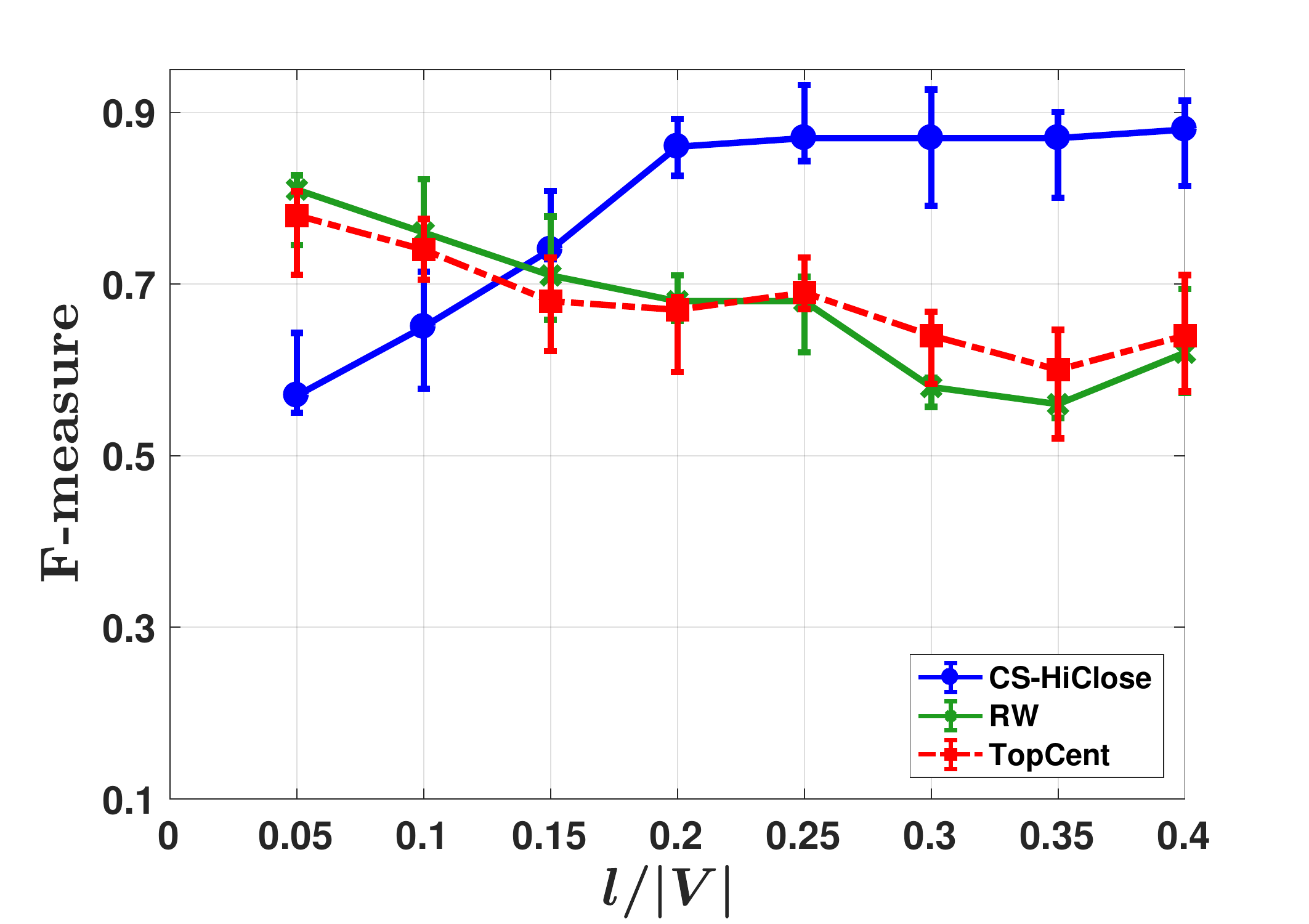}}
\subfigure[
ca-CondMat
]{\includegraphics[trim =3mm 0mm 10mm 0mm, clip, scale =0.18]{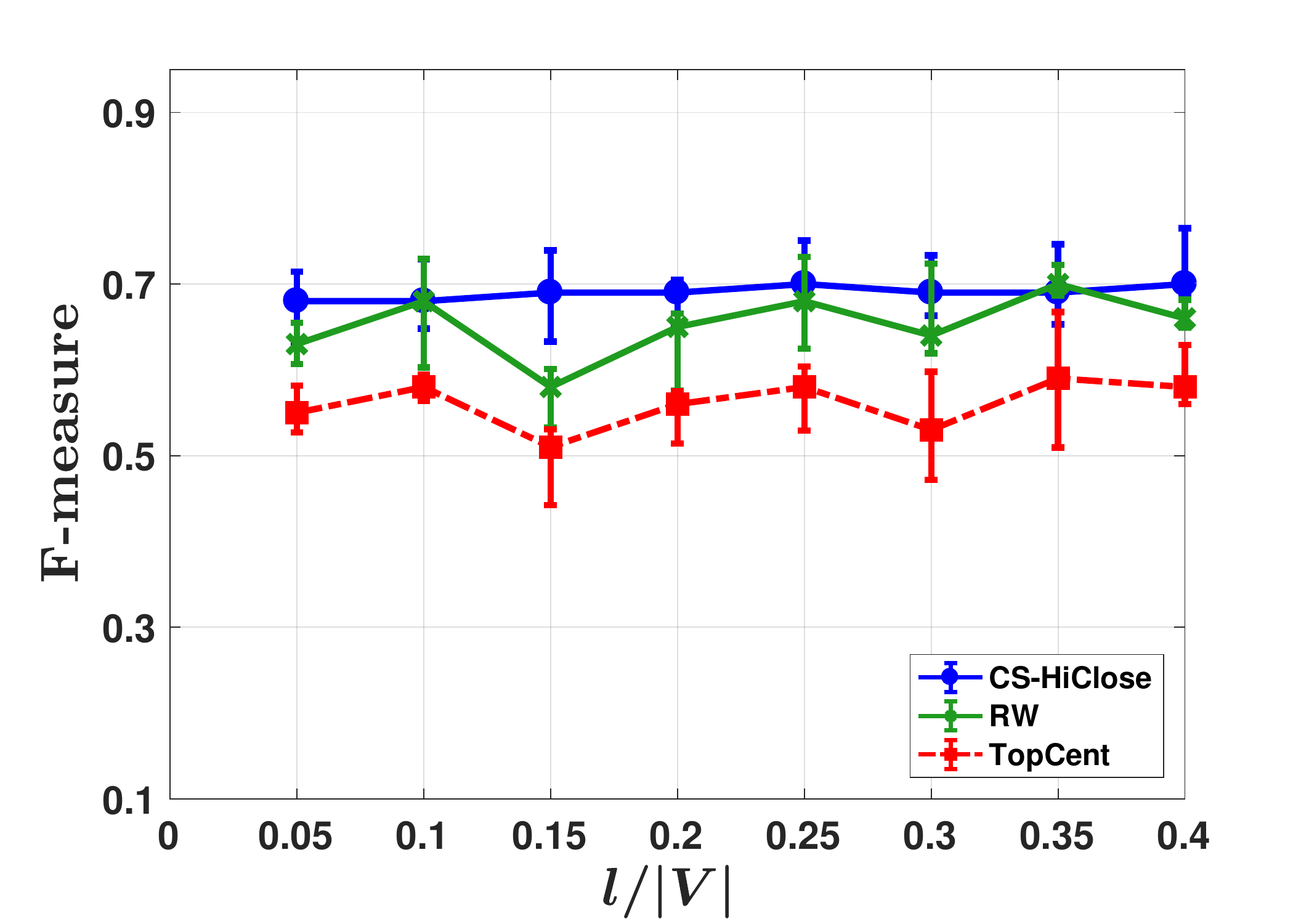}}
}
{
\subfigure[
ca-HepPh
]{\includegraphics[trim = 3mm 0mm 10mm 0mm, clip, scale =0.18]{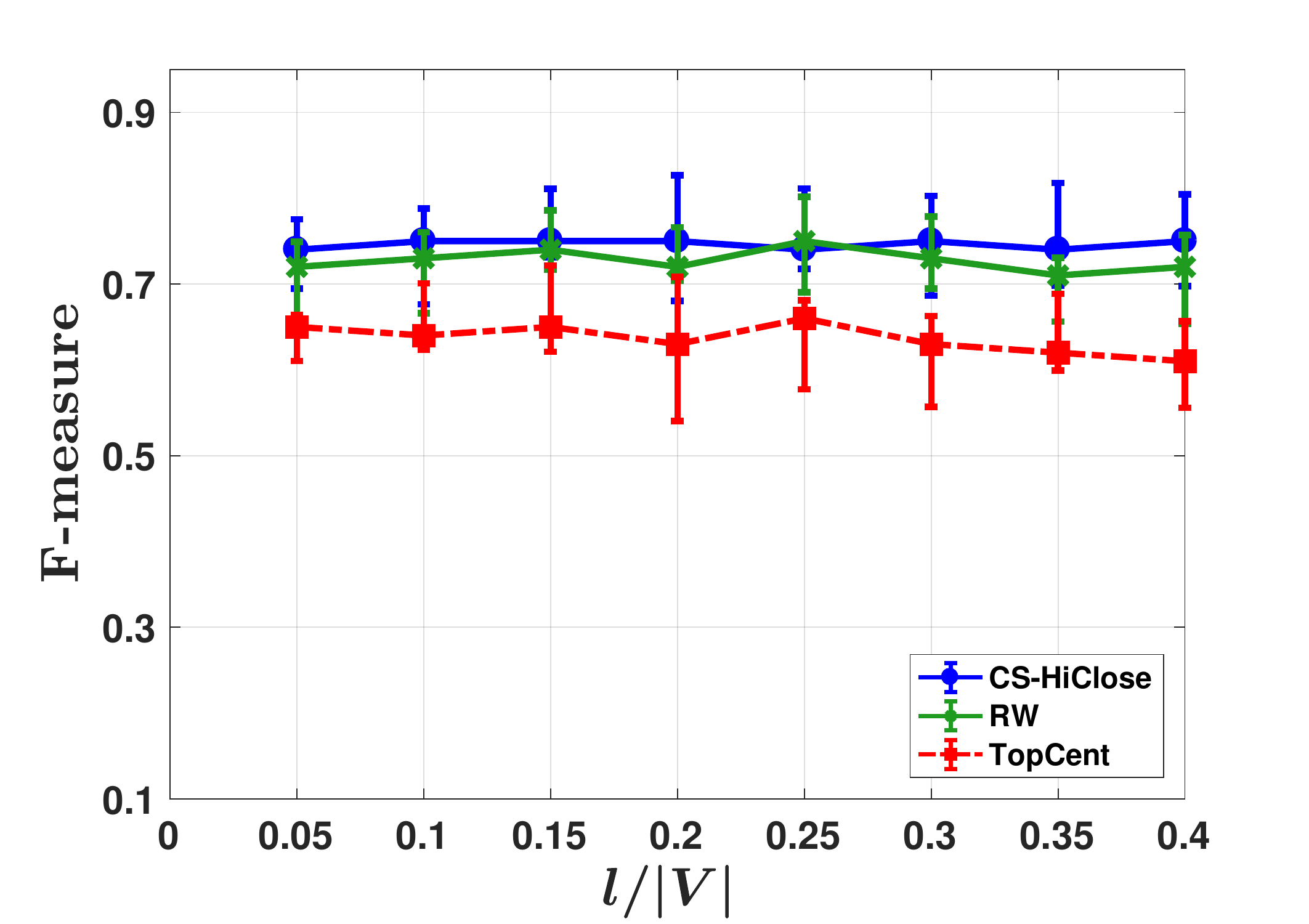}}
\subfigure[
DBLP
]{\includegraphics[trim = 3mm 0mm 10mm 0mm, clip, scale =0.18]{CondMat_L.pdf}}
\subfigure[
email-Enron
]{\includegraphics[trim = 3mm 0mm 10mm 0mm, clip, scale =0.18]{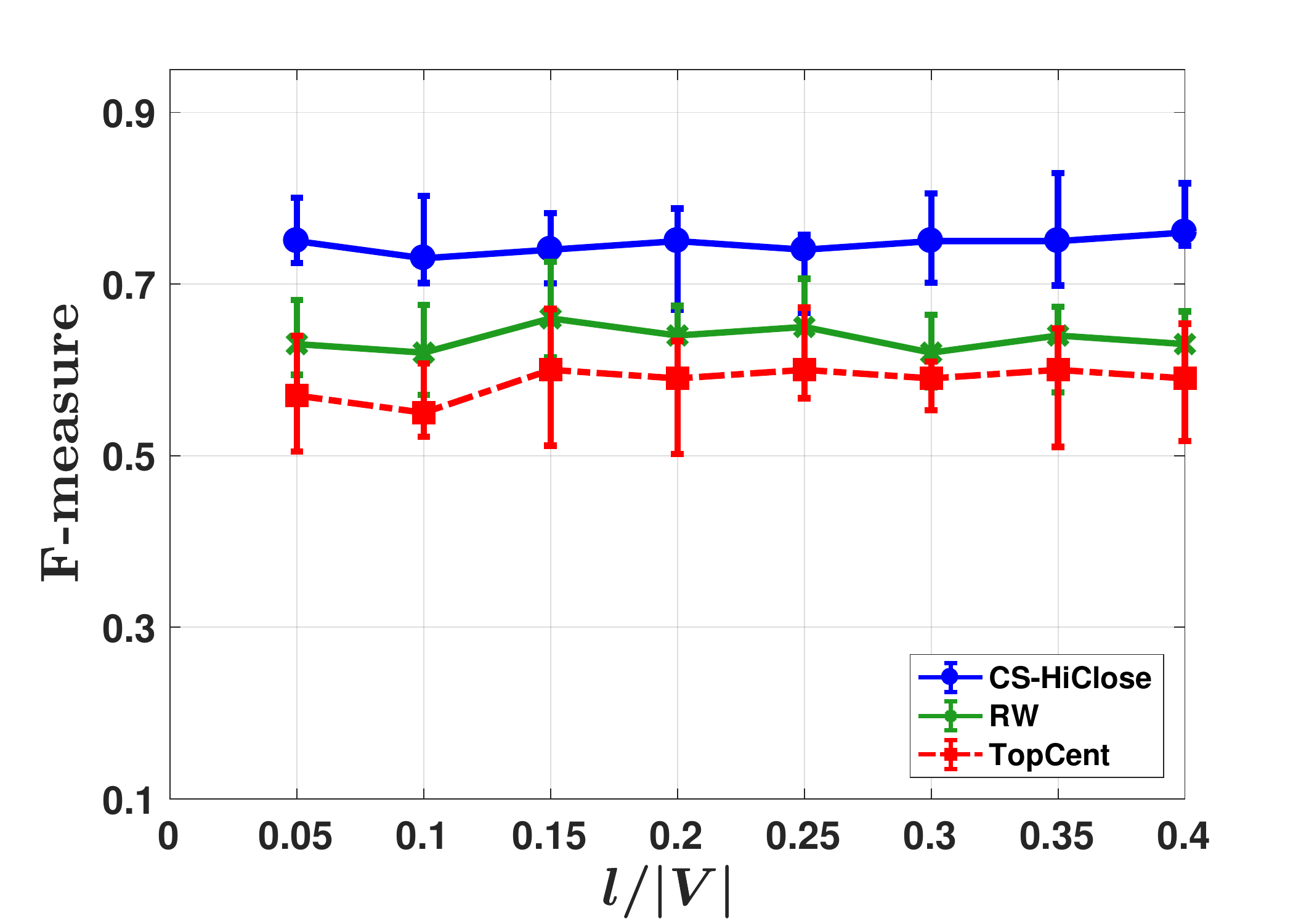}}
}
{
\subfigure[
BA
]{\includegraphics[trim = 3mm 0mm 10mm 0mm, clip, scale =0.18]{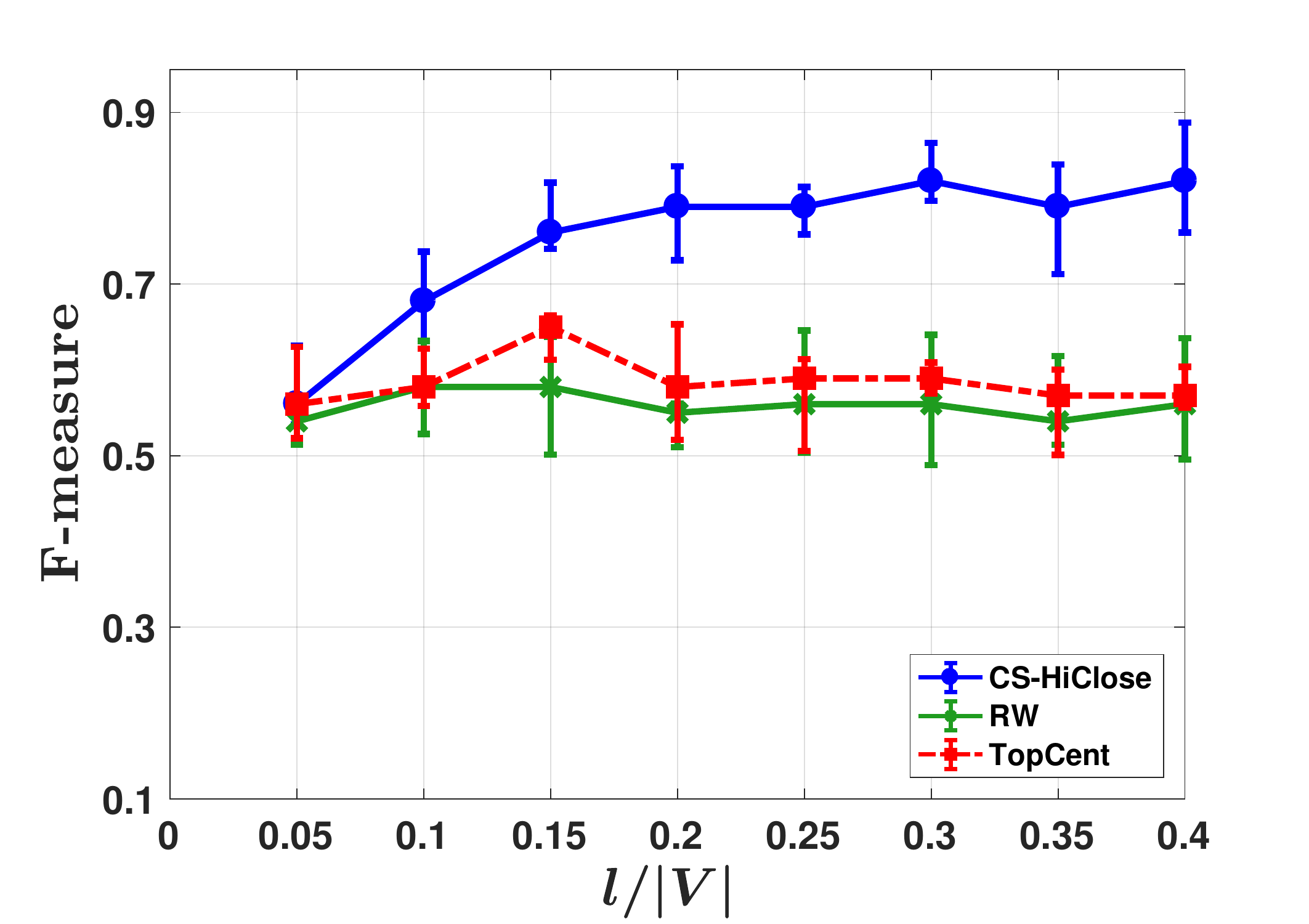}}
\subfigure[
ER
]{\includegraphics[trim = 3mm 0mm 10mm 0mm, clip, scale =0.18]{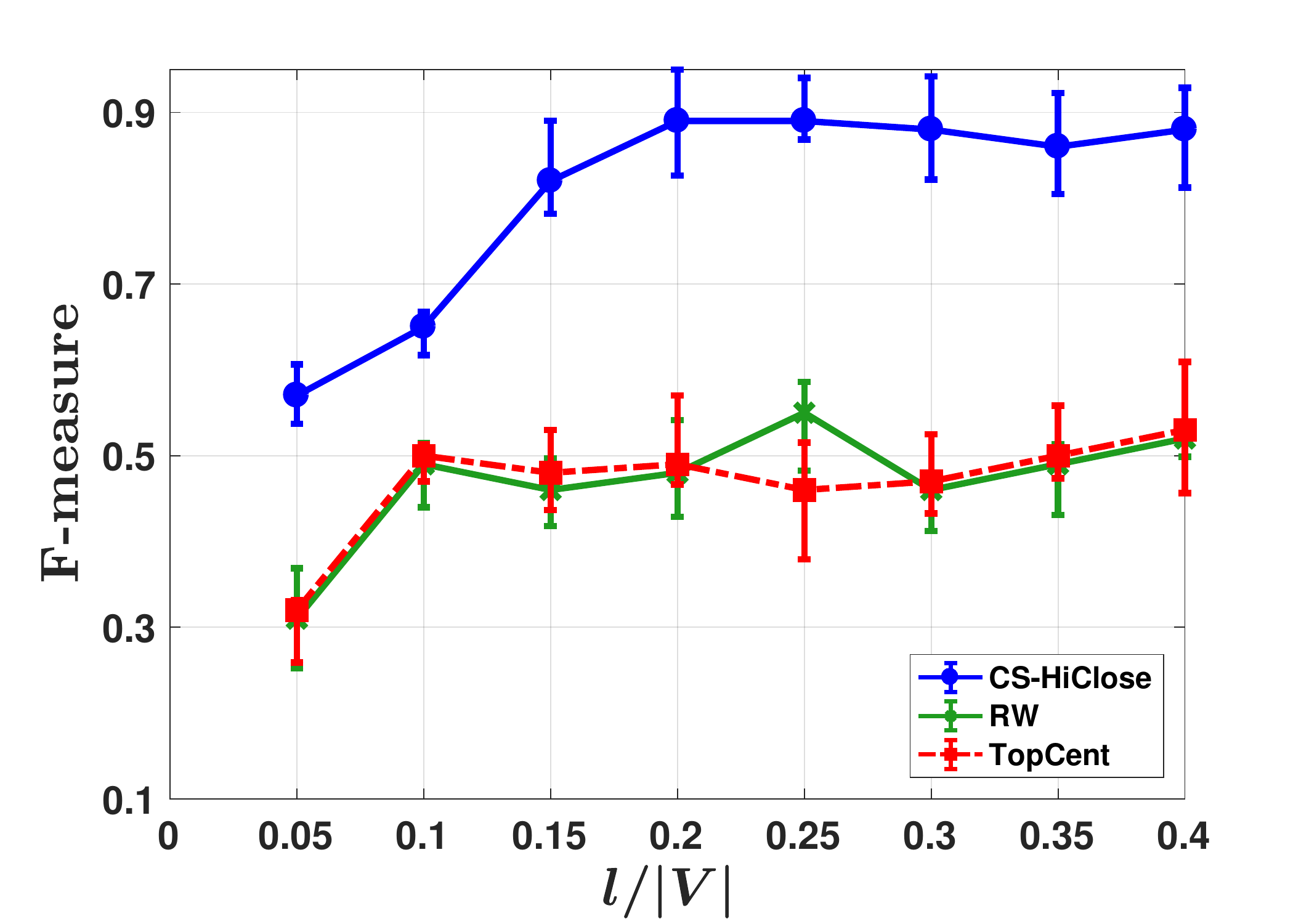}}
\subfigure[
SW
]{\includegraphics[trim = 3mm 0mm 10mm 0mm, clip, scale =0.18]{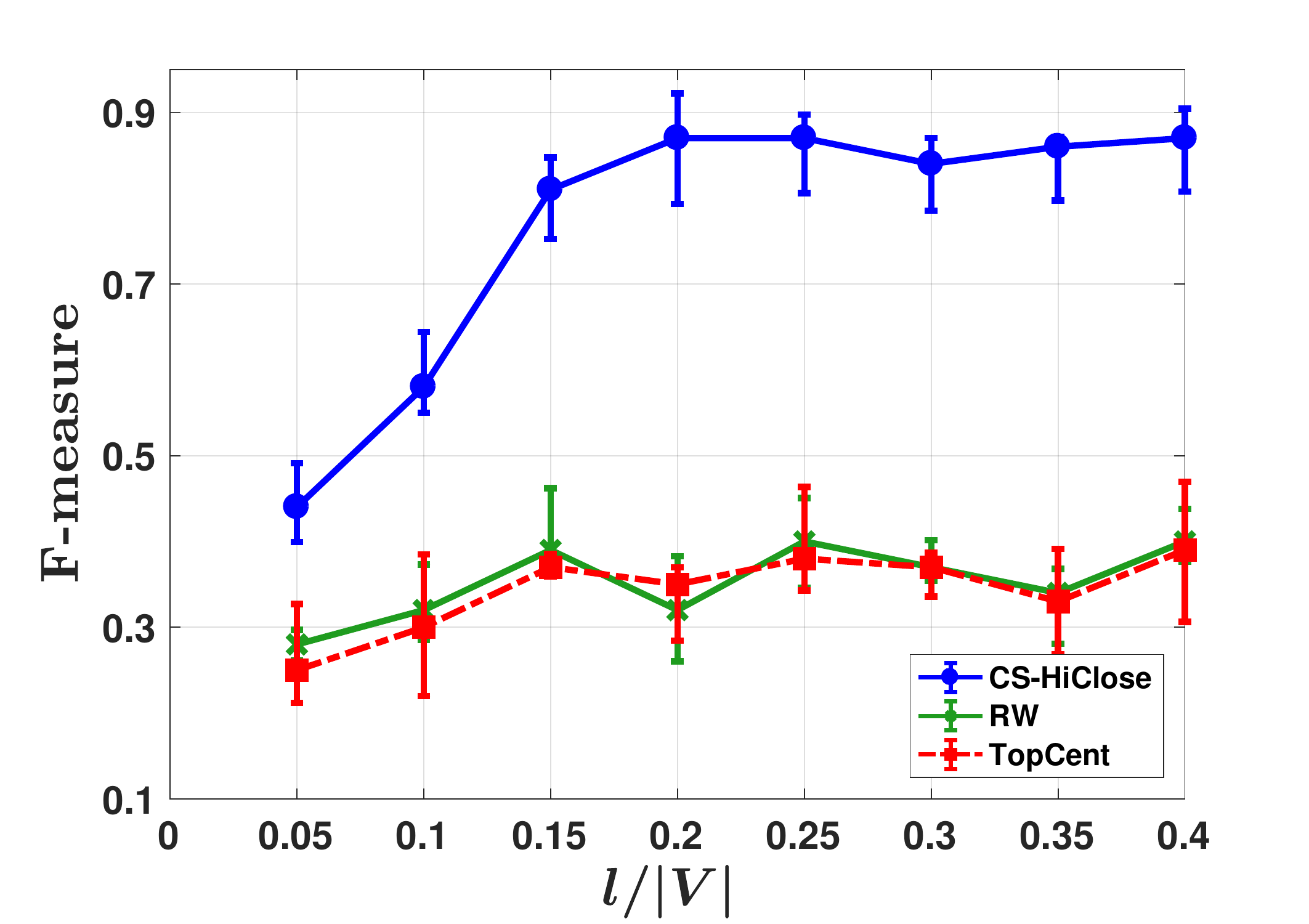}}
}
\caption{Effect of measurement length $l$ on the accuracy of \textsc{CS-HiClose} in terms of F-measure, compared to RW, TopCent, and DICeNod. For each method, we set the number of measurements to $0.4 |V|$ and the sparsity level to $0.2 |V|$ in a network with $|V|$ nodes.}
\label{EffectL}
\end{center}
\end{figure}

\subsubsection{Effect of Measurement Length $l$ on Accuracy}
Figure~\ref{EffectL} illustrates that \textsc{CS-HiClose} has higher F-measure for the most measurement lengths in all test cases, in comparison with the CS-based methods RW, TopCent and DICeNod.
Since the concept of measurement length is again irrelevant to the other competing methods, we only compared our accuracy with the CS-based approaches. The horizontal axis in Figure~\ref{EffectL} shows the measurement length $l$ divided by the total number of network nodes $|V|$ (\textit{i.e.} $\frac{l}{|V|}$). This experiment is performed over the network with $|V|$ nodes where the number of measurements sets to $m = 0.4 |V|$ and the sparsity level sets to $k = 0.2 |V|$ for all methods. We repeated each test 10 times to reduce the methods' randomness and the points in the figures show the mean value of these repetitions. In Figure~\ref{EffectL}, we can observe an increasing trend for F-measure in \textsc{CS-HiClose} when we increase the measurements length.

\section{Conclusion}\label{conclusions}
Closeness centrality has been utilized as a primary metric to measure the relative importance/influence of nodes in a given network. In this paper, we introduced a new ego-centric metric which has very low computational cost and correlates well with the global closeness centrality. Then, we proposed a compressive sensing framework for distributed detection of top-$k$ central nodes based on the ego-closeness metric using only indirect measurements. Extensive simulations experimental evaluations on both synthetic and real networks demonstrated that the proposed method outperforms the best existing methods to efficiently detect high closeness centrality nodes, in terms of having high F-measure and low complexity.


\begin{backmatter}




\bibliographystyle{bmc-mathphys}
\bibliography{Mahyar_Closeness_Journal}


\begin{thebibliography}{34}
\ifx \bisbn   \undefined \def \bisbn  #1{ISBN #1}\fi
\ifx \binits  \undefined \def \binits#1{#1}\fi
\ifx \bauthor  \undefined \def \bauthor#1{#1}\fi
\ifx \batitle  \undefined \def \batitle#1{#1}\fi
\ifx \bjtitle  \undefined \def \bjtitle#1{#1}\fi
\ifx \bvolume  \undefined \def \bvolume#1{\textbf{#1}}\fi
\ifx \byear  \undefined \def \byear#1{#1}\fi
\ifx \bissue  \undefined \def \bissue#1{#1}\fi
\ifx \bfpage  \undefined \def \bfpage#1{#1}\fi
\ifx \blpage  \undefined \def \blpage #1{#1}\fi
\ifx \burl  \undefined \def \burl#1{\textsf{#1}}\fi
\ifx \doiurl  \undefined \def \doiurl#1{\textsf{#1}}\fi
\ifx \betal  \undefined \def \betal{\textit{et al.}}\fi
\ifx \binstitute  \undefined \def \binstitute#1{#1}\fi
\ifx \binstitutionaled  \undefined \def \binstitutionaled#1{#1}\fi
\ifx \bctitle  \undefined \def \bctitle#1{#1}\fi
\ifx \beditor  \undefined \def \beditor#1{#1}\fi
\ifx \bpublisher  \undefined \def \bpublisher#1{#1}\fi
\ifx \bbtitle  \undefined \def \bbtitle#1{#1}\fi
\ifx \bedition  \undefined \def \bedition#1{#1}\fi
\ifx \bseriesno  \undefined \def \bseriesno#1{#1}\fi
\ifx \blocation  \undefined \def \blocation#1{#1}\fi
\ifx \bsertitle  \undefined \def \bsertitle#1{#1}\fi
\ifx \bsnm \undefined \def \bsnm#1{#1}\fi
\ifx \bsuffix \undefined \def \bsuffix#1{#1}\fi
\ifx \bparticle \undefined \def \bparticle#1{#1}\fi
\ifx \barticle \undefined \def \barticle#1{#1}\fi
\ifx \bconfdate \undefined \def \bconfdate #1{#1}\fi
\ifx \botherref \undefined \def \botherref #1{#1}\fi
\ifx \url \undefined \def \url#1{\textsf{#1}}\fi
\ifx \bchapter \undefined \def \bchapter#1{#1}\fi
\ifx \bbook \undefined \def \bbook#1{#1}\fi
\ifx \bcomment \undefined \def \bcomment#1{#1}\fi
\ifx \oauthor \undefined \def \oauthor#1{#1}\fi
\ifx \citeauthoryear \undefined \def \citeauthoryear#1{#1}\fi
\ifx \endbibitem  \undefined \def \endbibitem {}\fi
\ifx \bconflocation  \undefined \def \bconflocation#1{#1}\fi
\ifx \arxivurl  \undefined \def \arxivurl#1{\textsf{#1}}\fi
\csname PreBibitemsHook\endcsname

\bibitem{SaxenaGI17}
\begin{bchapter}
\bauthor{\bsnm{Saxena}, \binits{A.}},
\bauthor{\bsnm{Gera}, \binits{R.}},
\bauthor{\bsnm{Iyengar}, \binits{S.}}:
\bctitle{Fast estimation of closeness centrality ranking}.
In: \bbtitle{Proceedings of the 2017 IEEE/ACM ASONAM},
pp. \bfpage{80}--\blpage{85}
(\byear{2017})
\end{bchapter}
\endbibitem

\bibitem{taheri2017hellrank}
\begin{barticle}
\bauthor{\bsnm{Taheri}, \binits{S.M.}},
\bauthor{\bsnm{Mahyar}, \binits{H.}},
\bauthor{\bsnm{Firouzi}, \binits{M.}},
\bauthor{\bsnm{Ghalebi}, \binits{E.}},
\bauthor{\bsnm{Grosu}, \binits{R.}},
\bauthor{\bsnm{Movaghar}, \binits{A.}}:
\batitle{{HellRank}: a hellinger-based centrality measure for bipartite social
  networks}.
\bjtitle{Social Network Analysis and Mining}
\bvolume{7}(\bissue{1}),
\bfpage{22}
(\byear{2017})
\end{barticle}
\endbibitem

\bibitem{taheri2017extracting}
\begin{bchapter}
\bauthor{\bsnm{Taheri}, \binits{S.M.}},
\bauthor{\bsnm{Mahyar}, \binits{H.}},
\bauthor{\bsnm{Firouzi}, \binits{M.}},
\bauthor{\bsnm{Ghalebi~K}, \binits{E.}},
\bauthor{\bsnm{Grosu}, \binits{R.}},
\bauthor{\bsnm{Movaghar}, \binits{A.}}:
\bctitle{Extracting implicit social relation for social recommendation
  techniques in user rating prediction}.
In: \bbtitle{Proceedings of the 26th International Conference on World Wide Web
  Companion},
pp. \bfpage{1343}--\blpage{1351}
(\byear{2017})
\end{bchapter}
\endbibitem

\bibitem{wehmuth2012distributed}
\begin{bchapter}
\bauthor{\bsnm{Wehmuth}, \binits{K.}},
\bauthor{\bsnm{Ziviani}, \binits{A.}}:
\bctitle{Distributed assessment of the closeness centrality ranking in complex
  networks}.
In: \bbtitle{Simp. Comp. Net. for Pract.}
(\byear{2012})
\end{bchapter}
\endbibitem

\bibitem{you2017distributed}
\begin{barticle}
\bauthor{\bsnm{You}, \binits{K.}},
\bauthor{\bsnm{Tempo}, \binits{R.}},
\bauthor{\bsnm{Qiu}, \binits{L.}}:
\batitle{Distributed algorithms for computation of centrality measures in
  complex networks}.
\bjtitle{IEEE TAC}
\bvolume{62}(\bissue{5}),
\bfpage{2080}--\blpage{2094}
(\byear{2017})
\end{barticle}
\endbibitem

\bibitem{mahyar2018closeness}
\begin{bchapter}
\bauthor{\bsnm{Mahyar}, \binits{H.}},
\bauthor{\bsnm{Hasheminezhad}, \binits{R.}},
\bauthor{\bsnm{Ghalebi}, \binits{E.}},
\bauthor{\bsnm{Grosu}, \binits{R.}},
\bauthor{\bsnm{Stanley}, \binits{H.E.}}:
\bctitle{A compressive sensing framework for distributed detection of high
  closeness centrality nodes in networks}.
In: \bbtitle{International Conference on Complex Networks and Their
  Applications},
pp. \bfpage{91}--\blpage{103}
(\byear{2018})
\end{bchapter}
\endbibitem

\bibitem{xu2011related}
\begin{bchapter}
\bauthor{\bsnm{Xu}, \binits{W.}},
\bauthor{\bsnm{Mallada}, \binits{E.}},
\bauthor{\bsnm{Tang}, \binits{A.}}:
\bctitle{Compressive sensing over graphs}.
In: \bbtitle{IEEE INFOCOM},
pp. \bfpage{2087}--\blpage{2095}
(\byear{2011})
\end{bchapter}
\endbibitem

\bibitem{mahyar2013ucsnt}
\begin{bchapter}
\bauthor{\bsnm{Mahyar}, \binits{H.}},
\bauthor{\bsnm{Rabiee}, \binits{H.R.}},
\bauthor{\bsnm{Hashemifar}, \binits{Z.S.}}:
\bctitle{{UCS-NT: An Unbiased Compressive Sensing Framework for Network
  Tomography}}.
In: \bbtitle{IEEE ICASSP, Canada},
pp. \bfpage{4534}--\blpage{4538}
(\byear{2013})
\end{bchapter}
\endbibitem

\bibitem{Mahyar2017MLG}
\begin{bchapter}
\bauthor{\bsnm{Ghalebi}, \binits{E.}},
\bauthor{\bsnm{Mahyar}, \binits{H.}},
\bauthor{\bsnm{Grosu}, \binits{R.}},
\bauthor{\bsnm{Rabiee}, \binits{H.R.}}:
\bctitle{Compressive sampling for sparse recovery in networks}.
In: \bbtitle{Proc of the 23rd ACM SIGKDD Conference on Knowledge Discovery and
  Data Mining (KDD), 13th International Workshop on Mining and Learning with
  Graphs, Halifax, Nova Scotia, Canada},
pp. \bfpage{1}--\blpage{8}
(\byear{2017})
\end{bchapter}
\endbibitem

\bibitem{Mahyar2015CScomdet}
\begin{bchapter}
\bauthor{\bsnm{Mahyar}, \binits{H.}},
\bauthor{\bsnm{Rabiee}, \binits{H.R.}},
\bauthor{\bsnm{Movaghar}, \binits{A.}},
\bauthor{\bsnm{Ghalebi}, \binits{E.}},
\bauthor{\bsnm{Nazemian}, \binits{A.}}:
\bctitle{{CS-ComDet}: A compressive sensing approach for inter-community
  detection in social networks}.
In: \bbtitle{IEEE/ACM ASONAM, France},
pp. \bfpage{89}--\blpage{96}
(\byear{2015})
\end{bchapter}
\endbibitem

\bibitem{mahyar2018compressive}
\begin{barticle}
\bauthor{\bsnm{Mahyar}, \binits{H.}},
\bauthor{\bsnm{Hasheminezhad}, \binits{R.}},
\bauthor{\bsnm{Ghalebi}, \binits{E.}},
\bauthor{\bsnm{Nazemian}, \binits{A.}},
\bauthor{\bsnm{Grosu}, \binits{R.}},
\bauthor{\bsnm{Movaghar}, \binits{A.}},
\bauthor{\bsnm{Rabiee}, \binits{H.R.}}:
\batitle{Compressive sensing of high betweenness centrality nodes in networks}.
\bjtitle{Physica A: Statistical Mechanics and its Applications}
\bvolume{497},
\bfpage{166}--\blpage{184}
(\byear{2018})
\end{barticle}
\endbibitem

\bibitem{Mahyar2017ICML}
\begin{bchapter}
\bauthor{\bsnm{Mahyar}, \binits{H.}},
\bauthor{\bsnm{Ghalebi}, \binits{E.}},
\bauthor{\bsnm{Rabiee}, \binits{H.}},
\bauthor{\bsnm{Grosu}, \binits{R.}}:
\bctitle{The bottlenecks in biological networks}.
In: \bbtitle{Proc of the 34th International Conference on Machine Learning
  (ICML), Computational Biology Workshop, Sydney, Australia},
pp. \bfpage{1}--\blpage{5}
(\byear{2017})
\end{bchapter}
\endbibitem

\bibitem{wang2012related}
\begin{bchapter}
\bauthor{\bsnm{Wang}, \binits{M.}},
\bauthor{\bsnm{Xu}, \binits{W.}},
\bauthor{\bsnm{Mallada}, \binits{E.}},
\bauthor{\bsnm{Tang}, \binits{A.k.}}:
\bctitle{Sparse recovery with graph constraints: Fundamental limits and
  measurement construction}.
In: \bbtitle{IEEE INFOCOM},
pp. \bfpage{1871}--\blpage{1879}
(\byear{2012})
\end{bchapter}
\endbibitem

\bibitem{middya2017compressive}
\begin{barticle}
\bauthor{\bsnm{Middya}, \binits{R.}},
\bauthor{\bsnm{Chakravarty}, \binits{N.}},
\bauthor{\bsnm{Naskar}, \binits{M.K.}}:
\batitle{Compressive sensing in wireless sensor networks--a survey}.
\bjtitle{IETE technical review}
\bvolume{34}(\bissue{6}),
\bfpage{642}--\blpage{654}
(\byear{2017})
\end{barticle}
\endbibitem

\bibitem{Mahyar2015TopK}
\begin{bchapter}
\bauthor{\bsnm{Mahyar}, \binits{H.}}:
\bctitle{Detection of top-k central nodes in social networks: A compressive
  sensing approach}.
In: \bbtitle{IEEE/ACM ASONAM, Paris, France},
pp. \bfpage{902}--\blpage{909}
(\byear{2015})
\end{bchapter}
\endbibitem

\bibitem{Mahyar2015LSRweighted}
\begin{bchapter}
\bauthor{\bsnm{Mahyar}, \binits{H.}},
\bauthor{\bsnm{Rabiee}, \binits{H.R.}},
\bauthor{\bsnm{Movaghar}, \binits{A.}},
\bauthor{\bsnm{Hasheminezhad}, \binits{R.}},
\bauthor{\bsnm{Ghalebi}, \binits{E.}},
\bauthor{\bsnm{Nazemian}, \binits{A.}}:
\bctitle{A low-cost sparse recovery framework for weighted networks under
  compressive sensing}.
In: \bbtitle{IEEE SocialCom, Chengdu, China},
pp. \bfpage{183}--\blpage{190}
(\byear{2015})
\end{bchapter}
\endbibitem

\bibitem{grosu2018compressed}
\begin{bchapter}
\bauthor{\bsnm{Grosu}, \binits{R.}},
\bauthor{\bsnm{Ghalebi}, \binits{E.}},
\bauthor{\bsnm{Movaghar}, \binits{A.}},
\bauthor{\bsnm{Mahyar}, \binits{H.}}:
\bctitle{Compressed sensing in cyber physical social systems}.
In: \bbtitle{Principles of Modeling},
pp. \bfpage{287}--\blpage{305}
(\byear{2018})
\end{bchapter}
\endbibitem

\bibitem{mahyar2018dicenod}
\begin{barticle}
\bauthor{\bsnm{Mahyar}, \binits{H.}},
\bauthor{\bsnm{Hasheminezhad}, \binits{R.}},
\bauthor{\bsnm{Ghalebi}, \binits{E.}},
\bauthor{\bsnm{Nazemian}, \binits{A.}},
\bauthor{\bsnm{Grosu}, \binits{R.}},
\bauthor{\bsnm{Movaghar}, \binits{A.}},
\bauthor{\bsnm{Rabiee}, \binits{H.R.}}:
\batitle{Identifying central nodes for information flow in social networks
  using compressive sensing}.
\bjtitle{Social Network Analysis and Mining}
\bvolume{8}(\bissue{1}),
\bfpage{33}
(\byear{2018})
\end{barticle}
\endbibitem

\bibitem{wang2015distributed}
\begin{bchapter}
\bauthor{\bsnm{Wang}, \binits{W.}},
\bauthor{\bsnm{Tang}, \binits{C.Y.}}:
\bctitle{Distributed estimation of closeness centrality}.
In: \bbtitle{Decision and Control (CDC), 2015 IEEE 54th Annual Conference On},
pp. \bfpage{4860}--\blpage{4865}
(\byear{2015})
\end{bchapter}
\endbibitem

\bibitem{Kim2012WeightedVol}
\begin{bchapter}
\bauthor{\bsnm{Kim}, \binits{H.}},
\bauthor{\bsnm{Yoneki}, \binits{E.}}:
\bctitle{Influential neighbours selection for information diffusion in online
  social networks}.
In: \bbtitle{ICCCN},
pp. \bfpage{1}--\blpage{7}
(\byear{2012})
\end{bchapter}
\endbibitem

\bibitem{ref:tore1}
\begin{barticle}
\bauthor{\bsnm{Opsahl}, \binits{T.}},
\bauthor{\bsnm{Panzarasa}, \binits{P.}}:
\batitle{Clustering in weighted networks}.
\bjtitle{Soc Net}
\bvolume{31}(\bissue{2}),
\bfpage{155}--\blpage{163}
(\byear{2009})
\end{barticle}
\endbibitem

\bibitem{GephiTwitterDataset}
\begin{bchapter}
\bauthor{\bsnm{Twitter}}:
\bctitle{Gephi platform}.
In: \bbtitle{Http://rankinfo.pkqs.net/twittercrawl.dot.gz}
(\byear{2018})
\end{bchapter}
\endbibitem

\bibitem{Lescovec2007dataset}
\begin{barticle}
\bauthor{\bsnm{Leskovec}, \binits{J.}},
\bauthor{\bsnm{Kleinberg}, \binits{J.}},
\bauthor{\bsnm{Faloutsos}, \binits{C.}}:
\batitle{Graph evolution: Densification and shrinking diameters}.
\bjtitle{ACM TKDD}
\bvolume{1}(\bissue{1}),
\bfpage{2}
(\byear{2007})
\end{barticle}
\endbibitem

\bibitem{leskovec2009community}
\begin{barticle}
\bauthor{\bsnm{Leskovec}, \binits{J.}},
\bauthor{\bsnm{Lang}, \binits{K.J.}},
\bauthor{\bsnm{Dasgupta}, \binits{A.}},
\bauthor{\bsnm{Mahoney}, \binits{M.W.}}:
\batitle{Community structure in large networks: Natural cluster sizes and the
  absence of large well-defined clusters}.
\bjtitle{Internet Mathematics}
\bvolume{6}(\bissue{1}),
\bfpage{29}--\blpage{123}
(\byear{2009})
\end{barticle}
\endbibitem

\bibitem{yang2015defining}
\begin{barticle}
\bauthor{\bsnm{Yang}, \binits{J.}},
\bauthor{\bsnm{Leskovec}, \binits{J.}}:
\batitle{Defining and evaluating network communities based on ground-truth}.
\bjtitle{Knowledge and Information Systems}
\bvolume{42}(\bissue{1}),
\bfpage{181}--\blpage{213}
(\byear{2015})
\end{barticle}
\endbibitem

\bibitem{Leskovec2010WikiVote}
\begin{bchapter}
\bauthor{\bsnm{Leskovec}, \binits{J.}},
\bauthor{\bsnm{Huttenlocher}, \binits{D.}},
\bauthor{\bsnm{Kleinberg}, \binits{J.}}:
\bctitle{Predicting positive and negative links in online social networks}.
In: \bbtitle{WWW},
pp. \bfpage{641}--\blpage{650}
(\byear{2010})
\end{bchapter}
\endbibitem

\bibitem{ref:ba}
\begin{barticle}
\bauthor{\bsnm{Barabasi}, \binits{A.L.}},
\bauthor{\bsnm{Albert}, \binits{R.}}:
\batitle{Emregence of scaling in random networks}.
\bjtitle{Science}
\bvolume{286}(\bissue{5439}),
\bfpage{509}--\blpage{512}
(\byear{1999})
\end{barticle}
\endbibitem

\bibitem{ref:er}
\begin{bchapter}
\bauthor{\bsnm{Erdos}, \binits{P.}},
\bauthor{\bsnm{Renyi}, \binits{A.}}:
\bctitle{On the evolution of random graphs}.
In: \bbtitle{Publication of the Mathematical Institute of the Hungarian Academy
  of Science},
pp. \bfpage{17}--\blpage{61}
(\byear{1960})
\end{bchapter}
\endbibitem

\bibitem{ref:watts}
\begin{barticle}
\bauthor{\bsnm{Watts}, \binits{D.J.}},
\bauthor{\bsnm{Strogatz}, \binits{S.H.}}:
\batitle{Collective dynamics of small-world networks}.
\bjtitle{Nature}
\bvolume{393}(\bissue{6684}),
\bfpage{440}--\blpage{442}
(\byear{1998})
\end{barticle}
\endbibitem

\bibitem{POGS}
\begin{bchapter}
\bauthor{\bsnm{POGS}}:
\bctitle{Proximal operator graph solver}.
In: \bbtitle{Http://foges.github.io/pogs/}
(\byear{2018})
\end{bchapter}
\endbibitem

\bibitem{parikh2014block}
\begin{barticle}
\bauthor{\bsnm{Parikh}, \binits{N.}},
\bauthor{\bsnm{Boyd}, \binits{S.}}:
\batitle{Block splitting for distributed optimization}.
\bjtitle{Mathematical Programming Computation}
\bvolume{6}(\bissue{1}),
\bfpage{77}--\blpage{102}
(\byear{2014})
\end{barticle}
\endbibitem

\bibitem{benesty2009pearson}
\begin{botherref}
\oauthor{\bsnm{Benesty}, \binits{J.}},
\oauthor{\bsnm{Chen}, \binits{J.}},
\oauthor{\bsnm{Huang}, \binits{Y.}},
\oauthor{\bsnm{Cohen}, \binits{I.}}:
Pearson correlation coefficient.
Noise reduction in speech processing,
1--4
(2009)
\end{botherref}
\endbibitem

\bibitem{Schoch2015Posit-34821}
\begin{botherref}
\oauthor{\bsnm{Schoch}, \binits{D.}}:
A positional approach for network centrality.
PhD thesis,
Universität Konstanz,
Konstanz
(2015)
\end{botherref}
\endbibitem

\bibitem{mahyar2013ucswn}
\begin{bchapter}
\bauthor{\bsnm{Mahyar}, \binits{H.}},
\bauthor{\bsnm{Rabiee}, \binits{H.R.}},
\bauthor{\bsnm{Hashemifar}, \binits{Z.S.}},
\bauthor{\bsnm{Siyari}, \binits{P.}}:
\bctitle{{UCS-WN: An Unbiased Compressive Sensing Framework for Weighted
  Networks}}.
In: \bbtitle{CISS, USA}
(\byear{2013})
\end{bchapter}
\endbibitem

\end{thebibliography}

\newcommand{\BMCxmlcomment}[1]{}

\BMCxmlcomment{

<refgrp>

<bibl id="B1">
  <title><p>Fast estimation of closeness centrality ranking</p></title>
  <aug>
    <au><snm>Saxena</snm><fnm>A</fnm></au>
    <au><snm>Gera</snm><fnm>R</fnm></au>
    <au><snm>Iyengar</snm><fnm>SRS</fnm></au>
  </aug>
  <source>Proceedings of the 2017 IEEE/ACM ASONAM</source>
  <pubdate>2017</pubdate>
  <fpage>80</fpage>
  <lpage>-85</lpage>
</bibl>

<bibl id="B2">
  <title><p>{HellRank}: a Hellinger-based centrality measure for bipartite
  social networks</p></title>
  <aug>
    <au><snm>Taheri</snm><fnm>SM</fnm></au>
    <au><snm>Mahyar</snm><fnm>H</fnm></au>
    <au><snm>Firouzi</snm><fnm>M</fnm></au>
    <au><snm>Ghalebi</snm><fnm>E</fnm></au>
    <au><snm>Grosu</snm><fnm>R</fnm></au>
    <au><snm>Movaghar</snm><fnm>A</fnm></au>
  </aug>
  <source>Social Network Analysis and Mining</source>
  <pubdate>2017</pubdate>
  <volume>7</volume>
  <issue>1</issue>
  <fpage>22</fpage>
</bibl>

<bibl id="B3">
  <title><p>Extracting implicit social relation for social recommendation
  techniques in user rating prediction</p></title>
  <aug>
    <au><snm>Taheri</snm><fnm>SM</fnm></au>
    <au><snm>Mahyar</snm><fnm>H</fnm></au>
    <au><snm>Firouzi</snm><fnm>M</fnm></au>
    <au><snm>Ghalebi K</snm><fnm>E</fnm></au>
    <au><snm>Grosu</snm><fnm>R</fnm></au>
    <au><snm>Movaghar</snm><fnm>A</fnm></au>
  </aug>
  <source>Proceedings of the 26th International Conference on World Wide Web
  Companion</source>
  <pubdate>2017</pubdate>
  <fpage>1343</fpage>
  <lpage>-1351</lpage>
</bibl>

<bibl id="B4">
  <title><p>Distributed assessment of the closeness centrality ranking in
  complex networks</p></title>
  <aug>
    <au><snm>Wehmuth</snm><fnm>K</fnm></au>
    <au><snm>Ziviani</snm><fnm>A</fnm></au>
  </aug>
  <source>Simp. Comp. Net. for Pract.</source>
  <pubdate>2012</pubdate>
</bibl>

<bibl id="B5">
  <title><p>Distributed algorithms for computation of centrality measures in
  complex networks</p></title>
  <aug>
    <au><snm>You</snm><fnm>K</fnm></au>
    <au><snm>Tempo</snm><fnm>R</fnm></au>
    <au><snm>Qiu</snm><fnm>L</fnm></au>
  </aug>
  <source>IEEE TAC</source>
  <pubdate>2017</pubdate>
  <volume>62</volume>
  <issue>5</issue>
  <fpage>2080</fpage>
  <lpage>-2094</lpage>
</bibl>

<bibl id="B6">
  <title><p>A Compressive Sensing Framework for Distributed Detection of High
  Closeness Centrality Nodes in Networks</p></title>
  <aug>
    <au><snm>Mahyar</snm><fnm>H</fnm></au>
    <au><snm>Hasheminezhad</snm><fnm>R</fnm></au>
    <au><snm>Ghalebi</snm><fnm>E</fnm></au>
    <au><snm>Grosu</snm><fnm>R</fnm></au>
    <au><snm>Stanley</snm><fnm>HE</fnm></au>
  </aug>
  <source>International Conference on Complex Networks and their
  Applications</source>
  <pubdate>2018</pubdate>
  <fpage>91</fpage>
  <lpage>-103</lpage>
</bibl>

<bibl id="B7">
  <title><p>Compressive sensing over graphs</p></title>
  <aug>
    <au><snm>Xu</snm><fnm>W.</fnm></au>
    <au><snm>Mallada</snm><fnm>E.</fnm></au>
    <au><snm>Tang</snm><fnm>A.</fnm></au>
  </aug>
  <source>IEEE INFOCOM</source>
  <pubdate>2011</pubdate>
  <fpage>2087</fpage>
  <lpage>-2095</lpage>
</bibl>

<bibl id="B8">
  <title><p>{UCS-NT: An Unbiased Compressive Sensing Framework for Network
  Tomography}</p></title>
  <aug>
    <au><snm>Mahyar</snm><fnm>H.</fnm></au>
    <au><snm>Rabiee</snm><fnm>H. R.</fnm></au>
    <au><snm>Hashemifar</snm><fnm>Z. S.</fnm></au>
  </aug>
  <source>IEEE ICASSP, Canada</source>
  <pubdate>2013</pubdate>
  <fpage>4534</fpage>
  <lpage>-4538</lpage>
</bibl>

<bibl id="B9">
  <title><p>Compressive sampling for sparse recovery in networks</p></title>
  <aug>
    <au><snm>Ghalebi</snm><fnm>E</fnm></au>
    <au><snm>Mahyar</snm><fnm>H</fnm></au>
    <au><snm>Grosu</snm><fnm>R</fnm></au>
    <au><snm>Rabiee</snm><fnm>HR</fnm></au>
  </aug>
  <source>Proc of the 23rd ACM SIGKDD conference on knowledge discovery and
  data mining (KDD), 13th international workshop on mining and learning with
  graphs, Halifax, Nova Scotia, Canada</source>
  <pubdate>2017</pubdate>
  <fpage>1</fpage>
  <lpage>-8</lpage>
</bibl>

<bibl id="B10">
  <title><p>{CS-ComDet}: A Compressive Sensing Approach for Inter-Community
  Detection in Social Networks</p></title>
  <aug>
    <au><snm>Mahyar</snm><fnm>H.</fnm></au>
    <au><snm>Rabiee</snm><fnm>H. R.</fnm></au>
    <au><snm>Movaghar</snm><fnm>A.</fnm></au>
    <au><snm>Ghalebi</snm><fnm>E.</fnm></au>
    <au><snm>Nazemian</snm><fnm>A.</fnm></au>
  </aug>
  <source>IEEE/ACM ASONAM, France</source>
  <pubdate>2015</pubdate>
  <fpage>89</fpage>
  <lpage>-96</lpage>
</bibl>

<bibl id="B11">
  <title><p>Compressive sensing of high betweenness centrality nodes in
  networks</p></title>
  <aug>
    <au><snm>Mahyar</snm><fnm>H</fnm></au>
    <au><snm>Hasheminezhad</snm><fnm>R</fnm></au>
    <au><snm>Ghalebi</snm><fnm>E</fnm></au>
    <au><snm>Nazemian</snm><fnm>A</fnm></au>
    <au><snm>Grosu</snm><fnm>R</fnm></au>
    <au><snm>Movaghar</snm><fnm>A</fnm></au>
    <au><snm>Rabiee</snm><fnm>HR</fnm></au>
  </aug>
  <source>Physica A: Statistical Mechanics and its Applications</source>
  <publisher>Elsevier</publisher>
  <pubdate>2018</pubdate>
  <volume>497</volume>
  <fpage>166</fpage>
  <lpage>-184</lpage>
</bibl>

<bibl id="B12">
  <title><p>The bottlenecks in biological networks</p></title>
  <aug>
    <au><snm>Mahyar</snm><fnm>H</fnm></au>
    <au><snm>Ghalebi</snm><fnm>E</fnm></au>
    <au><snm>Rabiee</snm><fnm>HR</fnm></au>
    <au><snm>Grosu</snm><fnm>R</fnm></au>
  </aug>
  <source>Proc of the 34th international conference on machine learning (ICML),
  Computational Biology Workshop, Sydney, Australia</source>
  <pubdate>2017</pubdate>
  <fpage>1</fpage>
  <lpage>-5</lpage>
</bibl>

<bibl id="B13">
  <title><p>Sparse recovery with graph constraints: Fundamental limits and
  measurement construction</p></title>
  <aug>
    <au><snm>Wang</snm><fnm>M.</fnm></au>
    <au><snm>Xu</snm><fnm>W.</fnm></au>
    <au><snm>Mallada</snm><fnm>E.</fnm></au>
    <au><snm>Tang</snm><fnm>A</fnm></au>
  </aug>
  <source>IEEE INFOCOM</source>
  <pubdate>2012</pubdate>
  <fpage>1871</fpage>
  <lpage>-1879</lpage>
</bibl>

<bibl id="B14">
  <title><p>Compressive sensing in wireless sensor networks--a
  survey</p></title>
  <aug>
    <au><snm>Middya</snm><fnm>R</fnm></au>
    <au><snm>Chakravarty</snm><fnm>N</fnm></au>
    <au><snm>Naskar</snm><fnm>MK</fnm></au>
  </aug>
  <source>IETE technical review</source>
  <pubdate>2017</pubdate>
  <volume>34</volume>
  <issue>6</issue>
  <fpage>642</fpage>
  <lpage>-654</lpage>
</bibl>

<bibl id="B15">
  <title><p>Detection of Top-K Central Nodes in Social Networks: A Compressive
  Sensing Approach</p></title>
  <aug>
    <au><snm>Mahyar</snm><fnm>H.</fnm></au>
  </aug>
  <source>IEEE/ACM ASONAM, Paris, France</source>
  <pubdate>2015</pubdate>
  <fpage>902</fpage>
  <lpage>-909</lpage>
</bibl>

<bibl id="B16">
  <title><p>A Low-cost Sparse Recovery Framework for Weighted Networks under
  Compressive Sensing</p></title>
  <aug>
    <au><snm>Mahyar</snm><fnm>H.</fnm></au>
    <au><snm>Rabiee</snm><fnm>H. R.</fnm></au>
    <au><snm>Movaghar</snm><fnm>A.</fnm></au>
    <au><snm>Hasheminezhad</snm><fnm>R.</fnm></au>
    <au><snm>Ghalebi</snm><fnm>E.</fnm></au>
    <au><snm>Nazemian</snm><fnm>A.</fnm></au>
  </aug>
  <source>IEEE SocialCom, Chengdu, China</source>
  <pubdate>2015</pubdate>
  <fpage>183</fpage>
  <lpage>-190</lpage>
</bibl>

<bibl id="B17">
  <title><p>Compressed Sensing in Cyber Physical Social Systems</p></title>
  <aug>
    <au><snm>Grosu</snm><fnm>R</fnm></au>
    <au><snm>Ghalebi</snm><fnm>E</fnm></au>
    <au><snm>Movaghar</snm><fnm>A</fnm></au>
    <au><snm>Mahyar</snm><fnm>H</fnm></au>
  </aug>
  <source>Principles of Modeling</source>
  <pubdate>2018</pubdate>
  <fpage>287</fpage>
  <lpage>-305</lpage>
</bibl>

<bibl id="B18">
  <title><p>Identifying central nodes for information flow in social networks
  using compressive sensing</p></title>
  <aug>
    <au><snm>Mahyar</snm><fnm>H</fnm></au>
    <au><snm>Hasheminezhad</snm><fnm>R</fnm></au>
    <au><snm>Ghalebi</snm><fnm>E</fnm></au>
    <au><snm>Nazemian</snm><fnm>A</fnm></au>
    <au><snm>Grosu</snm><fnm>R</fnm></au>
    <au><snm>Movaghar</snm><fnm>A</fnm></au>
    <au><snm>Rabiee</snm><fnm>HR</fnm></au>
  </aug>
  <source>Social Network Analysis and Mining</source>
  <pubdate>2018</pubdate>
  <volume>8</volume>
  <issue>1</issue>
  <fpage>33</fpage>
</bibl>

<bibl id="B19">
  <title><p>Distributed estimation of closeness centrality</p></title>
  <aug>
    <au><snm>Wang</snm><fnm>W</fnm></au>
    <au><snm>Tang</snm><fnm>CY</fnm></au>
  </aug>
  <source>Decision and Control (CDC), 2015 IEEE 54th Annual Conference
  on</source>
  <pubdate>2015</pubdate>
  <fpage>4860</fpage>
  <lpage>-4865</lpage>
</bibl>

<bibl id="B20">
  <title><p>Influential neighbours selection for information diffusion in
  online social networks</p></title>
  <aug>
    <au><snm>Kim</snm><fnm>H.</fnm></au>
    <au><snm>Yoneki</snm><fnm>E</fnm></au>
  </aug>
  <source>ICCCN</source>
  <pubdate>2012</pubdate>
  <fpage>1</fpage>
  <lpage>-7</lpage>
</bibl>

<bibl id="B21">
  <title><p>Clustering in weighted networks</p></title>
  <aug>
    <au><snm>Opsahl</snm><fnm>T.</fnm></au>
    <au><snm>Panzarasa</snm><fnm>P.</fnm></au>
  </aug>
  <source>Soc Net</source>
  <pubdate>2009</pubdate>
  <volume>31</volume>
  <issue>2</issue>
  <fpage>155</fpage>
  <lpage>-163</lpage>
</bibl>

<bibl id="B22">
  <title><p>Gephi platform</p></title>
  <aug>
    <au><cnm>Twitter</cnm></au>
  </aug>
  <source>http://rankinfo.pkqs.net/twittercrawl.dot.gz</source>
  <pubdate>2018</pubdate>
</bibl>

<bibl id="B23">
  <title><p>Graph evolution: Densification and shrinking diameters</p></title>
  <aug>
    <au><snm>Leskovec</snm><fnm>J</fnm></au>
    <au><snm>Kleinberg</snm><fnm>J</fnm></au>
    <au><snm>Faloutsos</snm><fnm>C</fnm></au>
  </aug>
  <source>ACM TKDD</source>
  <publisher>ACM</publisher>
  <pubdate>2007</pubdate>
  <volume>1</volume>
  <issue>1</issue>
  <fpage>2</fpage>
</bibl>

<bibl id="B24">
  <title><p>Community structure in large networks: Natural cluster sizes and
  the absence of large well-defined clusters</p></title>
  <aug>
    <au><snm>Leskovec</snm><fnm>J.</fnm></au>
    <au><snm>Lang</snm><fnm>K. J.</fnm></au>
    <au><snm>Dasgupta</snm><fnm>A.</fnm></au>
    <au><snm>Mahoney</snm><fnm>M. W.</fnm></au>
  </aug>
  <source>Internet Mathematics</source>
  <pubdate>2009</pubdate>
  <volume>6</volume>
  <issue>1</issue>
  <fpage>29</fpage>
  <lpage>-123</lpage>
</bibl>

<bibl id="B25">
  <title><p>Defining and evaluating network communities based on
  ground-truth</p></title>
  <aug>
    <au><snm>Yang</snm><fnm>J.</fnm></au>
    <au><snm>Leskovec</snm><fnm>J.</fnm></au>
  </aug>
  <source>Knowledge and Information Systems</source>
  <pubdate>2015</pubdate>
  <volume>42</volume>
  <issue>1</issue>
  <fpage>181</fpage>
  <lpage>-213</lpage>
</bibl>

<bibl id="B26">
  <title><p>Predicting positive and negative links in online social
  networks</p></title>
  <aug>
    <au><snm>Leskovec</snm><fnm>J.</fnm></au>
    <au><snm>Huttenlocher</snm><fnm>D.</fnm></au>
    <au><snm>Kleinberg</snm><fnm>J.</fnm></au>
  </aug>
  <source>WWW</source>
  <pubdate>2010</pubdate>
  <fpage>641</fpage>
  <lpage>-650</lpage>
</bibl>

<bibl id="B27">
  <title><p>Emregence of scaling in random networks</p></title>
  <aug>
    <au><snm>Barabasi</snm><fnm>A. L.</fnm></au>
    <au><snm>Albert</snm><fnm>R.</fnm></au>
  </aug>
  <source>Science</source>
  <pubdate>1999</pubdate>
  <volume>286</volume>
  <issue>5439</issue>
  <fpage>509</fpage>
  <lpage>-512</lpage>
</bibl>

<bibl id="B28">
  <title><p>On the evolution of random graphs</p></title>
  <aug>
    <au><snm>Erdos</snm><fnm>P.</fnm></au>
    <au><snm>Renyi</snm><fnm>A.</fnm></au>
  </aug>
  <source>Publication of the Mathematical Institute of the Hungarian Academy of
  Science</source>
  <pubdate>1960</pubdate>
  <fpage>17</fpage>
  <lpage>-61</lpage>
</bibl>

<bibl id="B29">
  <title><p>Collective dynamics of small-world networks</p></title>
  <aug>
    <au><snm>Watts</snm><fnm>D. J.</fnm></au>
    <au><snm>Strogatz</snm><fnm>S. H.</fnm></au>
  </aug>
  <source>Nature</source>
  <pubdate>1998</pubdate>
  <volume>393</volume>
  <issue>6684</issue>
  <fpage>440</fpage>
  <lpage>-442</lpage>
</bibl>

<bibl id="B30">
  <title><p>Proximal Operator Graph Solver</p></title>
  <aug>
    <au><cnm>POGS</cnm></au>
  </aug>
  <source>http://foges.github.io/pogs/</source>
  <pubdate>2018</pubdate>
</bibl>

<bibl id="B31">
  <title><p>Block splitting for distributed optimization</p></title>
  <aug>
    <au><snm>Parikh</snm><fnm>N</fnm></au>
    <au><snm>Boyd</snm><fnm>S</fnm></au>
  </aug>
  <source>Mathematical Programming Computation</source>
  <publisher>Springer</publisher>
  <pubdate>2014</pubdate>
  <volume>6</volume>
  <issue>1</issue>
  <fpage>77</fpage>
  <lpage>-102</lpage>
</bibl>

<bibl id="B32">
  <title><p>Pearson correlation coefficient</p></title>
  <aug>
    <au><snm>Benesty</snm><fnm>J.</fnm></au>
    <au><snm>Chen</snm><fnm>J.</fnm></au>
    <au><snm>Huang</snm><fnm>Y.</fnm></au>
    <au><snm>Cohen</snm><fnm>I.</fnm></au>
  </aug>
  <source>Noise reduction in speech processing</source>
  <pubdate>2009</pubdate>
  <fpage>1</fpage>
  <lpage>-4</lpage>
</bibl>

<bibl id="B33">
  <title><p>A Positional Approach for Network Centrality</p></title>
  <aug>
    <au><snm>Schoch</snm><fnm>D</fnm></au>
  </aug>
  <source>PhD thesis</source>
  <publisher>Universität Konstanz</publisher>
  <pubdate>2015</pubdate>
</bibl>

<bibl id="B34">
  <title><p>{UCS-WN: An Unbiased Compressive Sensing Framework for Weighted
  Networks}</p></title>
  <aug>
    <au><snm>Mahyar</snm><fnm>H.</fnm></au>
    <au><snm>Rabiee</snm><fnm>H. R.</fnm></au>
    <au><snm>Hashemifar</snm><fnm>Z. S.</fnm></au>
    <au><snm>Siyari</snm><fnm>P.</fnm></au>
  </aug>
  <source>CISS, USA</source>
  <pubdate>2013</pubdate>
</bibl>

</refgrp>
} 

\end{backmatter}

\end{document}